\documentclass{article}

\usepackage[dvips]{graphicx}
\usepackage{amssymb,amsfonts,amsmath}
\usepackage{xcolor}
\usepackage{cite} 
\usepackage{afterpage}
\usepackage{placeins}
\usepackage{fullpage}

\begin{document}

\title{An explanatory evo-devo model for the developmental hourglass}

\author{Saamer Akhshabi\\ School of Computer Science, Georgia Tech, GA, USA
\and Shrutii Sarda\\ Biology Department, University of Maryland, MD, USA
\and Constantine Dovrolis \\ School of Computer Science, Georgia Tech, GA, USA
\and Soojin Yi \\School of Biology, Georgia Tech, GA, USA}

\maketitle

%\begin{article}
\begin{abstract}
The ``developmental hourglass'' describes a pattern of increasing
morphological divergence towards earlier and later embryonic development, 
separated by a period of significant conservation across distant species
(the ``phylotypic stage'').
Recent studies have found evidence in support of the hourglass effect
at the genomic level. 
For instance, the phylotypic stage expresses the oldest and most 
conserved transcriptomes.
However, the regulatory mechanism that causes the hourglass pattern 
remains an open question. 
Here, we use an evolutionary model of regulatory gene interactions during 
development to identify the conditions under which the hourglass effect can emerge
in a general setting. 
The model focuses on the hierarchical gene regulatory network 
that controls the developmental process, 
and on the evolution of a population under random perturbations in the 
structure of that network.
The model predicts, under fairly general assumptions, the emergence of an hourglass 
pattern in the structure of a temporal representation of
the underlying gene regulatory network.  
The evolutionary age of the corresponding genes also follows
an hourglass pattern, with the oldest genes concentrated at the hourglass waist.
The key behind the hourglass effect is that developmental
regulators should have an increasingly specific 
function as development progresses.
Analysis of developmental gene expression profiles from {\em Drosophila melanogaster} 
and {\em Arabidopsis thaliana} provide consistent results with our theoretical predictions.
\end{abstract}

%\keywords{embryonic development | transcriptional control system | computational modeling | network science}

%\abbreviations{}

The evolutionary mechanism of conservation during embryogenesis,
and its connection to the gene regulatory networks that control development, 
are fundamental questions in systems biology 
\cite{raff1996shape, richardson2002haeckel, davidson2010regulatory}. 
Several models have been presented in the context of morphological, molecular, and 
genetic developmental patterns.
The most widely discussed model is the ``developmental hourglass,'' 
which places the strongest conservation across species in the ``phylotypic stage.''  
The first observations supporting the hourglass model go back to von Baer 
when he noticed that there exists a mid-developmental stage in which 
embryos of different animals look similar 
\cite{duboule1994temporal}.  
On the other hand, the ``developmental funnel'' model of 
conservation predicts increasing diversification 
as development progresses \cite{rasmussen1987new, roux2008developmental}. 

Recently, the hourglass model has come under new light. 
Multiple studies have observed the hourglass pattern across diverse biological processes, 
including transcriptome divergence 
\cite{irie2011comparative, irie2007vertebrate, kalinka2010gene, levin2012developmental, quint2012transcriptomic}, 
transcriptome age \cite{domazet2010phylogenetically, hazkani2005search, irie2011comparative}, 
molecular interaction \cite{galis2001testing}, and evolutionary selective constraints 
\cite{cruickshank2008microevolutionary, galis2001testing, levin2012developmental}. 
Despite these observations the genomic basis and even the existence of the
developmental hourglass effect have been the subject of an intense debate 
\cite{comte2010molecular, hall1997phylotypic, hazkani2005search, kalinka2012evolution, piasecka2013hourglass, raff1996shape, richardson1998phylotypic, roux2008developmental, rp2003inverting}. 
More importantly, the underlying mechanism that can shape the 
developmental process in the hourglass or funnel forms is still unknown. 

We aim to understand 
the conditions under which the hourglass effect can emerge
in a general setting, based on an abstract model 
for the evolution of embryonic development.
The model focuses on a hierarchical network that represents 
the temporal ``execution'' of the underlying Gene Regulatory Network (GRN)
during development.
Each layer of the network corresponds to a developmental stage.
The nodes at each layer represent regulatory genes (i.e., genes
encoding transcription factors or signaling molecules) that undergo
significant activity change at that corresponding stage.
The edges from genes at one layer to genes at the next layer represent
regulatory interactions that cause those activity changes.
We refer to this hierarchical network as Developmental Gene Execution Network (DGEN)
to distinguish it from the corresponding GRN.
A DGEN is subject to evolutionary perturbations (e.g., gene 
deletions, rewiring, duplication) that may be lethal, or that may impede 
development, for the corresponding organism.

The model predicts that the evolutionary process shapes the DGENs of
a population in the form of an hourglass, under fairly general assumptions.
Specifically, the number of genes at each developmental stage
follows an hourglass pattern, with the smallest number at the ``waist'' of the hourglass.
The main condition for the appearance of the hourglass pattern 
is that the DGEN should gradually get sparser as development progresses,
with general-purpose regulatory genes at the earlier developmental stages 
and highly specialized regulatory genes at the later stages. 
Another model prediction is that the evolutionary age of DGEN genes
also follows an hourglass pattern, with the oldest genes concentrated at the waist. 

We have examined the aforementioned predictions using transcriptome data
from the development of {\em Drosophila melanogaster} and {\em Arabidopsis thaliana}. 
This data is insufficient to reconstruct the complete DGEN of these species but it allows
to estimate the number of genes at each developmental stage, given an 
activity variation threshold.
Under a wide range of this threshold, the inferred DGEN shape 
follows an hourglass pattern, the waist of that hourglass roughly 
coincides with the previously reported phylotypic stage for these species,
and the age of the corresponding genes follows the predicted hourglass pattern. 

%We believe that the DevoNet model provides new insights into understanding 
%and interrogating the evolution of hierarchical gene interactions that are 
%fundamental to biological processes such as development. 

\begin{figure*}[t]
\centering
\includegraphics[width=160mm,height=40mm,angle=0]{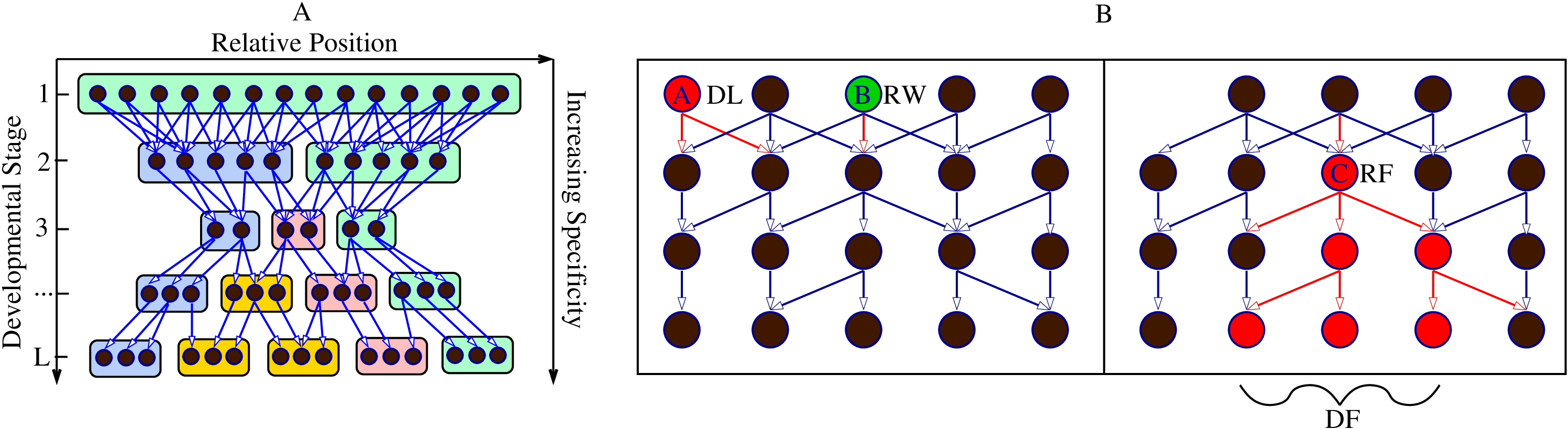}
\caption{(A) An abstract DGEN.
The circles denote state-transitioning genes,
edges represent directed regulatory interactions, 
and colored boxes refer to spatial domains that form during development.
If regulatory genes become increasingly function-specific as development
progresses, the network gradually becomes sparser in that direction.
(B) Evolutionary perturbations on a DGEN's structure: 
Gene A is deleted (DL), while gene B is rewired (RW) losing an outgoing edge. 
This RW event causes the regulatory failure (RF) of gene C, which then 
causes a cascade of five more RF events. This cascade causes developmental failure (DF). 
Note that the removal of some upstream regulators does not always cause an RF event 
(e.g., genes regulated by A). 
}
\label{fig:embryo}
\end{figure*}

\vspace{0.1in}
\noindent {\bf Developmental Gene Execution Networks}\\
As a first-order approximation, a regulatory gene can be modeled
in one of several discrete {\em functional states} \cite{peter2012predictive}.
In the simplest case, a regulatory gene can act as a binary 
switch (``on'' and ``off'') but in general a gene may have more than two
functional states.   
The transition of a regulatory gene X from one functional state to another 
is often (but not always) caused by one or more upstream regulators of X that 
go through a functional state transition before X. 
%For example, even though a gene X may have two 
%regulators A and B, it could be that at a particular 
%point in developmental time and space 
%only gene B has transitioned from one state to another 
%causing the subsequent transition of gene X.
We use the term {\em transitioning gene} to refer to a regulatory
gene that goes through a functional state transition at a given 
developmental time anywhere at the developing embryo.

A DGEN is a directed and acyclic network; see Fig.\ref{fig:embryo}A
for an abstract example.
The vertical direction refers to developmental time, from the zygote
at the top to the developed organism at the bottom. 
In the horizontal direction we can represent different spatial
domains, even though this is not necessary and it is not done
in our model. 
For instance, the zygote at the top of the DGEN 
would be a single domain, while the organism at the bottom 
of the DGEN would have the largest number of spatial domains. 

Development is often approximated (conceptually and experimentally)
as a succession of discrete developmental stages.
The duration of a developmental stage can be thought of as the 
typical time that is required for a gene's functional state transition,
and it does not need to be the same for all stages.
Each layer of a DGEN refers to a developmental stage,
and it includes only the transitioning genes during that 
stage anywhere in the embryo.
The same gene can appear in more than one stage if 
it goes through several functional state transitions during development.
Additionally, a DGEN edge from a gene $X$ at stage $l$ to 
a gene $Y$ at stage $l$+1 implies that the functional transition
of $X$ {\em caused} the functional transition of $Y$ at the next stage.
If gene $Y$ has more than one incoming edge, its functional
state transition was caused by the coupled effect of more than one 
transitioning genes at the earlier stage.
Any upstream regulators of $Y$ that remained at the same
functional state at stage $l$ are not included in that stage of the DGEN.
%even though such regulators may be necessary for the transition of
%$Y$ they do not trigger that transition since their functional
%state has not changed.

%Metaphorically speaking, think of a network of toppling dominoes.
%The starting point may be one or more dominoes that fall at 
%the same time due to an external force. 
%At any point in time $l$, one or more dominoes topple 
%(a functional state transition) causing one or more dominoes
%to then topple at time $l$+1.
%The resulting cascade may proceed in parallel at more than one location. 
%A DGEN would capture in that case both the dominoes that fall
%at any time but also for each toppling domino 
%the set of dominoes that caused that transition.

The sequence of developmental stages is denoted by $l$=$1\dots L$. 
The set of transitioning genes at stage $l$ is $G(l)$.
A gene $g$ at stage $l$$<$$L$ regulates a set of {\em downstream} 
genes at stage $l$+1 denoted by $D(g)$ (outgoing edges from g). 
Similarly, a gene $g$ at stage $l$$>$1 is regulated by
a set of {\em upstream} genes $U(g)$ at stage $l$-1 (incoming edges to g).
The functional transitions at the first stage are
assumed to be caused by regulatory maternal genes that initiate 
the developmental process.

\vspace{0.1in}
\noindent {\bf Model Description}\\
The model captures certain aspects of both the developmental process, 
in the form of a given DGEN for each embryo,
and of the evolutionary process, as random  
perturbations in the structure of individual DGENs in a population.
%The modeling events occur either
%in developmental time (namely the success or failure
%of an embryo's development) or in evolutionary
%time (namely deletions, duplication or rewiring of genes).
The model does not need to capture the actual functional 
state transitions or the regulatory input function of each gene.
It does capture however the dynamic and stochastic
effect of structural network perturbations (gene deletion, rewiring
and duplication) on the  success of the developmental process,
as explained in the following. 

Similar to the Wright-Fisher model, we consider a population of $N$ individuals,
each represented by a DGEN. In each generation, individuals 
reproduce asexually, inheriting the DGEN of their parent.
Various evolutionary events can cause
structural changes in the DGEN of an individual
that may result in ``developmental failure.'' Such individuals
(and their DGENs) are replaced with developmentally successful
individuals so that the population size remains constant.

The model is meant to be as general as possible
and so the regulatory interactions between genes
of successive stages are determined probabilistically, as follows.
Each stage $l$ is assigned a {\em regulatory specificity}, or simply
{\em specificity} $s(l)$ with $0\leq s(l)\leq 1$.  
A gene $g$ at stage $l$ acts as upstream regulator for a gene $g'$ at
stage $l+1$ with probability $s'(l)=1-s(l)$. 
So, the specificity of a developmental stage determines how
likely it is for regulatory genes of that stage to cause
a state transition of the genes at the next stage; 
a higher specificity decreases that likelihood.

Our major assumption is that the regulatory 
specificity increases substantially as development progresses.
In other words, the DGEN becomes gradually sparser along the developmental time axis,
starting with $s(1)$$\approx$0 and ending with $s(L)$$\approx$1. 
This assumption is plausible for the 
following reasons. First, as development progresses the embryo grows
in size forming distinct spatial domains. So, extracellular 
gene regulation becomes more difficult, especially 
across different domains. Additionally, as development progresses
the transitioning genes become more organ- or tissue-specific,  
implying that their downstream interactions become sparser.  
Unfortunately, an empirical investigation of the increasing specificity
assumption requires knowledge of the complete DGEN for a given species; 
this is currently not feasible for even the most well-studied model organisms.

The DGEN structural changes we consider are gene deletions,
gene duplications, and gene rewiring:

\noindent
{\bf Deletions (DL):} This event removes a gene from the DGEN, 
including its incoming and outgoing edges. There are many 
genetic mechanisms that may cause such events.
A DL event deletes each gene of an individual and at each generation
with probability $P_{DL}$.

\noindent
{\bf Duplication (DP):} This event creates an identical copy
of a gene $g$ with the same downstream and upstream regulators
and at the same developmental stage as $g$.
The two genes may have different fates if one
of them is subject later to deletion or rewiring.
Otherwise, the two genes are considered identical.
A DP event duplicates each gene of an individual and at each generation
with probability $P_{DP}$.

\noindent
{\bf Rewiring (RW):} This event changes
the upstream and/or downstream regulators of a gene.
Changes in the upstream versus downstream regulators may have
different biological basis.
The former occur, for instance, as a result of mutations in the transcription factor 
binding sites in a gene's promoter or mutations in distal regulatory elements such as enhancers,
while the latter may be mostly caused by coding sequence mutations.
%Changes in the upstream versus downstream regulators may have 
%different biological basis.
%The former occur, for instance, as a result of mutations in the 
%binding sites in a gene's promoter or enhancer 
%or any other distal regulatory element, 
%while the latter may be caused by coding sequence mutations. 
The details of the rewiring process do not
affect the results qualitatively as long as the
average density of edges in each stage remains consistent with 
the specificity of that stage.
The specific rewiring mechanisms we use are presented next.

Suppose that a RW event affects gene $g$ at stage $l$. 
The upstream regulators of $g$ are recomputed 
based on the specificity of the previous stage, i.e., by choosing each 
distinct gene of stage $l-1$ with probability $s'(l-1)$.   
For the downstream regulators of $g$, 
we randomly remove $N_{-}$ existing outgoing edges of $g$,
and then add $N_{+}$ outgoing edges to randomly chosen genes of stage $l+1$
that $g$ is not already connected to.
Both $N_{-}$ and $N_{+}$ follow a Binomial distribution
with $|D(g)|$ trials and success probability $s'(l)$.
This captures that the downstream regulators of $g$ are
derived by incremental changes in $D(g)$, instead of 
giving $g$ a completely new network configuration (thereby, new regulatory function). 
The higher the regulatory specificity of a stage, the less likely
these incremental changes are.
An RW event rewires each gene of an individual and at each generation
with probability $P_{RW}$.

A gene deletion or rewiring event at stage $l$ can remove an  
upstream regulator from genes at stage $l+1$. A loss of incoming 
edges may trigger the {\em regulatory failure} of a gene, as 
described next. 

\noindent
{\bf Regulatory failures (RF):} A gene $g$ may not be able to 
change functional state if some of its 
upstream regulators $U(g)$ are lost due to DL or RW 
events. Even though regulatory networks are often robust to structural 
perturbations, even a partial gene loss in $U(g)$ may disable $g$
causing a {\em regulatory failure}. 
It is plausible that the probability of a regulatory failure 
increases with the fraction of lost upstream regulators.
So, if $U'(g)$ is the new set of upstream regulators
and $|U(g)|>|U'(g)|>0$, gene $g$ is removed with probability:\\
\begin{center}
$P_{RF}(r)=1-e^{\frac{-z \, r}{1-r}},\quad 0<r=1-\frac{|U'(g)|}{|U(g)|}< 1$
\end{center}
while if $|U'(g)|=0$ we set $P_{RF}(1)$=1.
$z$ is the {\em RF parameter} and it depends on the
robustness of regulatory interactions to gene loss. 
%Fig. \ref{fig:rfFunction} shows the RF probability 
%as a function of the fraction $r$ for different values of $z$.

When a DL or RW event causes one or more RF events, the latter can trigger
additional RF events in subsequent developmental stages, 
leading to {\em cascades of regulatory failures}
(see Fig.\ref{fig:embryo}B).
Such RF cascades may cause {\em developmental
failure}, meaning that the developed embryo is unable to survive or reproduce.  

\noindent
{\bf Developmental failure (DF):}
The last stage of a DGEN represents the fully developed embryo.
If that stage includes $\Gamma$ transitioning genes at a successfully 
developed embryo, 
the simplest assumption is that an individual with less than $\Gamma$ genes 
at stage-$L$ has failed to develop properly.  
Such DGENs are removed from the population and they are 
replaced with randomly chosen but successfully developed DGENs.
We have also experimented with two variations of the DF event: 
first, the individual is removed if its last stage has less than $\Gamma-\gamma$
genes, where $\gamma$ is small relative to $\Gamma$, 
and second, the probability of a DF event increases
as the number of genes at stage-$L$ decreases below $\Gamma$.
The qualitative results, as described next, 
do not change with these two model variations. 

\FloatBarrier
%\afterpage{\clearpage}
%\afterpage{\cleardoublepage}

\begin{figure*}[t]
\centering
\includegraphics[width=30mm,height=44mm,angle=270]{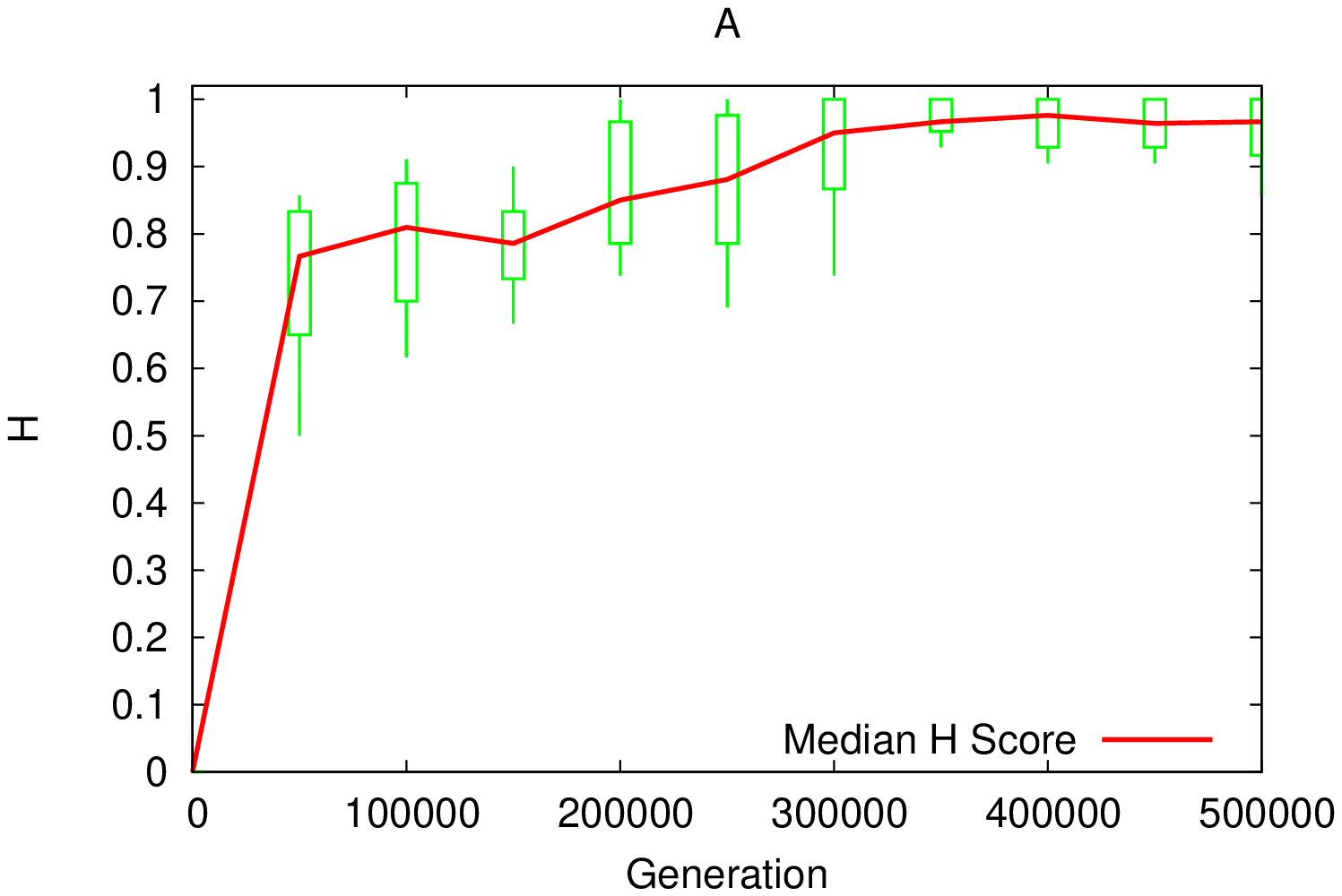}
\includegraphics[width=30mm,height=44mm,angle=270]{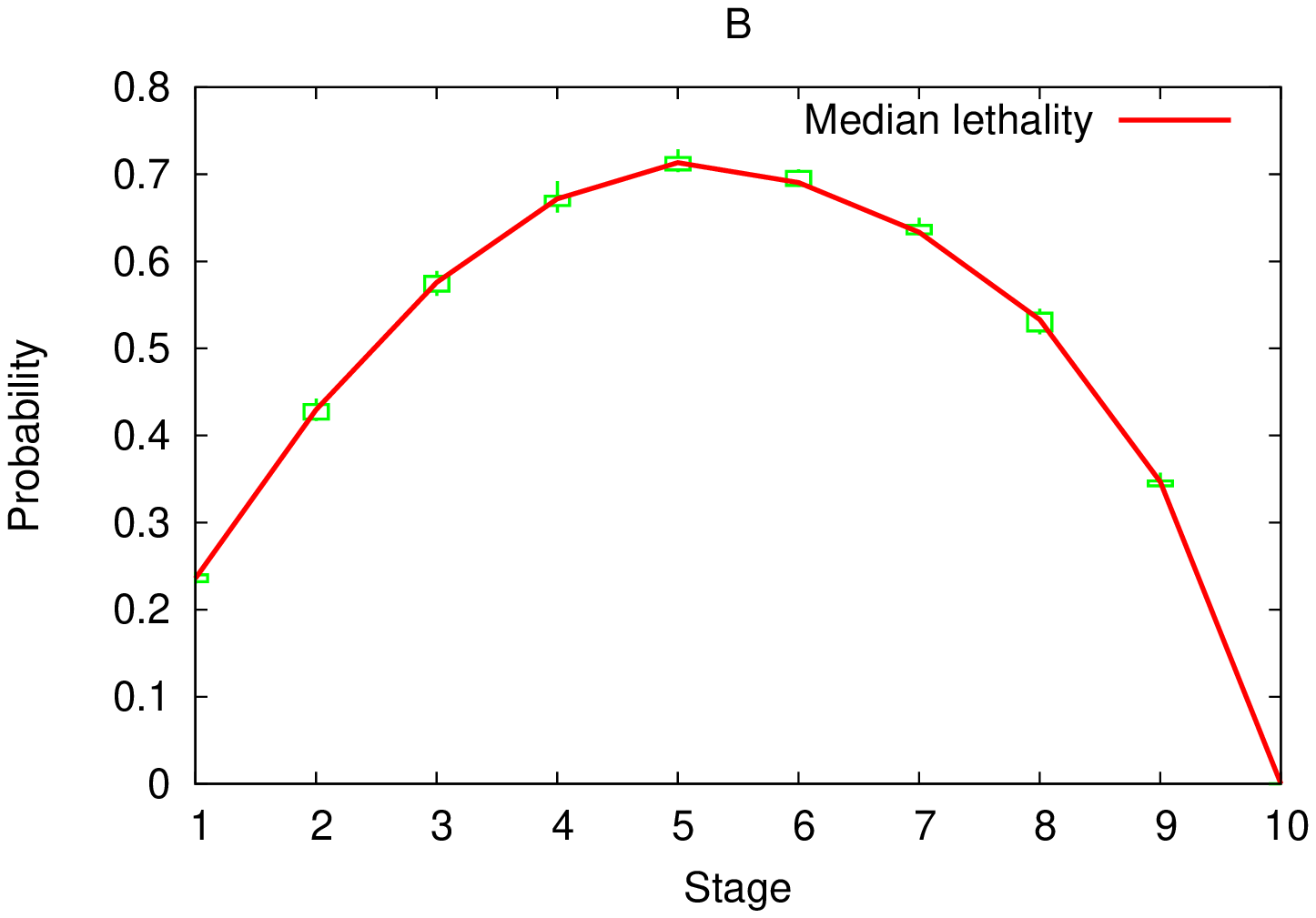}
\includegraphics[width=30mm,height=44mm,angle=270]{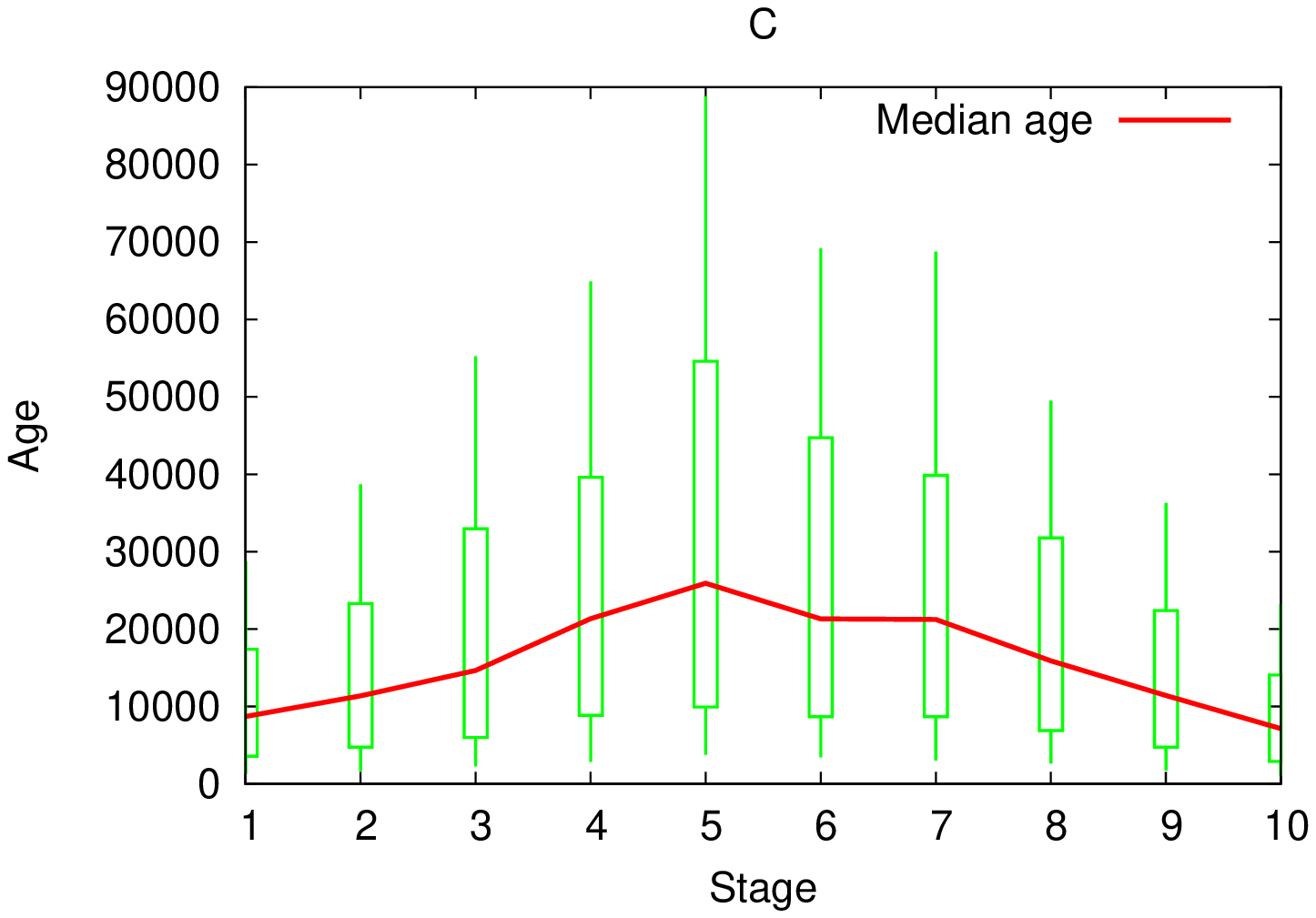}
\caption{Computational results with {\bf Model-2}.
Parameters: 10 runs with different initial populations, 
$N$=1000 individuals, $L$=10 stages, 
specificity function $s(l)$=$l/L$ ($l$=$1\dots L-1$), 
$\Gamma$=100 genes at each stage initially, 
RF parameter $z$=4, 500,000 generations, 
probability of RW event $P_{RW}$=$10^{-4}$. 
The red line is the median and the green boxes are the 10th, 25th, 75th, 
and 90th percentiles, across all individuals and all runs.
(A) The hourglass score $H$ across evolutionary time. 
(B) Lethality probability at each stage. 
(C) Age of existing genes at the last generation.}
\label{fig:main_results}
\end{figure*}

\vspace{0.1in}
\noindent {\bf Computational Results}\\
We simulate the presented model to examine the properties
of the surviving DGENs as evolutionary time progresses.  
The initial population consists of $N$ identical DGENs
with $\Gamma$ genes at each stage. 
The edges between genes are constructed probabilistically based on the 
specificity of each stage, as described previously. 
Simulating the complete model would not show
the significance of individual mechanisms such as
the increasing specificity assumption. 
For this reason we construct a sequence of four models
with increasing complexity, presenting results separately for each of them:  

\noindent
{\bf Model-1: Constant specificity.}
Each stage has the same specificity, $s(l)$=0.5 for $l=1\dots L-1$. 
Further, this model does not include gene deletion and duplication.
Gene rewiring can cause RF and DF events even if there are no DL or DP events. 
 
\noindent
{\bf Model-2: Increasing specificity.}
The difference from Model-1 is that the specificity 
is gradually increasing across developmental stages.
Unless noted otherwise, the specificity 
is linear, $s(l)=l/L$ for $l=1\dots (L-1)$;
a nonlinear specificity function in considered in the SI 
(see Fig.\ref{fig:sigmoid_specificity}). 

\noindent
{\bf Model-3: With gene duplications.}
Model-3 adds DP events in Model-2. 
The duplication probability $P_{DP}$ is set so 
that the average size of a DGEN, across the entire population, 
stays within a given range (70\%-80\% of $L \times \Gamma$ genes). 

\noindent
{\bf Model-4: With gene deletions.}
Model-4 adds DL events in Model-3 (complete model).
The deletion probability $P_{DL}$ is set so 
that the average size of a DGEN, across the entire population, 
stays within the same range as in Model-3. 

In Model-1 and Model-2, genes can be only removed (due to 
RW events, potentially followed by RF cascades) and so the 
average DGEN size decreases as evolutionary time progresses, 
which is unrealistic. 
Model-3 and Model-4 are more realistic because
they can maintain a roughly constant DGEN size in the long-term.
However, as will be shown next, all aspects of the developmental
hourglass effect can already be seen with Model-2 (but not with Model-1).
This highlights the increasing specificity assumption as the
key property behind the developmental hourglass effect.

\noindent
{\bf Hourglass shape.}
A first observation is that as evolutionary time progresses, 
DGENs acquire an ``hourglass-like'' shape in Models-2,3,4.
This means that the  width of each stage
first decreases until a certain stage
(referred to as the waist of the hourglass) and then gradually increases. 
The hourglass may not be symmetric with respect to the waist. 
To quantify this observation, we define an ``hourglass score'' $H$
(see Methods and Fig.\ref{fig:hScore}) that is equal to 1 
if the sequence of $L$ stage widths
consists of two segments: a decreasing sub-sequence of
$k\geq 2$ stages followed by an increasing sub-sequence of $L-k\geq 2$ stages. 
Fig.\ref{fig:main_results}A shows the hourglass score 
for the population of DGENs in Model-2.
Similar graphs for the three other models are shown in the SI
(Fig.\ref{fig:model-1_results}A, Fig.\ref{fig:model-3_results}A, and Fig.\ref{fig:model-4_results}A). 
The $H$ score quickly increases in the three models
that have increasing specificity, and it fluctuates close to 1 afterwards. 
%An analysis of the variations of $H$ across
%the population of DGENs shows that the hourglass shape is a property 
%of almost all surviving invividuals. The occassional drops of $H$ are due
%to RW or DL events that cause the removal of high-$H$ individuals, 
%and they become less pronounced and frequent as the 
%population size increases. 
 
What is the reason behind the hourglass shape of DGENs?
When a gene $g$ is rewired at stage $l$, it may trigger RF events
in stage $l+1$ depending on the number of its lost outgoing edges.
In the first few stages, where specificity is low, it is 
unlikely that a gene would lose a large fraction of its (typically 
many) incoming edges.
In the last few stages, where specificity is high, 
edges are unlikely to get rewired in the first place.
In the mid-stages however, where the specificity is close to 50\%, 
there is higher variability in the number of outgoing edges lost or gained due 
to RW events. 
The loss of several outgoing edges due to an RW event at stage $l$ 
can trigger several RF events and gene removals in the subsequent stage.
Thus, the probability of RF events in mid-stages 
is higher than in early/late stages, 
making the removal of genes more likely in the former.

%In the first few stages, where the specificity is low, a gene $g$ 
%typically has a large number of outgoing edges, and most of them 
%can be rewired upon an RW event. However, due to the low specificity,
%only few genes would lose a large fraction of incoming edges, and
%so the probability of RF events is low.
%On the other hand,
%in the last few stages, where the specificity is high, a gene $g$
%typically has a small number of outgoing edges, and it is unlikely
%that any of them will be rewired. So,
%an RW event at the last few stages is also unlikely to cause RF events. 
%In the mid-stages, however, where the specificity 
%is close to 50\%, the variability in the number of outgoing 
%edges that will be lost after an RW event is maximized (recall 
%that the variance of a Bernoulli trial is maximized when 
%the success probability is 50\%).  
%So, a single RW event at a stage of medium specificity
%can trigger several RF events and gene removals in the subsequent stages.
%Putting the previous three cases together we see that, even 
%though the probability of a RW event is the same at all 
%stages, the probability of RF events and gene loss due to RW 
%events is higher in intermediate stages; thus, 
%the width of those intermediate stages is decreased, giving an hourglass
%shape to the DGEN.   

The constant specificity of Model-1 does
not result in an hourglass pattern (see Fig.\ref{fig:model-1_results}A)
for the following reason. 
RW events at stage $l$ can cause RF events at the next stage  
with the same probability, independent of $l$. 
However, after the occurrence of an RF event, the size of the
potential cascade increases as $l$ decreases simply 
because there are more subsequent stages to affect.
This gives DGENs a ``funnel-like'' shape with a gradually increasing
number of genes after stage-1; $H$ fluctuates around 0.5, as expected
for an increasing sequence.

\begin{figure*}[!t]
\centering
\includegraphics[width=45mm,height=70mm,angle=270]{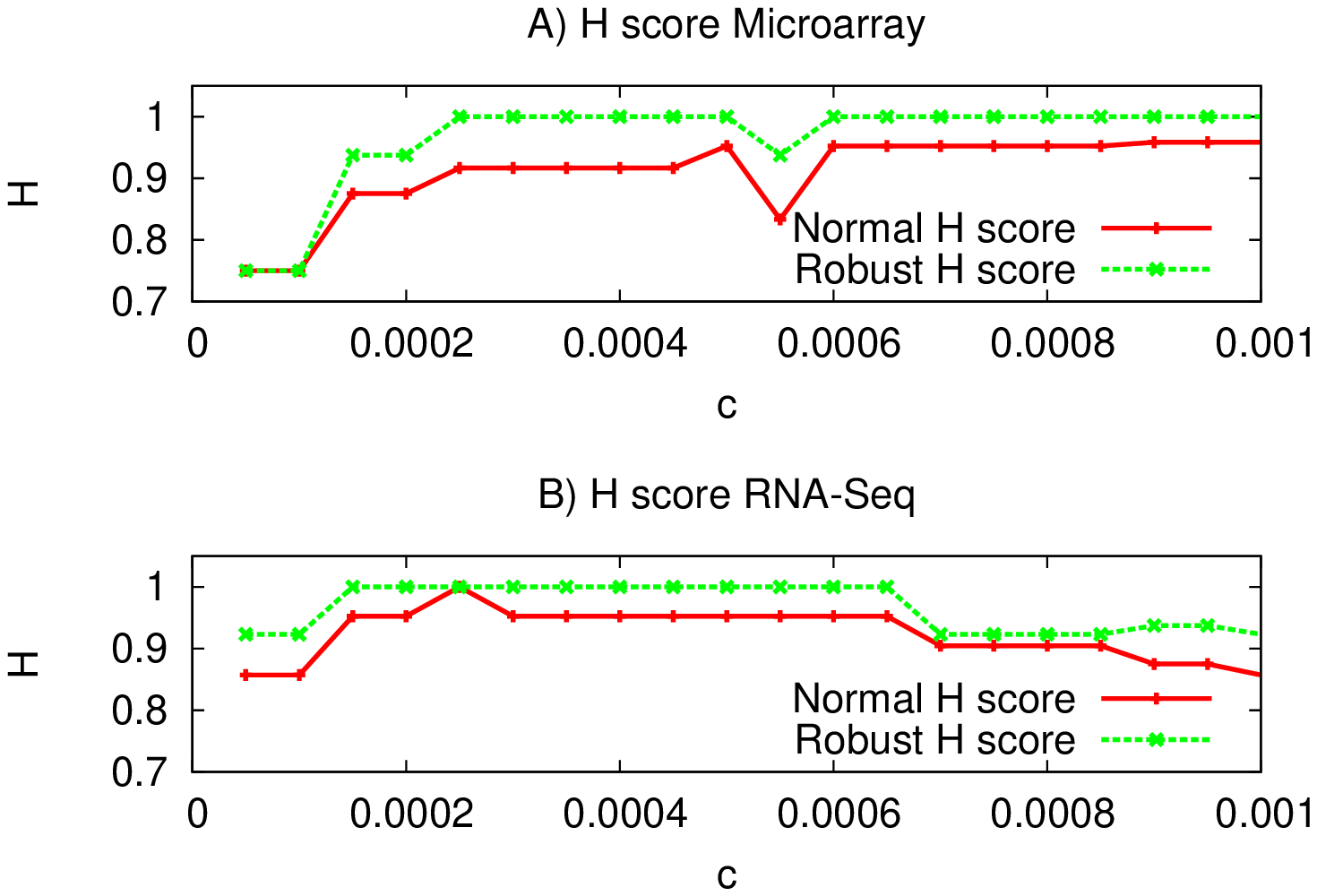}
\includegraphics[width=45mm,height=70mm,angle=270]{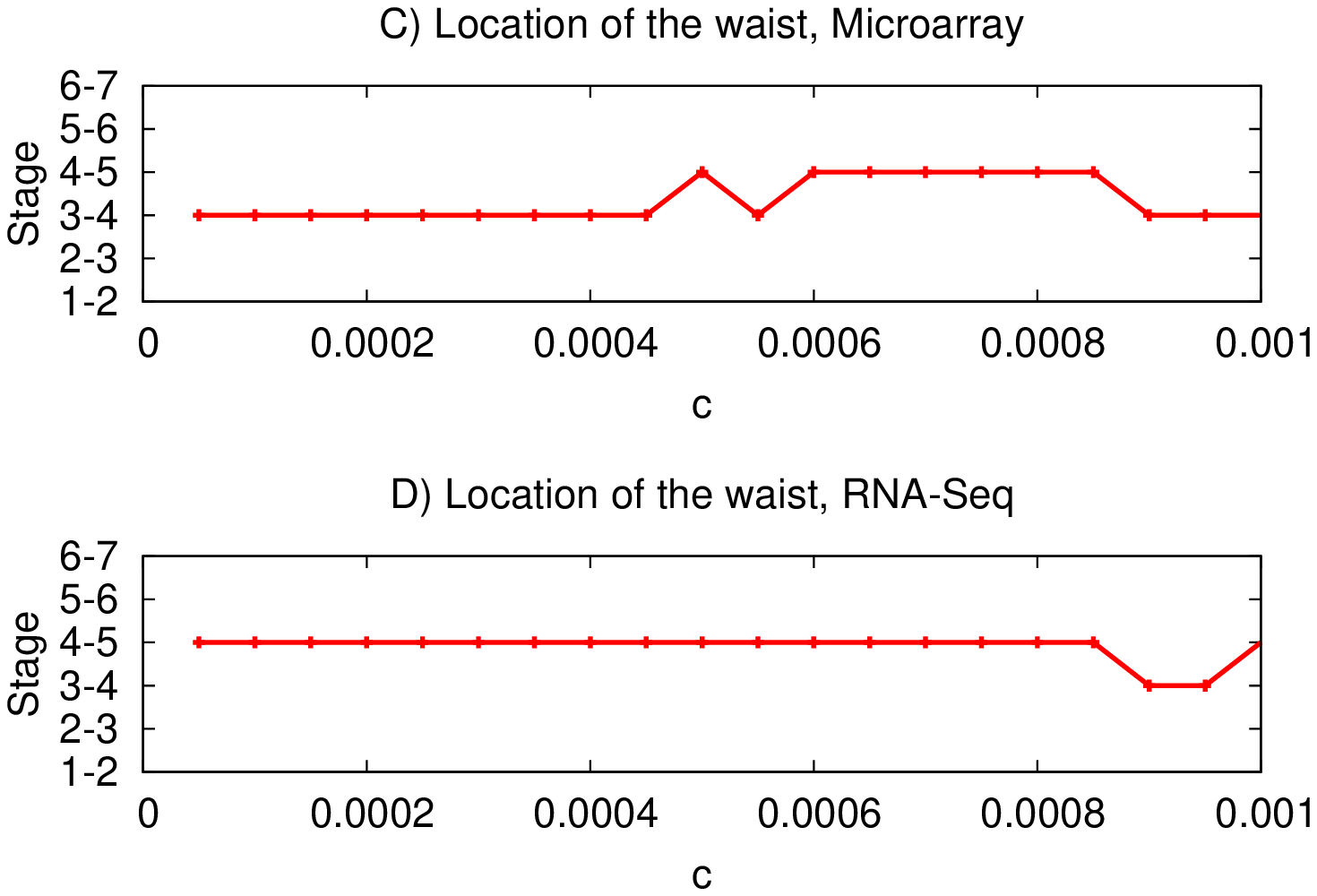}
\includegraphics[width=40mm,height=55mm,angle=270]{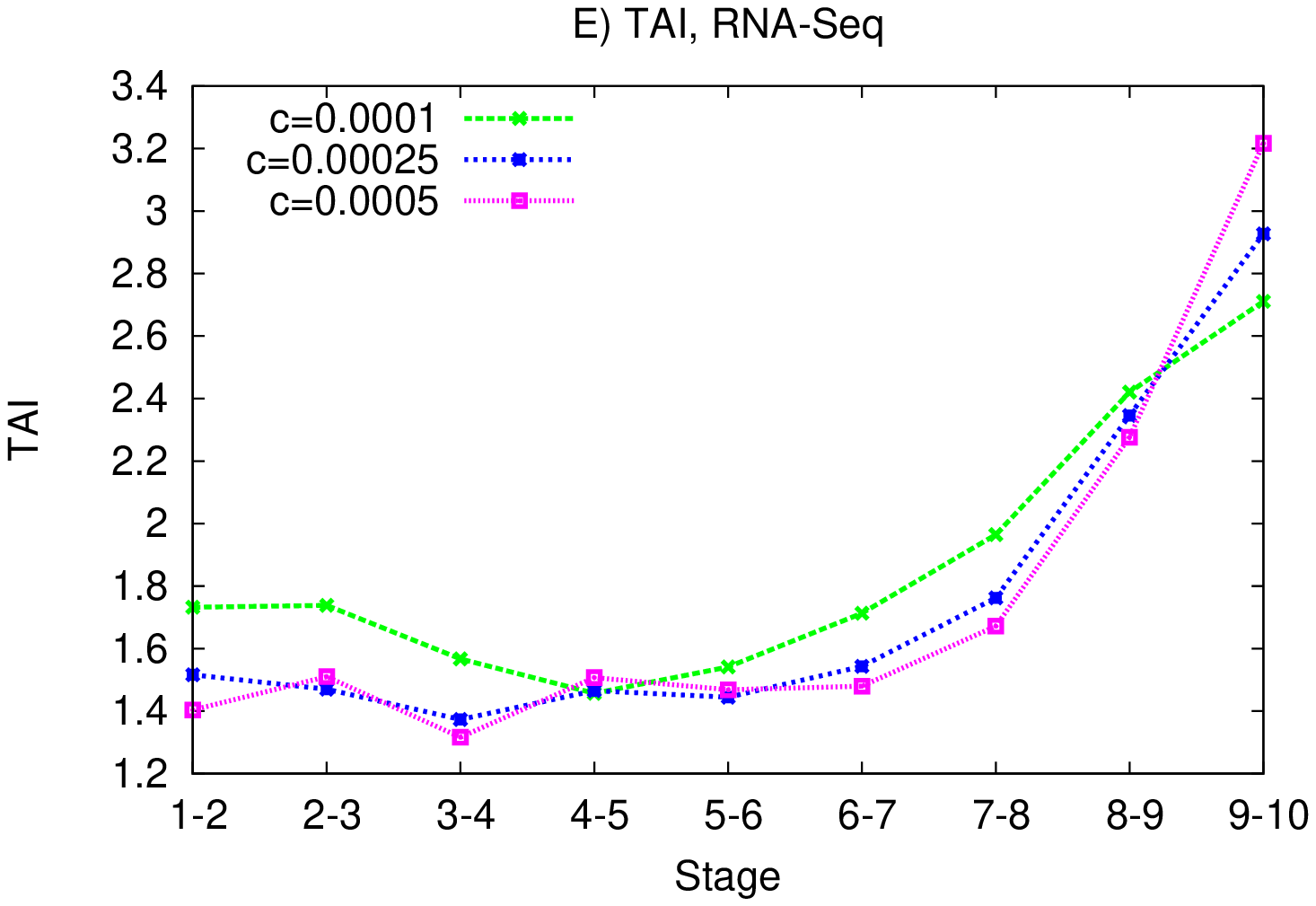}
\includegraphics[width=40mm,height=55mm,angle=270]{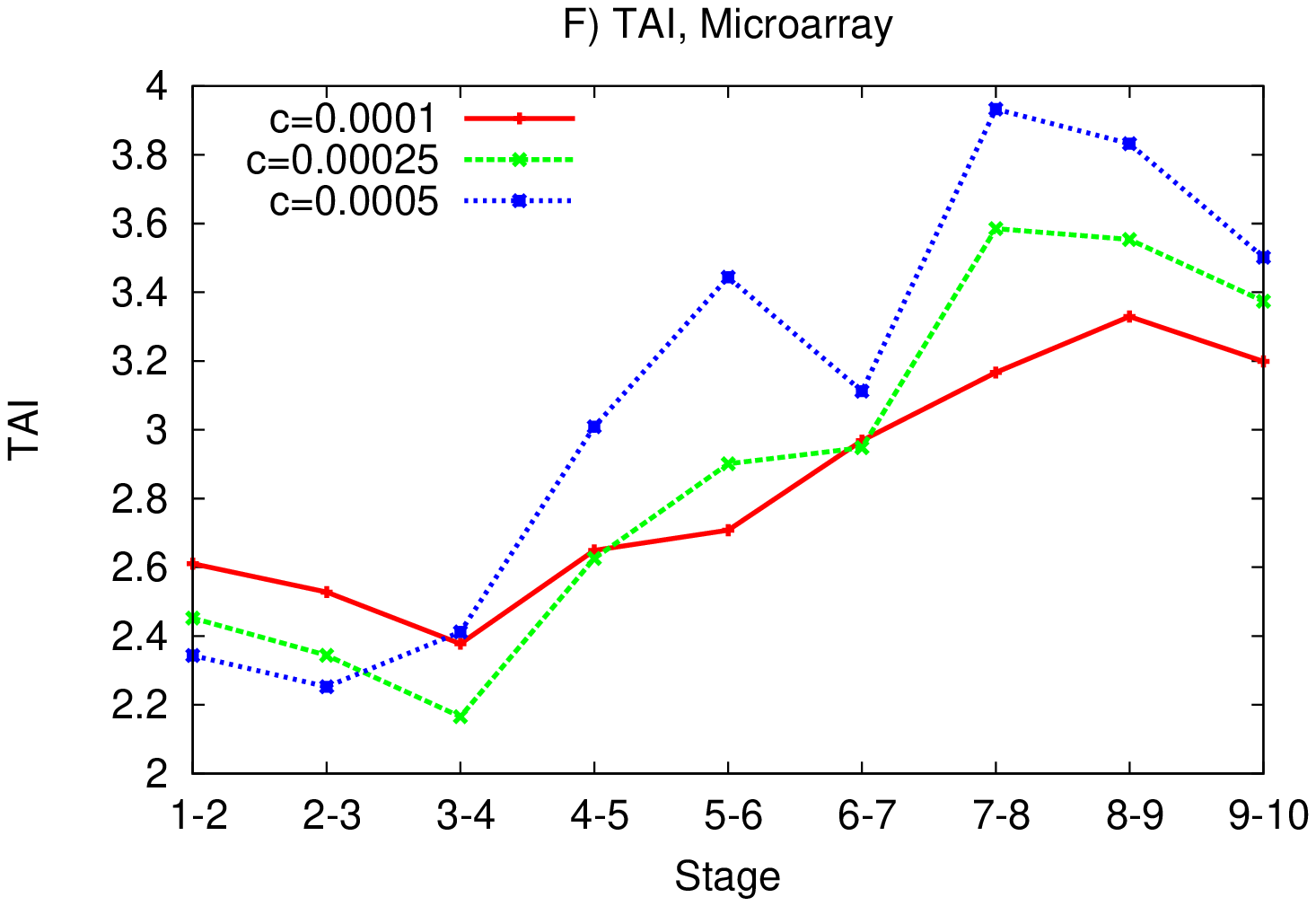}
\includegraphics[width=40mm,height=55mm,angle=270]{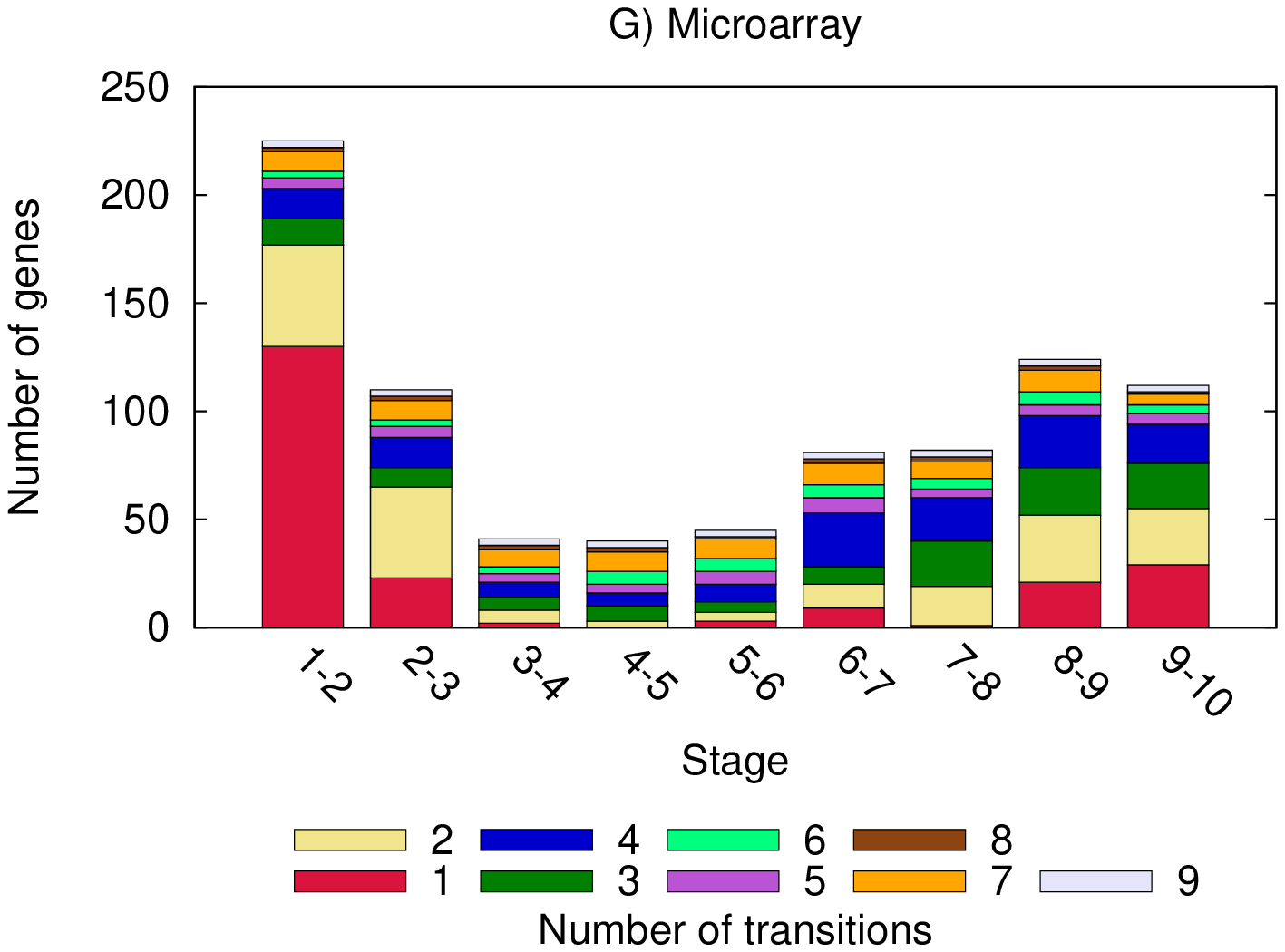}
\includegraphics[width=40mm,height=55mm,angle=270]{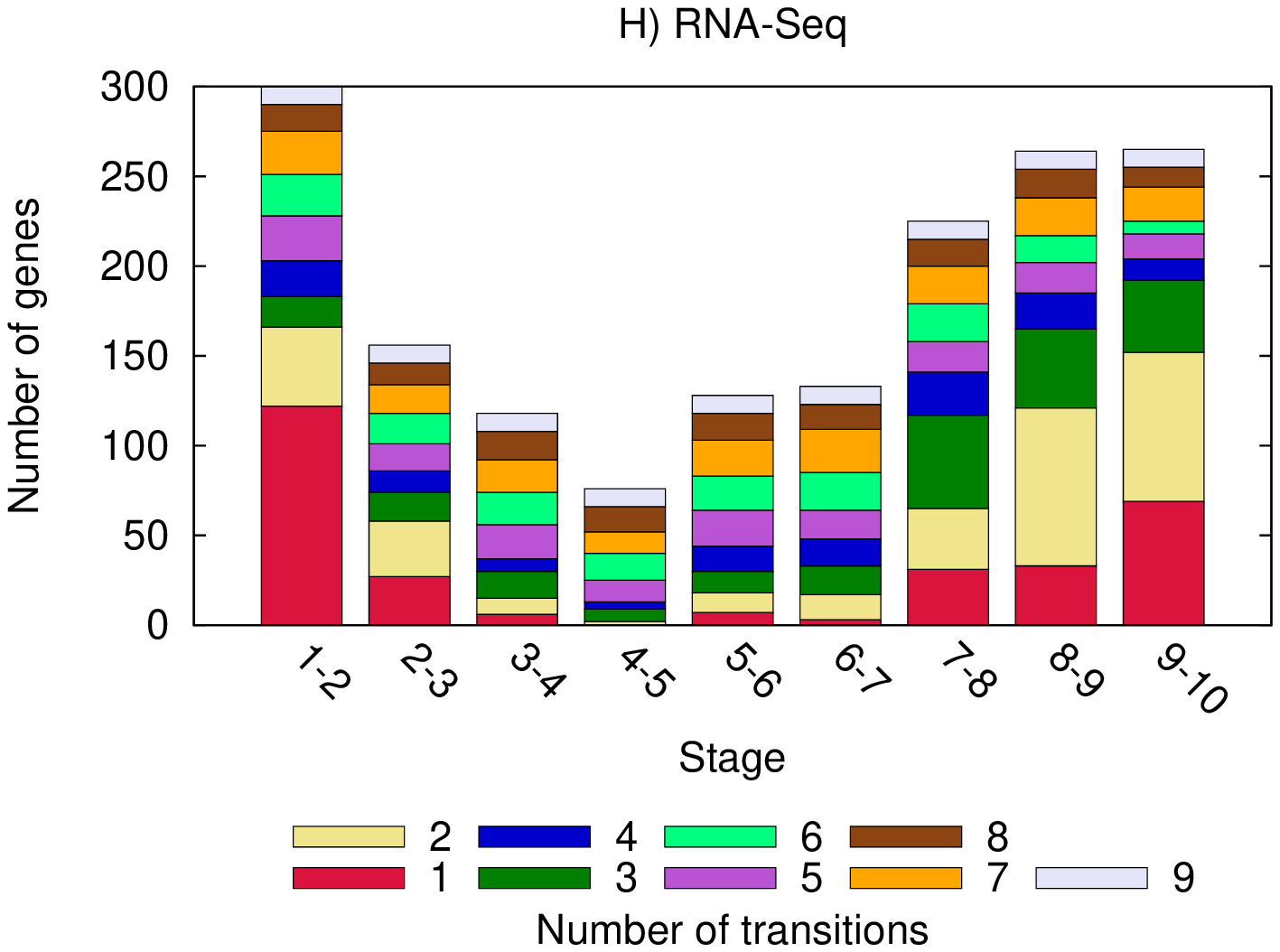}
\caption{{\em Drosophila} results using normalized expression levels. 
Graphs (A) and (B) show the hourglass score (normal and robust) 
as a function of the transition threshold $c$ for the two datasets.
Graphs (C) and (D) show the location of the hourglass waist (stage-pair) 
as a function of the transition threshold $c$ for the two datasets.
Graphs (E) and (F) show the Transcriptome Age Index 
of transitioning genes for three different values of $c$ (chosen so that the 
number of genes with known age index assigned to each stage is at least 25) 
for the two datasets. 
Graph (G) shows the transitioning genes for the Microarray dataset
with $c$=0.0005.
The transitioning genes constitute 11\% of all genes in that dataset.
53\% of those genes transition in a single stage-pair. 
Of the remaining, 64\% transition only in consecutive stage-pairs.
Note that if a gene transitions $n$ times, it is counted in $n$ 
stage-pairs.
Similarly, graph (H) shows the transitioning genes for the RNA-Seq dataset
with $c$=0.00025. 
The transitioning genes constitute 5\% of all genes in that dataset.
45\% of those genes transition in a single stage-pair. 
Of the remaining, 52\% transition only in consecutive stage-pairs.
}
\label{fig:data_analysis_results}
\end{figure*}

\noindent
{\bf Stage lethality.}
Another aspect of the developmental hourglass is in terms 
of the significance of each stage for the survival of the embryo. 
We define {\em lethality of stage $l$} as the probability
that a RW or DL event at stage $l$ 
starts a RF cascade that eventually leads to a DF event. 
We estimate this probability at generation $i$ 
as the fraction of RW and DL events,
during the past $i$ generations,
that occurred at stage $l$ and led to a DF. 

In Model-1, there is no clear trend for the stage lethality 
probability (see Fig.\ref{fig:model-1_results}B); with the exception of
the last stage (in which RW events cannot result in gene loss),
the lethality probability is roughly the same at all stages. 
In the three models with increasing specificity, however,
we observe a clear pattern: the lethality gradually increases until
the waist of the hourglass, and then it 
decreases (see Fig.\ref{fig:main_results}B, Fig.\ref{fig:model-3_results}B, 
and Fig.\ref{fig:model-4_results}B). 
The reason, as explained earlier, is that the probability of RF events 
in mid-stages is higher that in early/late stages.
Additionally, after the formation of the hourglass shape
the mid-stages have relatively few genes and so an RF event
in those genes is more likely to initiate a lethal RF cascade.

\noindent
{\bf Age of genes.}
A third aspect of the developmental hourglass effect 
is related to the evolutionary age of genes. 
The age of a gene $g$ at generation $i$ is defined as
$A(g)=i-t_0(g)$, where $t_0(g)$ is the generation at which 
$g$ was most recently rewired (and 0, if it was not rewired earlier). 
The rationale behind this definition is that a 
rewiring event may give that gene a new function, at least
in terms of its upstream and downstream regulators. 

In the case of Model-2, 
Fig.\ref{fig:main_results}C shows the median 
age of the genes at each stage, considering the population
of all genes across all individuals at a given generation. 
See Fig.\ref{fig:model-1_results}C, 
Fig.\ref{fig:model-3_results}C, and Fig.\ref{fig:model-3_results}C 
for the same results with the three other models.
The evolutionary age at stage $l$ follows the
same pattern as the lethality probability:
it gradually increases until
we reach the waist of the hourglass, and then it gradually decreases.
Genes at intermediate stages tend to be older because, 
as explained earlier, they are fewer and their rewiring 
is more likely to be lethal. 
When one of those genes $g$ is rewired or deleted from a DGEN,
the corresponding individual is often replaced (DF event)
by another individual that has the same gene $g$.
So, the genes at the waist of a DGEN tend to be  
more conserved than genes at earlier or later stages. 
%The rewiring of a gene at an early or late stage, 
%or the rewiring of a gene at a late stage, 
%is rarely lethal, 
%and so those genes tend to not be conserved for long.

%\input{tex_files/modeling_variations.tex}

\vspace{0.1in}
\noindent {\bf Data Analysis}\\
We have examined the predictions of the previous 
model using transcriptome data for 
{\em Drosophila melanogaster} and {\em Arabidopsis thaliana}.
We summarize the {\em Drosophila} results here; the corresponding
figures for {\em A.thaliana} can be found in the SI 
(Fig.\ref{fig:data_analysis_plant1} and Fig.\ref{fig:data_analysis_plant2}).
For {\em Drosophila}, we analyze Microarray \cite{kalinka2010gene}
and RNA-Seq \cite{graveley2010developmental} temporal 
expression profiles during the first 20 hours of development,
taken at 10 stages of 2-hr intervals.
We examine whether a) the number of transitioning 
genes follows an hourglass pattern, b) the waist 
of that hourglass coincides with the {\em Drosophila} phylotypic
stage, and c) the evolutionary age of the transitioning 
genes follows a similar hourglass pattern. 
The two datasets are described in more detail in the Methods section. 
With such limited data, we cannot infer the regulatory edges between 
transitioning genes and we cannot reconstruct the underlying DGEN.
However, we can identify the transitioning genes at each developmental stage
given a ``transition threshold'' $c$ (see Methods).
Even though the correct value of this threshold is not known
the following results are robust in a wide interval of $c$,
which includes most of the expression variation range
across successive developmental stages (see Fig.\ref{fig:CDFs} for the CDFs of 
expression level variations across successive stages). 

Fig.\ref{fig:data_analysis_results}A and Fig.\ref{fig:data_analysis_results}B 
show the hourglass resemblance score H (and its more robust variant) as function of $c$. 
Note that the H score is close to 1 for a wide range of $c$,
confirming the presence of an hourglass-like structure in terms of 
the number of transitioning genes. 
Fig.\ref{fig:data_analysis_results}G and Fig.\ref{fig:data_analysis_results}H
exhibit this pattern more clearly in the number of transitioning genes 
for a specific value of $c$.
The two datasets also show reasonable agreement in terms of the 
assignment of transitioning genes to stage-pairs (see Fig.\ref{fig:jacks}). 

Second, the location of the waist in this hourglass pattern, 
shown in Fig.\ref{fig:data_analysis_results}C and Fig.\ref{fig:data_analysis_results}D, 
occurs at the stage-pair (3,4) or (4,5), depending on $c$.
This is roughly 8 hours after the formation 
of the zygote, and it includes the phylotypic 
stage for {\em Drosophila melanogaster} \cite{kalinka2010gene}. 

We have also estimated the evolutionary age of most of the transitioning 
genes at each developmental stage-pair using  the 
Transcriptome Age Index (TAI) metric \cite{domazet2010phylogenetically}
(see Methods).  TAI is lower for older genes. 
Fig.\ref{fig:data_analysis_results}E and Fig.\ref{fig:data_analysis_results}F
show the average TAI for transitioning genes, weighted by the 
expression level of each gene, at each stage-pair and for each dataset using
three values of $c$.
The TAI index follows the pattern that the model predicts,
with older genes (lower TAI values) close to the waist of the hourglass.
This result appears consistent with
the main observation of Domazet-Loso and Tautz \cite{domazet2010phylogenetically},
even though that study did not analyze transitioning genes.

%Finally, we evaluated the agreement between the two datasets 
%in terms of the transitioning genes assigned to each stage-pair, 
%considering only those genes that appear in both datasets.
%Because the appropriate transition threshold may be different at
%each dataset, we use a different threshold for each dataset,
%say $c_1$ and $c_2$. For each pair $(c_1,c_2)$, we determine
%the transitioning genes at each stage with the corresponding dataset 
%(i.e., $L$ pairs of gene sets), 
%and then calculate the average Jaccard similarity across these $L$ pairs. 
%Fig. \ref{fig:data_analysis_results}-G shows that, when the two
%thresholds are roughly equal, the average Jaccard similarity is 
%as high as 50\%; this means that about 2/3 of the genes assigned to 
%a certain stage-pair using one dataset are also assigned to the same stage-pair
%using the other dataset. 

\pagebreak

\vspace{0.1in}
\noindent {\bf Discussion}\\
Early studies of the developmental hourglass effect 
mostly analyzed morphological and phenotypic similarities across species 
\cite{richardson1997there, richardson2002haeckel}. 
Recently, the focus has shifted towards genomic 
and molecular comparative studies \cite{domazet2010phylogenetically, irie2007vertebrate, kalinka2010gene, piasecka2013hourglass, quint2012transcriptomic} that
investigate conservation of gene expression variation, sequence conservation, 
selective constraint on coding sequences, and evolutionary gene ``age.''
These studies often report contradicting observations: 
some support strong conservation in earlier developmental stages 
\cite{rasmussen1987new, roux2008developmental, piasecka2013hourglass}, 
while others support that strongest conservation occurs at a mid-developmental stage 
\cite{irie2011comparative, irie2007vertebrate, kalinka2010gene, levin2012developmental, quint2012transcriptomic, hazkani2005search, domazet2010phylogenetically, galis2001testing, cruickshank2008microevolutionary}. 
Nevertheless, the fact that the hourglass effect is observed in highly divergent 
species across deep phylogenetic scales (including fish, flies and plants), 
suggests that this observed pattern of conservation may stem from fundamental 
organization principles. 

What these principles are has remained elusive.  
Earlier stages may be conserved because any changes therein 
could have large cascading effects in later stages 
\cite{riedl1978order, arthur1984mechanisms, rasmussen1987new}.  
Later stages may experience less constraint because 
as development progresses gene interactions become more modular, and so
it is plausible that perturbations there have only local effects \cite{raff1996shape}.  
We refer to them as the ``temporal'' constraint model
and the ``spatial'' constraint model, respectively, 
following Tian et al. \cite{tian2013dictyostelium}.

%Based upon these ideas we introduced the concept of a DGEN,
%which focuses on the temporal execution of a gene regulatory network,
%the specific genes that change functional state at each developmental stage,
%and the interactions that cause such state transitions.
%The evolutionary aspect of DevoNet captures structural 
%changes in the DGEN of an organism.

In this paper, we developed an evolutionary model of development 
that combines some aspects of the previous two models. 
Regulatory perturbations at a certain stage can cause cascades of 
regulatory failures at subsequent stages (temporal model), while
the likelihood that a gene regulates genes at a subsequent stage
decreases as development progresses (spatial model). 
Our computational results lead to the following testable predictions:
a) the number of transitioning genes during development follows an 
hourglass pattern,
b) the evolutionary age of the transitioning genes also follows an 
hourglass pattern, with the oldest genes being at the waist of 
the hourglass, and
c) the genes at the waist of that hourglass are the most essential,
in the sense that their deletion maximizes the probability of developmental
failure.
We have relied on developmental gene expression profiles of 
{\em Drosophila melanogaster} and 
{\em Arabidopsis thaliana} to examine the predictions of the model. 
The analysis of that data agrees with the first two theoretical model predictions.
The increased conservation of genes at the waist
provides an indirect confirmation of the model's third prediction,
regarding the essentiality of different genes. 
Further, our simulations confirm that the details of these
regulatory perturbations, such as the probabilities of gene duplication and 
deletion or the parameter $z$ in the regulatory failure probability,  
do not affect the results of the model, at least at the qualitative level.

%The number of transitioning genes follows an 
%hourglass pattern for a wide range of the involved transition threshold. 
%Further, the waist of this hourglass seems to coincide with the 
%previously reported phylotypic stage for these two species.
%The estimated evolutionary age of the transitioning genes also 
%follows the expected hourglass pattern, with the oldest genes
%residing at the hourglass waist. This is consistent with 
%the main observation of Domazet-Loso and Tautz \cite{domazet2010phylogenetically},  
%even though that study did not analyze transitioning genes.

The use of DGENs in this work was only as an abstract tool to study 
the effect of gene regulatory perturbations in the developmental process.
In future work, it is important to infer the actual DGEN of model
organisms. This will require information about gene regulatory
interactions across time and space, but it should be possible for
at least some developmentally well studied species \cite{peter2012predictive}. 
Such DGENs would help to identify the specific genes that form 
the hourglass waist and their function. Additionally, an inferred
DGEN would allow to directly test the increasing specificity assumption.

Finally, we note that the hourglass effect (sometimes referred to as the 
``bow tie'' effect) has also been observed in other complex 
biological and technological systems that exhibit hierarchical 
modularity and that are subject to evolutionary pressure or optimization 
trade-offs 
\cite{beutler2004inferences, csete2004bow, doyle2011architecture, tieri2010network}. 
For instance, the Internet ``protocol stack'' is organized in an hourglass structure
\cite{saamer-sigcomm11}; this pattern was not designed but it emerged
through the competition between protocols that serve roughly the same function
at each communication layer, during the last 30-40 years. 
In earlier work, we proposed an abstract model (EvoArch) that 
captures the evolution of protocol architectures and that predicts
the emergence of an hourglass structure. 
Interestingly, both EvoArch and the model of this paper share the same principle:
the underlying hierarchical networks that control both systems should be 
increasingly sparser as complexity increases, i.e., the specificity of
each complexity stage (or layer) should be increasing.
In the future, we will further investigate this common organization
principle between biological and technological systems.

%\begin{materials}
\vspace{0.1in}
\noindent {\bf Materials and Methods}\\
\noindent
{\bf Hourglass score H.}
Suppose that $w(l)$ denotes the number of transitioning genes in stage $l$
and let $b$ be the stage with the minimum number of such genes.
We construct the sequences $X$=$\{w(l)\},l=1,\dots b\}$ and 
$Y$=$\{w(l)\},l=b,\dots L\}$. 
$\tau_X$ and $\tau_Y$ denote the normalized univariate Mann-Kendall statistic 
for monotonic trend in each sequence, respectively ($\tau$ is -1 for 
decreasing, +1 for increasing and almost 0 for random sequences). 
The H score is defined as $H=(\tau_Y-\tau_X)/2$. 
See Fig.\ref{fig:hScore} for an illustration, and for the definition
of a more robust version of H.

%Let $w(l)$ be the width of stage $l$, 
%i.e., the number of transitioning genes in that stage.
%Let $w_b$ be the minimum width across all stages, 
%and suppose that this minimum occurs at stage $l=b$;
%this is the {\em waist} of the network 
%(ties are broken so that the waist is closer to $\lfloor L/2 \rfloor$).
%Consider the sequence $X=\{w(l)\},l=1,\dots b\}$ and the sequence
%$Y=\{w(l)\},l=b,\dots L\}$. 
%We calculate the normalized univariate Mann-Kendall statistic for monotonic trend
%on the sequences $X$ and $Y$ as coefficients $\tau_{X}$
%and $\tau_{Y}$, respectively. The coefficients
%vary between -1 (strictly decreasing) and 1 (strictly increasing),
%while they are approximately zero for random samples.
%We define $H=(\tau_Y-\tau_X) / 2$; $H$ is referred to as the hourglass score. 
%$H = 1$ if the DGEN is structured as an hourglass, with a strictly
%decreasing sequence of $b$ stages followed by a strictly increasing
%sequence of $L-b$ stages. See Fig.\ref{fig:hScore} for an illustration. 
%In the computational modeling results, we do not consider
%the width of the first stage because it can never decrease in models-1 to 3.
%We also use a variation of the hourglass score
%in which we do not take into account adjacent stages in 
%calculating the Mann-Kendall statistics. That score is denoted by $\tilde{H}$
%and is referred to as the ``robust hourglass score.'' 

\noindent
{\bf Drosophila data and treatment.} 
Developmental gene expression profiles are obtained from two sources. 
First, microarray data from Kalinka et al. \cite{kalinka2010gene} for 3,610 genes. 
The expression level of each gene is calculated as the median 
of probes mapping to that gene. 
Each stage represents a 2-hr interval during the first 20 hours of
embryogenesis (10 stages).
The second source is RNA-Seq data from 
Graveley et al. \cite{graveley2010developmental}.  Raw data are 
processed to RPKM values. Each stage represents a 2-hr interval
during the first 24 hours (12 stages). 
Genes with zero RPKM value in all stages are discarded,
resulting in 14,110 genes. 

\noindent
{\bf Transitioning gene identification.}
Suppose that the reported expression value of gene $i$ at stage $l$ is $e_{i,l}$.
We analyze both these ``absolute'' expression values 
as well as the normalized expression values, given by
$e'_{i,l}$=$e_{i,l}/\sum_{j}e_{j,l}$.
The identification of transitioning genes follows the same method for 
both absolute and normalized expression levels. 
In the case of normalized expressions, 
we calculate $\delta_{i,l}=e'_{i,l}-e'_{i,l-1}$ for each gene and at each stage 
$l$=$2\dots L$.
Gene $i$ is considered ``transitioning'' at the stage-pair $(l-1,l)$ if 
$|\delta_{i,l}|>c$, where $c$ is a given transition threshold.
This condition is more robust to noise than a 
ratio-based rule ($e'_{i,l}/e'_{i,l-1}$) for the identification of transitioning genes.
Note that a gene may be transitioning at more than one stage-pair, but it
may also not be transitioning at any stage-pair.

\noindent
{\bf Transcriptome age index (TAI).}
We collected the groups of orthologs for each gene in Drosophila using 
two databases, 
OrthoDB \cite{waterhouse2013orthodb} and OrthoMCL \cite{li2003orthomcl}. 
The Eumetazoa data were taken from OrthoDB, while Fungi  and  Plants  species  
were  retrieved  from  OrthoMCL, and the two datasets were merged.  
Using these orthologs 
we then assigned an age index to each gene based on its absence and presence 
in a phylogenetic tree of 24 well-diverged species 
\cite{domazet2010phylogenetically, quint2012transcriptomic}
(see Fig.\ref{fig:phylo_tree}).
%Following this procedure we assigned each gene one of the following six age levels:\\
%Level 1: Common ancestor to Fungi, Plants and Eumetazoa.\\
%Level 2: Common ancestor to Fungi and Eumetazoa.\\
%Level 3: Common ancestor to all Eumetazoa.\\
%Level 4: Common ancestor to all Bilateria.\\
%Level 5: Common ancestor to all Arthropoda.\\
%Level 6: Common ancestor to all Dipteria.

\noindent
{\bf Age index for each stage-pair.} 
Suppose that we identify transitioning genes based on the normalized expression 
levels, and that $n(l)$ genes are assigned to stage-pair $(l-1,l)$. 
Denote by $p_i$ the phylogenetic rank (TAI value) of gene $i$.
The age index assigned to that stage-pair  is given by
TAI(l)=$(\Sigma_{i=1}^{n(l)} p_i e'_{i,l})/\Sigma_{i=1}^{n(l)} e'_{i,l}$.
The same method is used when transitioning genes are identified based on 
absolute expression levels.

\bibliographystyle{pnas}
\bibliography{papers}

\begin{thebibliography}{10}

\bibitem{carroll2005endless}
Carroll SB
\newblock (2005) \emph{Endless Forms Most Beautifull: The New Science of Evo
  Devo and the Making of the Animal Kingdom}
\newblock (WW Norton \& Company) No.{}~54.

\bibitem{raff1996shape}
Raff RA
\newblock (1996) The shape of life: genes, development, and the evolution of
  animal form.

\bibitem{richardson2002haeckel}
Richardson MK, Keuck G
\newblock (2002) Haeckel's abc of evolution and development.
\newblock \emph{Biological Reviews} 77:495--528.

\bibitem{davidson2010regulatory}
Davidson EH
\newblock (2010) \emph{The regulatory genome: gene regulatory networks in
  development and evolution}
\newblock (Academic Press), 2nd edition.

\bibitem{duboule1994temporal}
Duboule D
\newblock (1994) Temporal colinearity and the phylotypic progression: a basis
  for the stability of a vertebrate bauplan and the evolution of morphologies
  through heterochrony.
\newblock \emph{Development} 1994:135--142.

\bibitem{von1828entwicklungsgeschichte}
von Baer CE
\newblock (1828) \emph{{\"U}ber Entwicklungsgeschichte der Thiere. Beobachtung
  und Reflexion.-K{\"o}nigsberg, Gebr{\"u}der Borntr{\"a}ger 1828-1837}
\newblock (Gebr{\"u}der Borntr{\"a}ger) Vol.{}~1.

\bibitem{rasmussen1987new}
Rasmussen N
\newblock (1987) A new model of developmental constraints as applied to the
  drosophila system.
\newblock \emph{J Theor Biol} 127:271--299.

\bibitem{roux2008developmental}
Roux J, Robinson-Rechavi M
\newblock (2008) Developmental constraints on vertebrate genome evolution.
\newblock \emph{PLoS Genet} 4:e1000311.

\bibitem{irie2011comparative}
Irie N, Kuratani S
\newblock (2011) Comparative transcriptome analysis reveals vertebrate
  phylotypic period during organogenesis.
\newblock \emph{Nat Commun} 2:248.

\bibitem{irie2007vertebrate}
Irie N, Sehara-Fujisawa A
\newblock (2007) The vertebrate phylotypic stage and an early
  bilaterian-related stage in mouse embryogenesis defined by genomic
  information.
\newblock \emph{BMC Biol} 5:1.

\bibitem{kalinka2010gene}
Kalinka AT, {et~al.}
\newblock (2010) Gene expression divergence recapitulates the developmental
  hourglass model.
\newblock \emph{Nature} 468:811--814.

\bibitem{levin2012developmental}
Levin M, Hashimshony T, Wagner F, Yanai I
\newblock (2012) Developmental milestones punctuate gene expression in the
  caenorhabditis embryo.
\newblock \emph{Dev Cell} 22:1101--1108.

\bibitem{quint2012transcriptomic}
Quint M, {et~al.}
\newblock (2012) A transcriptomic hourglass in plant embryogenesis.
\newblock \emph{Nature} 490:98--101.

\bibitem{domazet2010phylogenetically}
Domazet-Lo{\v{s}}o T, Tautz D
\newblock (2010) A phylogenetically based transcriptome age index mirrors
  ontogenetic divergence patterns.
\newblock \emph{Nature} 468:815--818.

\bibitem{hazkani2005search}
Hazkani-Covo E, Wool D, Graur D
\newblock (2005) In search of the vertebrate phylotypic stage: a molecular
  examination of the developmental hourglass model and von baer's third law.
\newblock \emph{J Exp Zool B Mol Dev Evol} 304:150--158.

\bibitem{galis2001testing}
Galis F, Metz JA
\newblock (2001) Testing the vulnerability of the phylotypic stage: on
  modularity and evolutionary conservation.
\newblock \emph{J Exp Zool} 291:195--204.

\bibitem{cruickshank2008microevolutionary}
Cruickshank T, Wade MJ
\newblock (2008) Microevolutionary support for a developmental hourglass: gene
  expression patterns shape sequence variation and divergence in drosophila.
\newblock \emph{Evol Dev} 10:583--590.

\bibitem{comte2010molecular}
Comte A, Roux J, Robinson-Rechavi M
\newblock (2010) Molecular signaling in zebrafish development and the
  vertebrate phylotypic period.
\newblock \emph{Evol Dev} 12:144--156.

\bibitem{hall1997phylotypic}
Hall BK
\newblock (1997) Phylotypic stage or phantom: is there a highly conserved
  embryonic stage in vertebrates?
\newblock \emph{Trends Ecol Evol} 12:461--463.

\bibitem{kalinka2012evolution}
Kalinka AT, Tomancak P
\newblock (2012) The evolution of early animal embryos: conservation or
  divergence?
\newblock \emph{Trends Ecol Evol} 27:385--393.

\bibitem{piasecka2013hourglass}
Piasecka B, Lichocki P, Moretti S, Bergmann S, Robinson-Rechavi M
\newblock (2013) The hourglass and the early conservation models—co-existing
  patterns of developmental constraints in vertebrates.
\newblock \emph{PLoS Genet} 9:e1003476.

\bibitem{richardson1998phylotypic}
Richardson MK, Minelli A, Coates M, Hanken J
\newblock (1998) Phylotypic stage theory.
\newblock \emph{Trends Ecol Evol} 13:158.

\bibitem{rp2003inverting}
RP BEO, Richardson MK, {et~al.}
\newblock (2003) Inverting the hourglass: quantitative evidence against the
  phylotypic stage in vertebrate development.
\newblock \emph{Proc R Soc Lond B Biol Sci} 270:341--346.

\bibitem{peter2012predictive}
Peter IS, Faure E, Davidson EH
\newblock (2012) Predictive computation of genomic logic processing functions
  in embryonic development.
\newblock \emph{Proc Natl Acad Sci USA} 109:16434--16442.

\bibitem{graveley2010developmental}
Graveley BR, {et~al.}
\newblock (2010) The developmental transcriptome of drosophila melanogaster.
\newblock \emph{Nature} 471:473--479.

\bibitem{richardson1997there}
Richardson MK, {et~al.}
\newblock (1997) There is no highly conserved embryonic stage in the
  vertebrates: implications for current theories of evolution and development.
\newblock \emph{Anat Embryol (Berl)} 196:91--106.

\bibitem{riedl1978order}
Riedl R, Jefferies RPS
\newblock (1978) \emph{Order in living organisms: a systems analysis of
  evolution}
\newblock (Wiley New York).

\bibitem{arthur1984mechanisms}
Arthur W
\newblock (1984) \emph{Mechanisms of morphological evolution: a combined
  genetic, developmental, and ecological approach}
\newblock (Wiley New York).

\bibitem{tian2013dictyostelium}
Tian X, Strassmann JE, Queller DC
\newblock (2013) Dictyostelium development shows a novel pattern of
  evolutionary conservation.
\newblock \emph{Mol Biol Evol} 30:977--984.

\bibitem{beutler2004inferences}
Beutler B
\newblock (2004) Inferences, questions and possibilities in toll-like receptor
  signalling.
\newblock \emph{Nature} 430:257--263.

\bibitem{csete2004bow}
Csete M, Doyle J
\newblock (2004) Bow ties, metabolism and disease.
\newblock \emph{Trends Biotechnol} 22:446--450.

\bibitem{doyle2011architecture}
Doyle JC, Csete M
\newblock (2011) Architecture, constraints, and behavior.
\newblock \emph{Proc Natl Acad Sci USA} 108:15624--15630.

\bibitem{tieri2010network}
Tieri P, {et~al.}
\newblock (2010) Network, degeneracy and bow tie integrating paradigms and
  architectures to grasp the complexity of the immune system.
\newblock \emph{Theor Biol Med Model} 7:32.

\bibitem{zhao2006hierarchical}
Zhao J, Yu H, Luo JH, Cao ZW, Li YX
\newblock (2006) Hierarchical modularity of nested bow-ties in metabolic
  networks.
\newblock \emph{BMC Bioinformatics} 7:386.

\bibitem{saamer-sigcomm11}
Akhshabi S, Dovrolis C
\newblock (2011) \emph{{The Evolution of Layered Protocol Stacks Leads to an
  Hourglass-Shaped Architecture}}.

\bibitem{akhshabi2013evolution}
Akhshabi S, Dovrolis C
\newblock (2013) in \emph{Dynamics On and Of Complex Networks, Volume 2}
\newblock (Springer), pp 55--88.

\bibitem{waterhouse2013orthodb}
Waterhouse RM, Tegenfeldt F, Li J, Zdobnov EM, Kriventseva EV
\newblock (2013) Orthodb: a hierarchical catalog of animal, fungal and
  bacterial orthologs.
\newblock \emph{Nucleic Acids Res} 41:D358--D365.

\bibitem{li2003orthomcl}
Li L, Stoeckert CJ, Roos DS
\newblock (2003) Orthomcl: identification of ortholog groups for eukaryotic
  genomes.
\newblock \emph{Genome Res} 13:2178--2189.

\bibitem{xiang2011genome}
Xiang D, {et~al.}
\newblock (2011) Genome-wide analysis reveals gene expression and metabolic
  network dynamics during embryo development in arabidopsis.
\newblock \emph{Plant Physiol} 156:346--356.

\end{thebibliography}

%\end{article}
%\input{tex_files/figures.tex}

\pagebreak
\setcounter{figure}{0}
\makeatletter 
\renewcommand{\thefigure}{S\@arabic\c@figure}
\section{Supporting Information}

%\begin{figure*}[!htb]
%\centering
%\includegraphics[width=50mm,height=50mm,angle=270]{Figures/Fig1/b/specificity.eps}
%\caption{Two types of specificity function: linear and sigmoid}
%\label{fig:specificity}
%\end{figure*}

\begin{figure*}[!htb]
\centering
\includegraphics[width=50mm,height=70mm,angle=270]{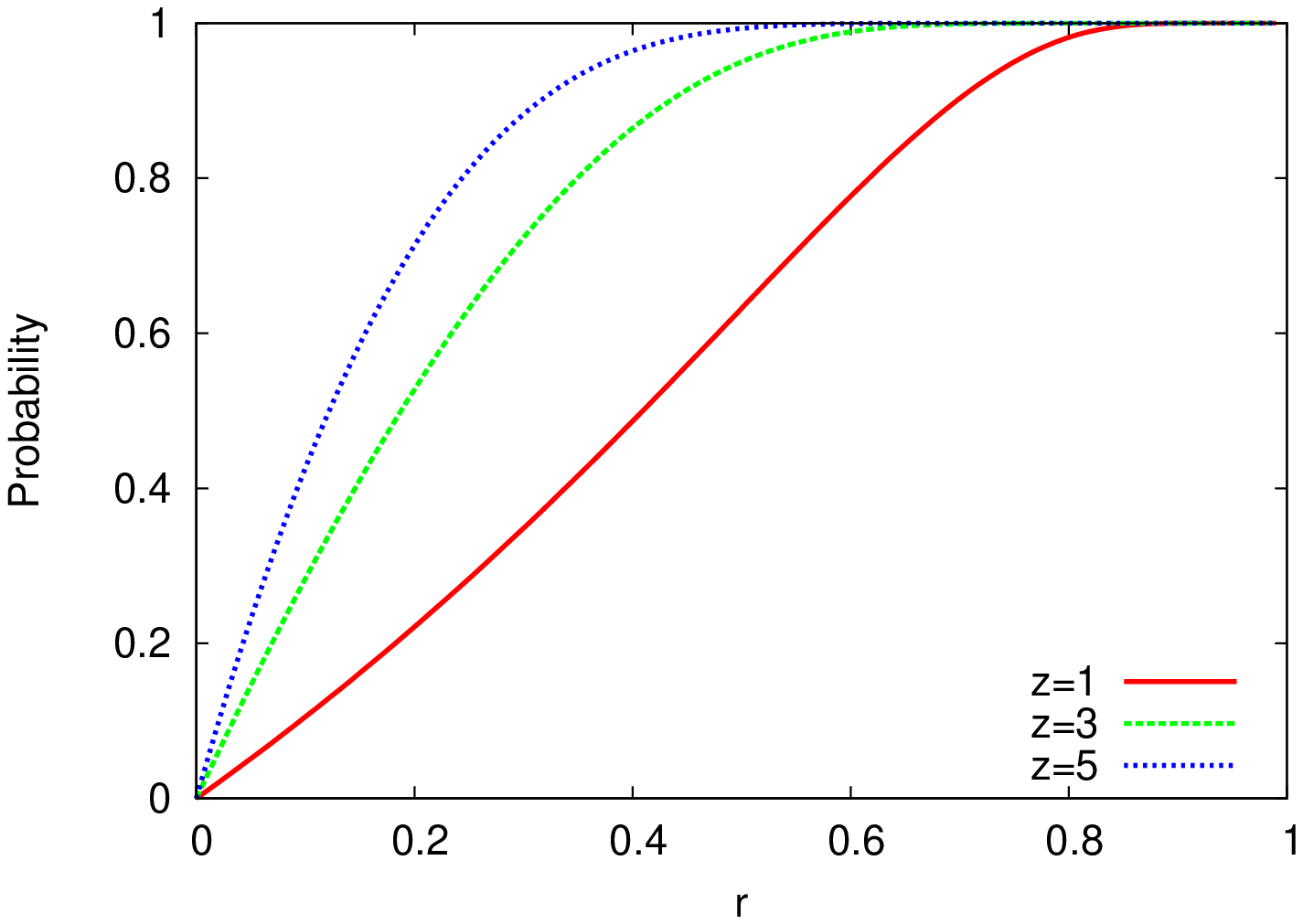}
\caption{Probability of Regulatory Failure (RF) for three values of the 
paramater $z$.
$r$ is the fraction of upstream regulating edges that are lost due to 
a DL or RW event.}
\label{fig:rfFunction}
\end{figure*}

\begin{figure*}[!htb]
\centering
\includegraphics[width=80mm,height=58mm,angle=0]{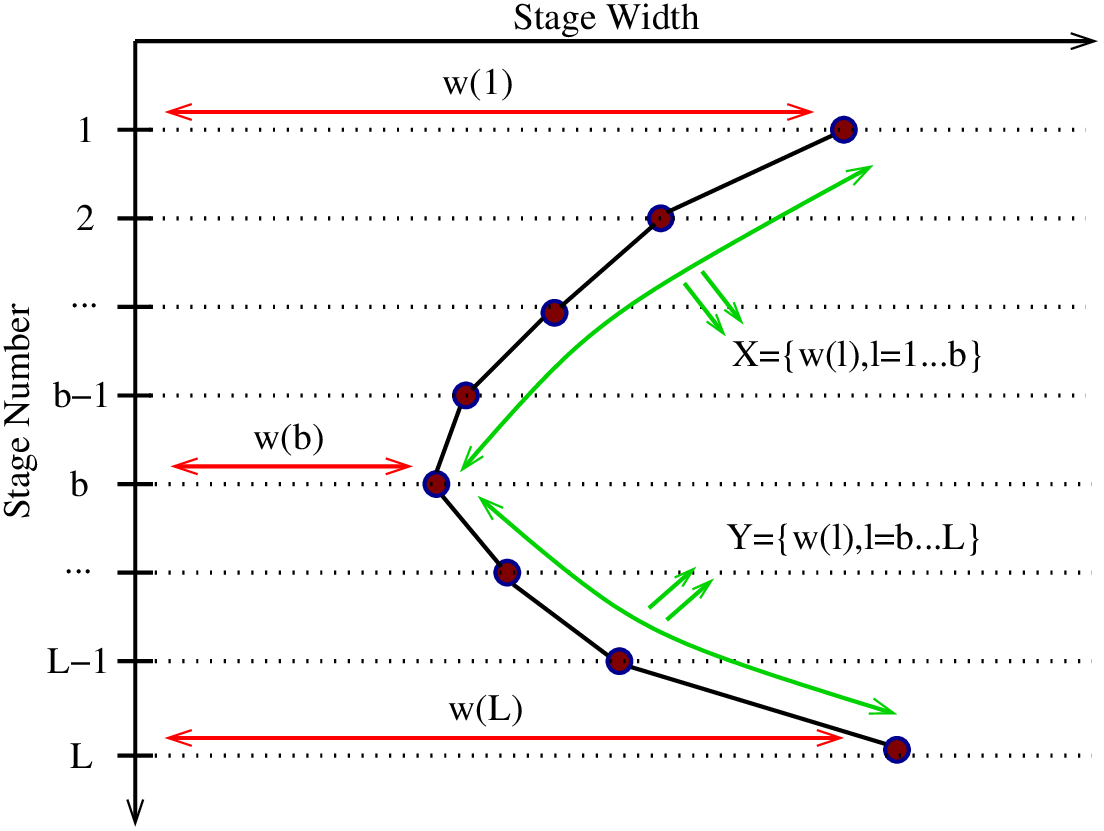}
\caption{Illustration of the $H$ score calculation. 
Let $w(l)$ be the width of stage $l$. 
Let $w_b$ be the minimum width across all stages, 
and suppose that this minimum occurs at stage $l=b$;
this is the {\em waist} of the network 
(ties are broken so that the waist is closer to $\lfloor L/2 \rfloor$).
Consider the sequence $X=\{w(l)\},l=1,\dots b\}$ and the sequence
$Y=\{w(l)\},l=b,\dots L\}$. 
We denote the normalized univariate Mann-Kendall 
statistic for monotonic trend
on the sequences $X$ and $Y$ as $\tau_{X}$
and $\tau_{Y}$, respectively. The Mann-Kendall statistic 
varies between -1 (decreasing) and 1 (increasing),
and it is approximately zero for a random sequence.
We define $H=(\tau_Y-\tau_X) / 2$; $H$ is referred to as the hourglass score. 
$H = 1$ if the DGEN is structured as an hourglass, with a 
decreasing sequence of $b$ stages followed by an increasing
sequence of $L-b$ stages.
In the computational modeling results, we do not consider
the width of the first stage because it can never decrease in Models-1,2,3.
We also define a variation of the hourglass score
in which we do not take into account adjacent stages in 
calculating the two Mann-Kendall statistics. That statistic is 
denoted by $\tilde{H}$ and is referred to as the robust hourglass score.} 
\label{fig:hScore}
\end{figure*}

\begin{figure*}[!htb]
\centering
\includegraphics[width=34mm,height=54mm,angle=270]{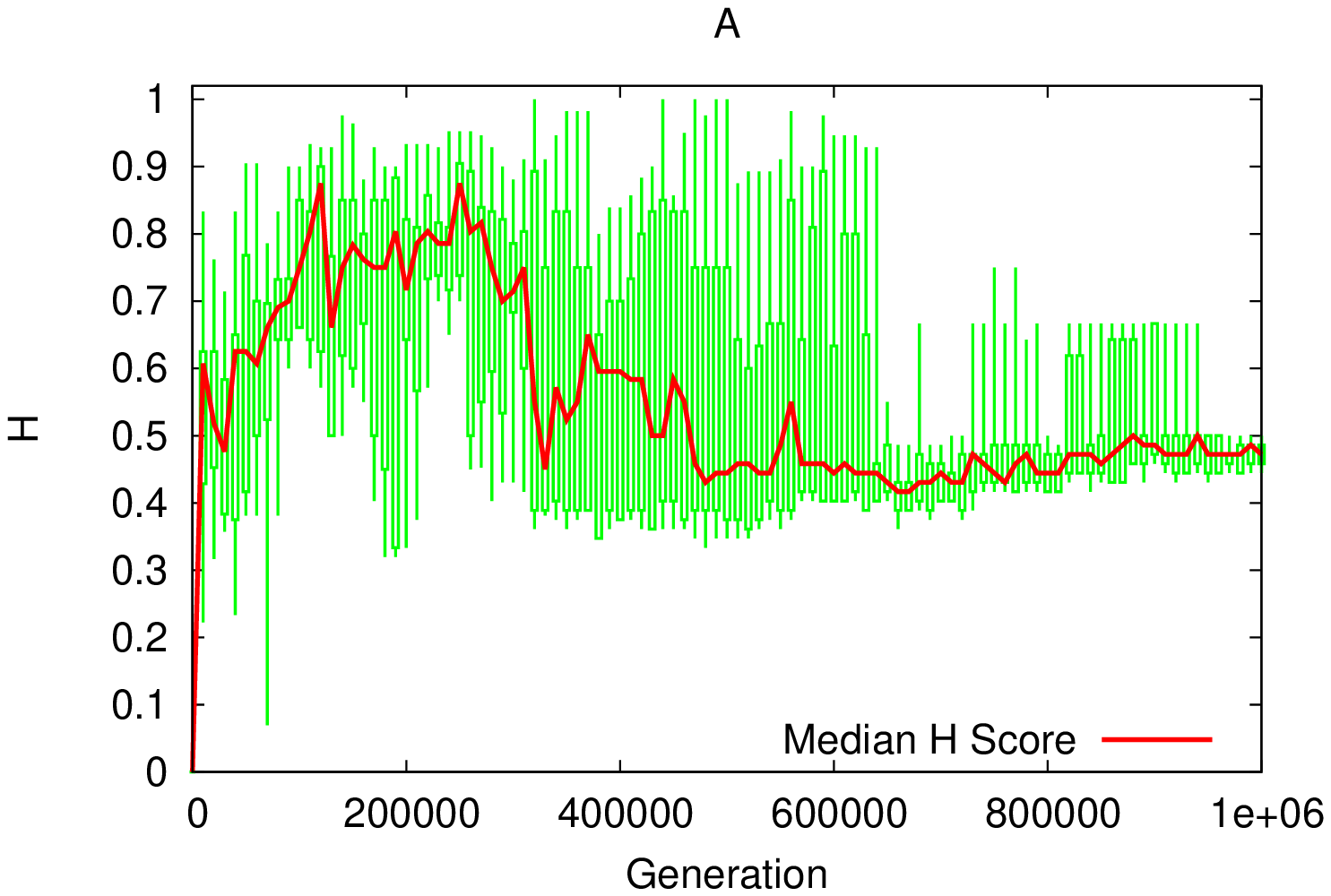}
\includegraphics[width=34mm,height=54mm,angle=270]{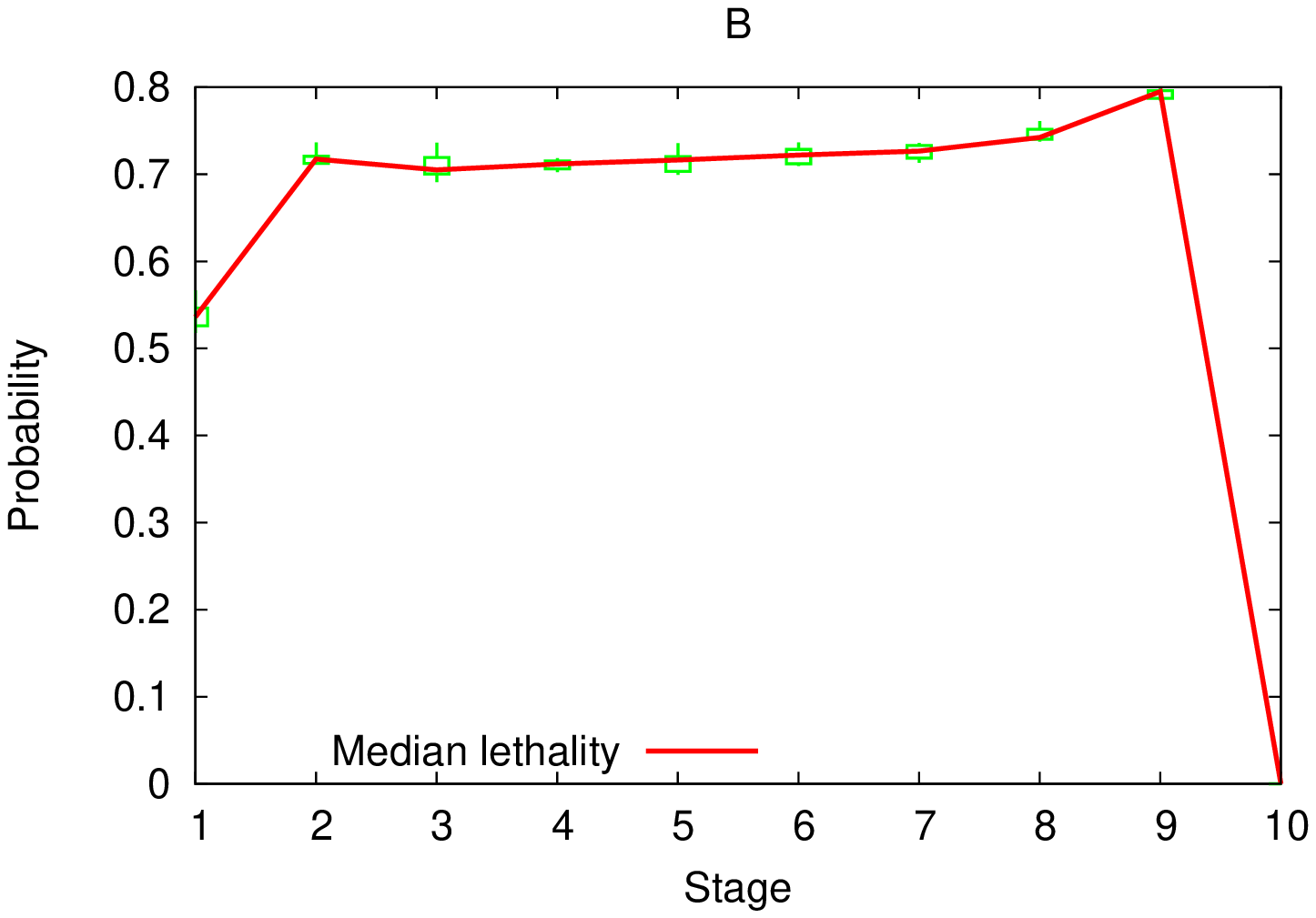}
\includegraphics[width=34mm,height=54mm,angle=270]{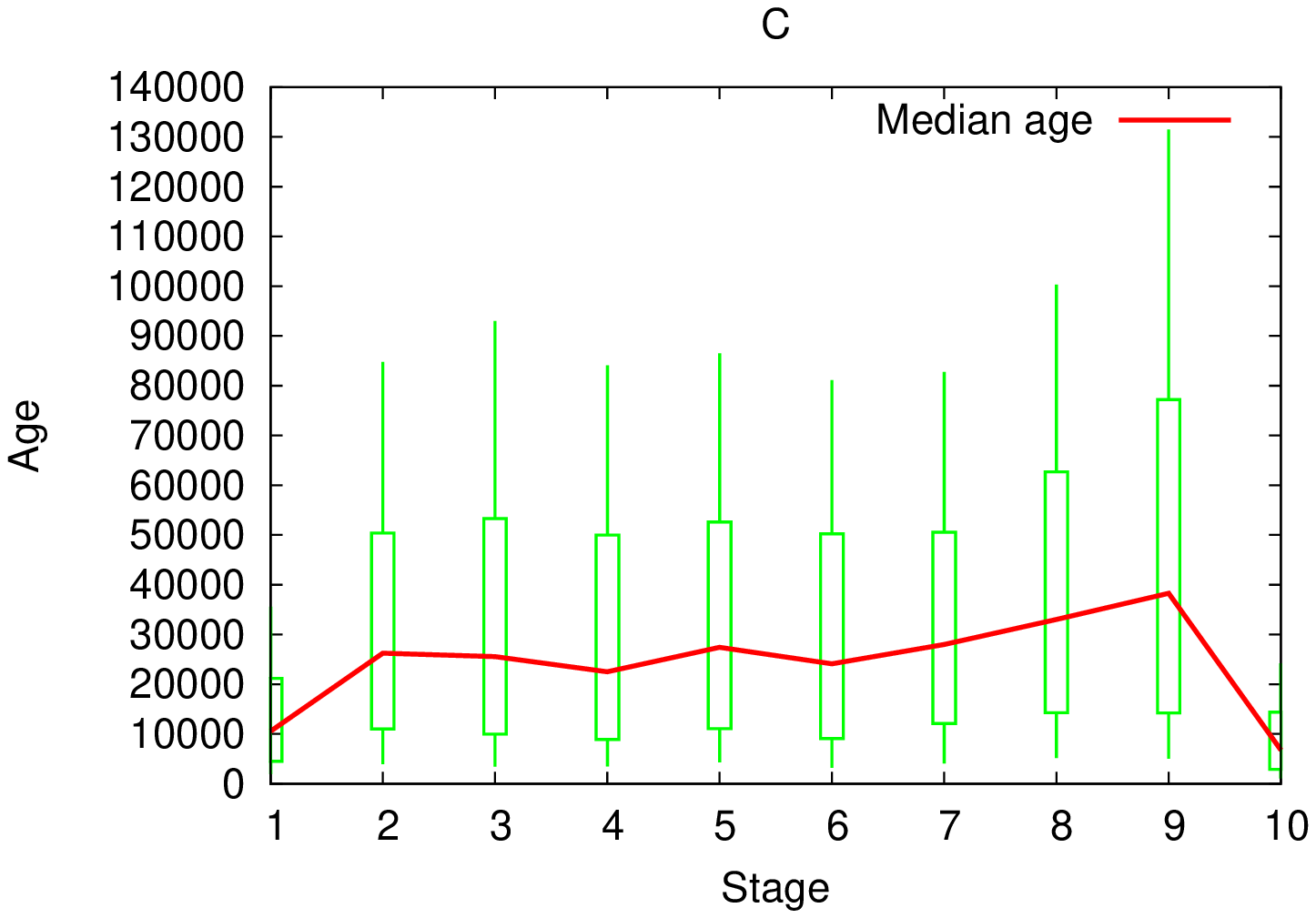}
\caption{Computational results with {\bf Model-1}.
Parameters: 10 runs with different initial populations,
$N$=10 individuals, $L$=10 stages,
specificity function $s(l)$=0.5 for all stages, 
$\Gamma$=100 genes at each stage initially,
RF parameter $z$=4, 1,000,000 generations,
probability of RW event $P_{RW}$=$10^{-4}$.
The red line is the median and the green boxes are the 10th, 25th, 75th,
and 90th percentiles, across all individuals and all runs.
(A) The hourglass score $H$ across evolutionary time.
(B) Lethality probability at each stage.
(C) Age of existing genes at the last generation.}
\label{fig:model-1_results}
\end{figure*}

\begin{figure*}[!htb]
\centering
\includegraphics[width=34mm,height=54mm,angle=270]{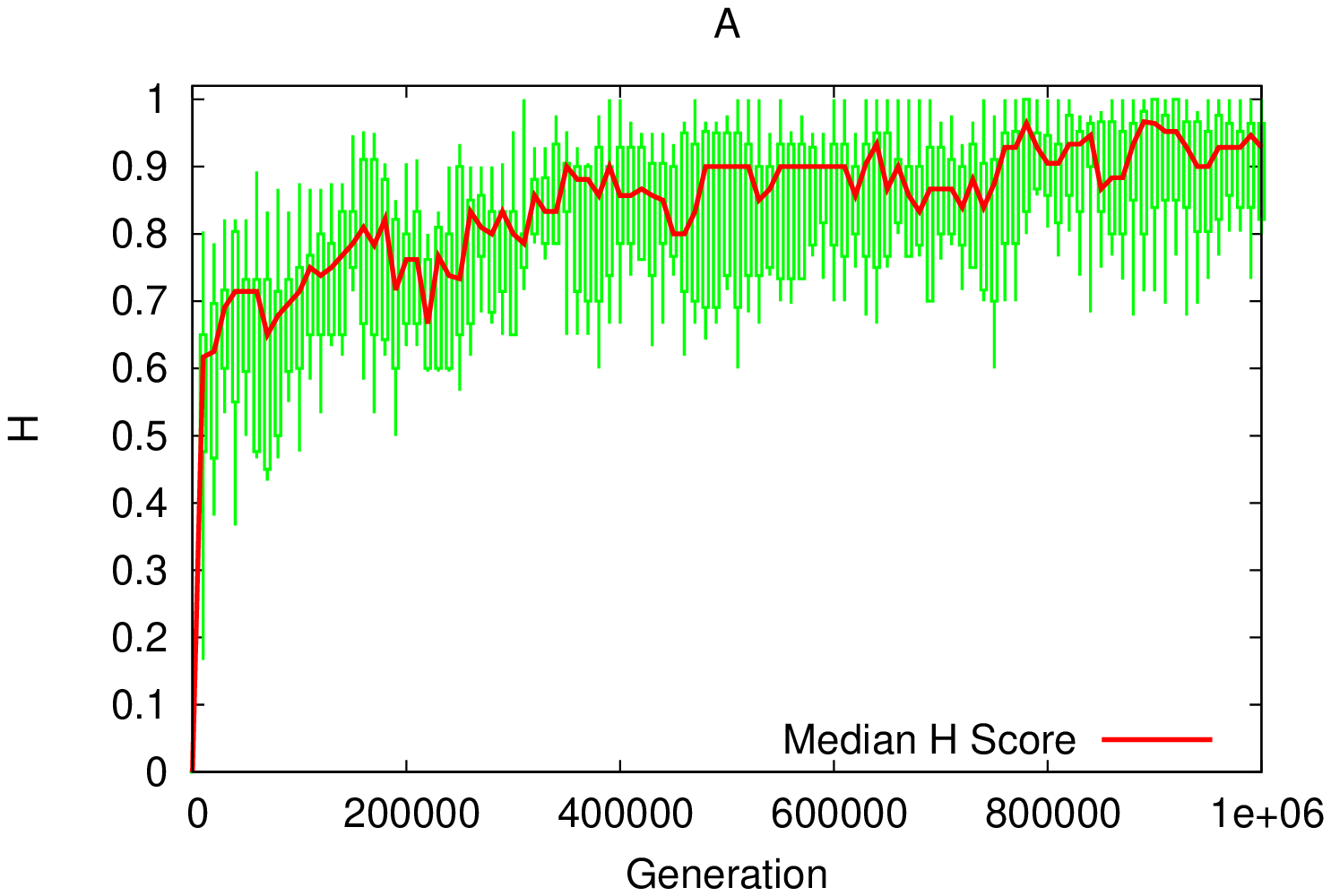}
\includegraphics[width=34mm,height=54mm,angle=270]{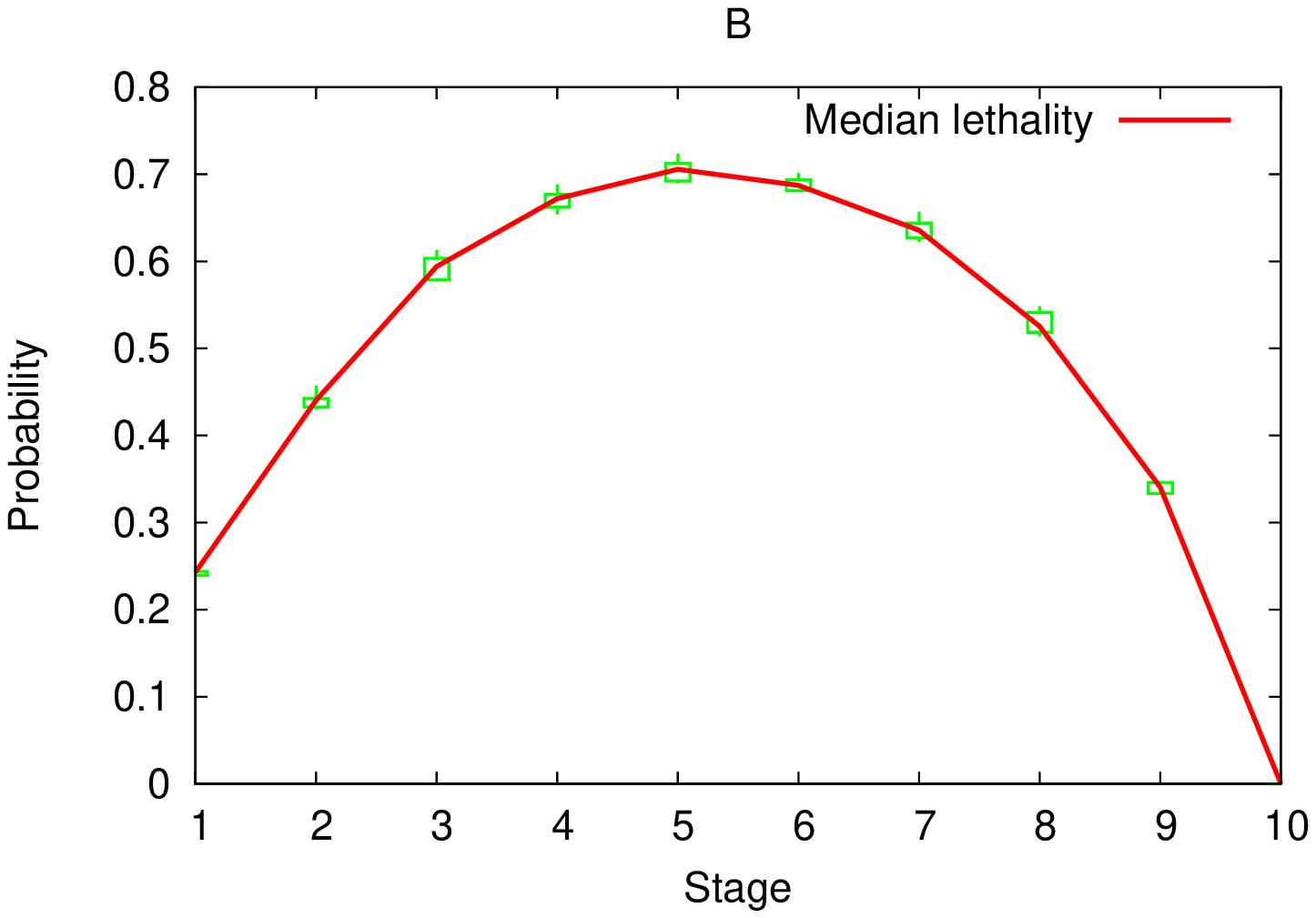}
\includegraphics[width=34mm,height=54mm,angle=270]{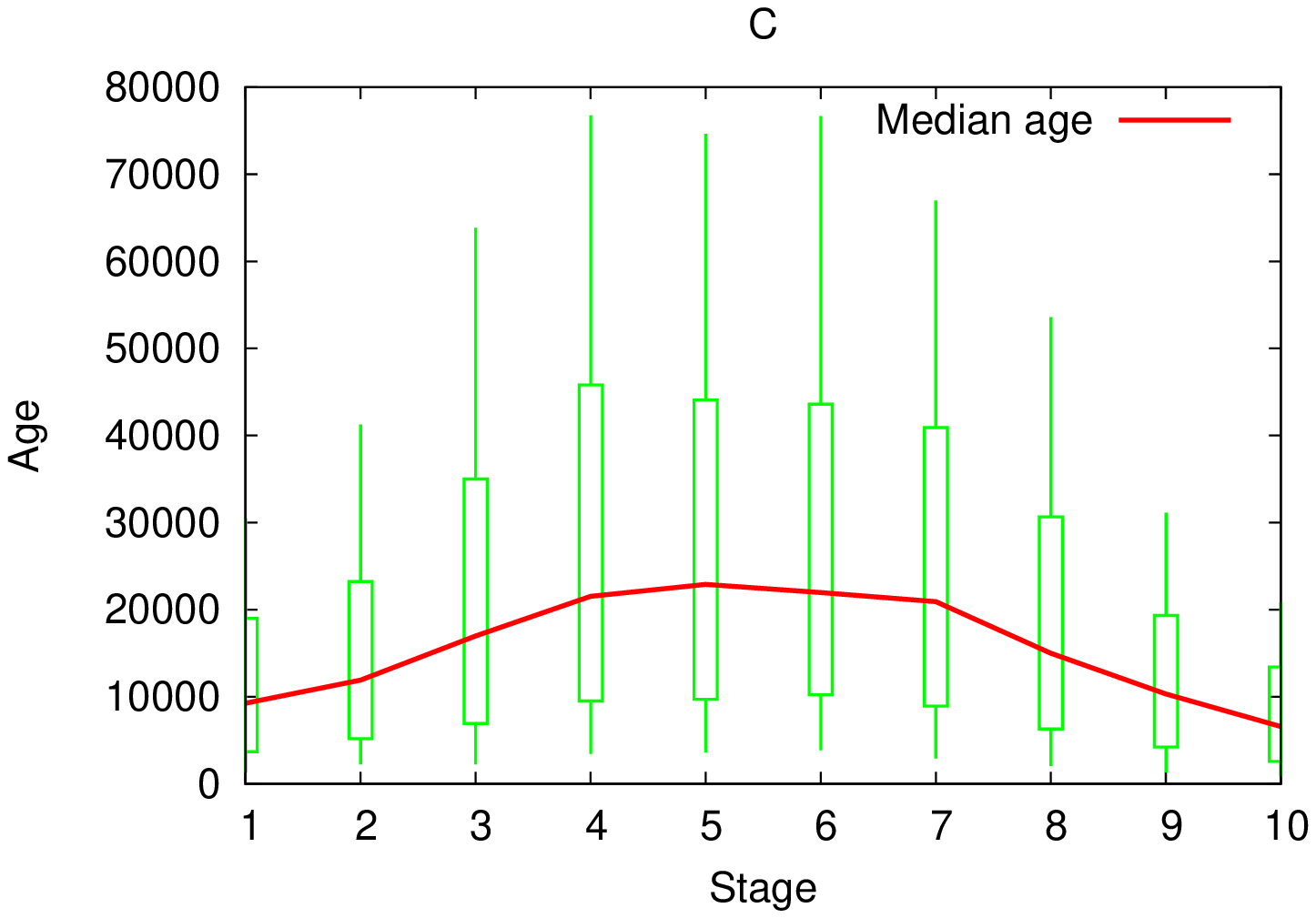}
\caption{Computational results with {\bf Model-3}.
Parameters: 10 runs with different initial populations,
$N$=10 individuals, $L$=10 stages,
specificity function $s(l)$=$l/L$ ($l$=$1\dots L-1$),
$\Gamma$=100 genes at each stage initially,
RF parameter $z$=4, 1,000,000 generations,
probability of RW event $P_{RW}$=$10^{-4}$.
The probability of gene duplication $P_{DP}$ is adjusted dynamically
so that the average DGEN size stays between 700 and 800 genes.
The red line is the median and the green boxes are the 10th, 25th, 75th,
and 90th percentiles, across all individuals and all runs.
(A) The hourglass score $H$ across evolutionary time.
(B) Lethality probability at each stage.
(C) Age of existing genes at the last generation.}
\label{fig:model-3_results}
\end{figure*}

\begin{figure*}[!htb]
\centering
\includegraphics[width=34mm,height=54mm,angle=270]{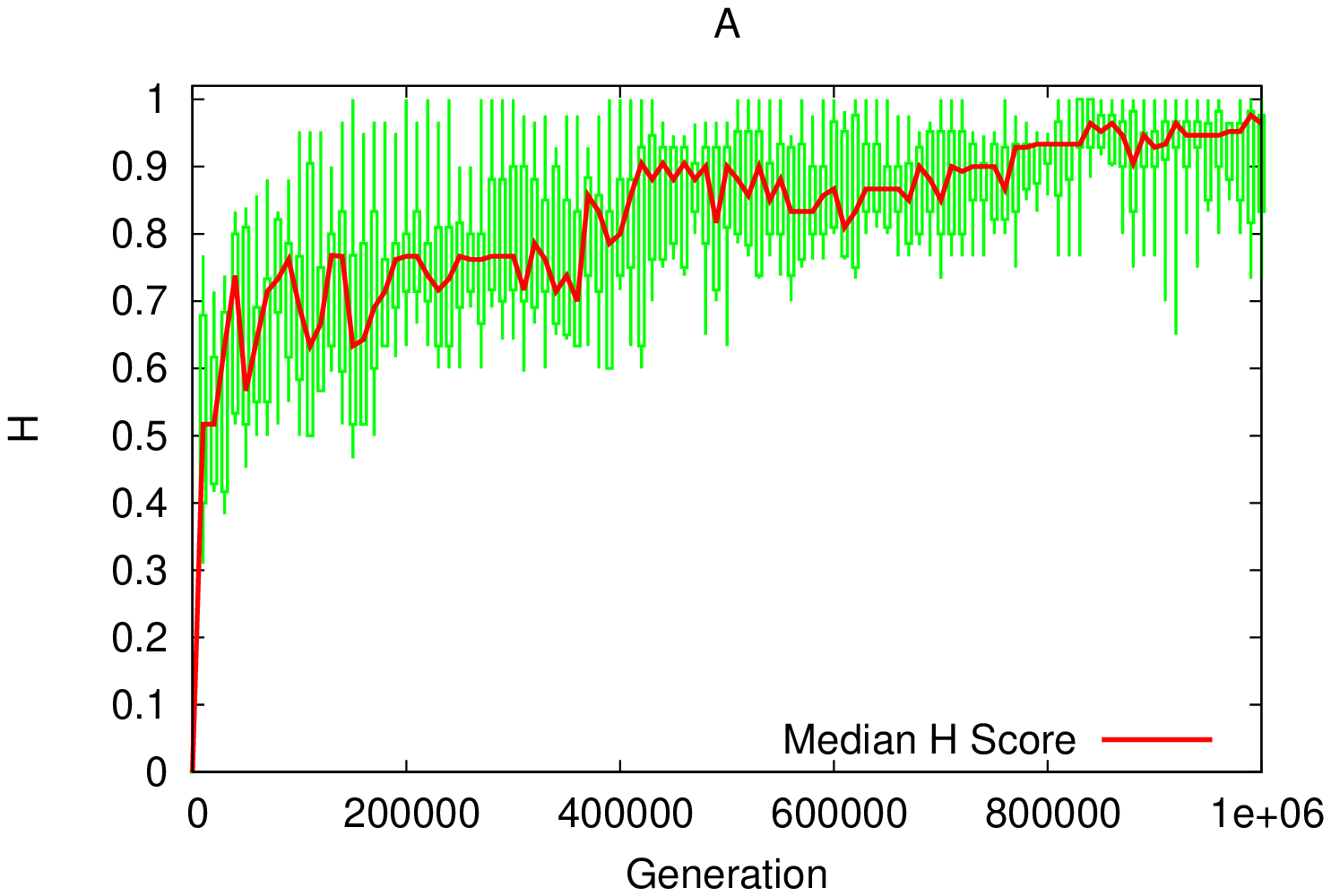}
\includegraphics[width=34mm,height=54mm,angle=270]{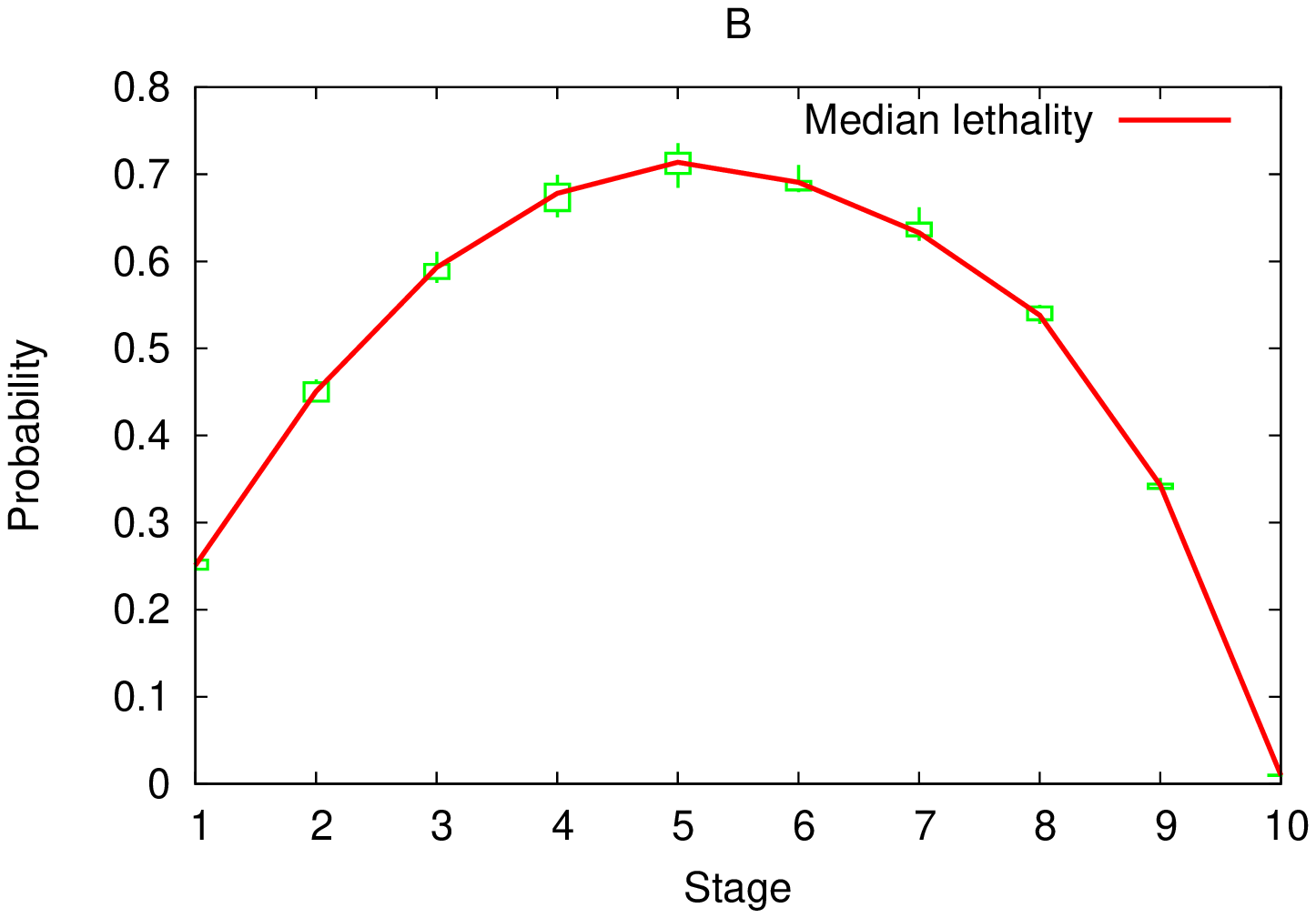}
\includegraphics[width=34mm,height=54mm,angle=270]{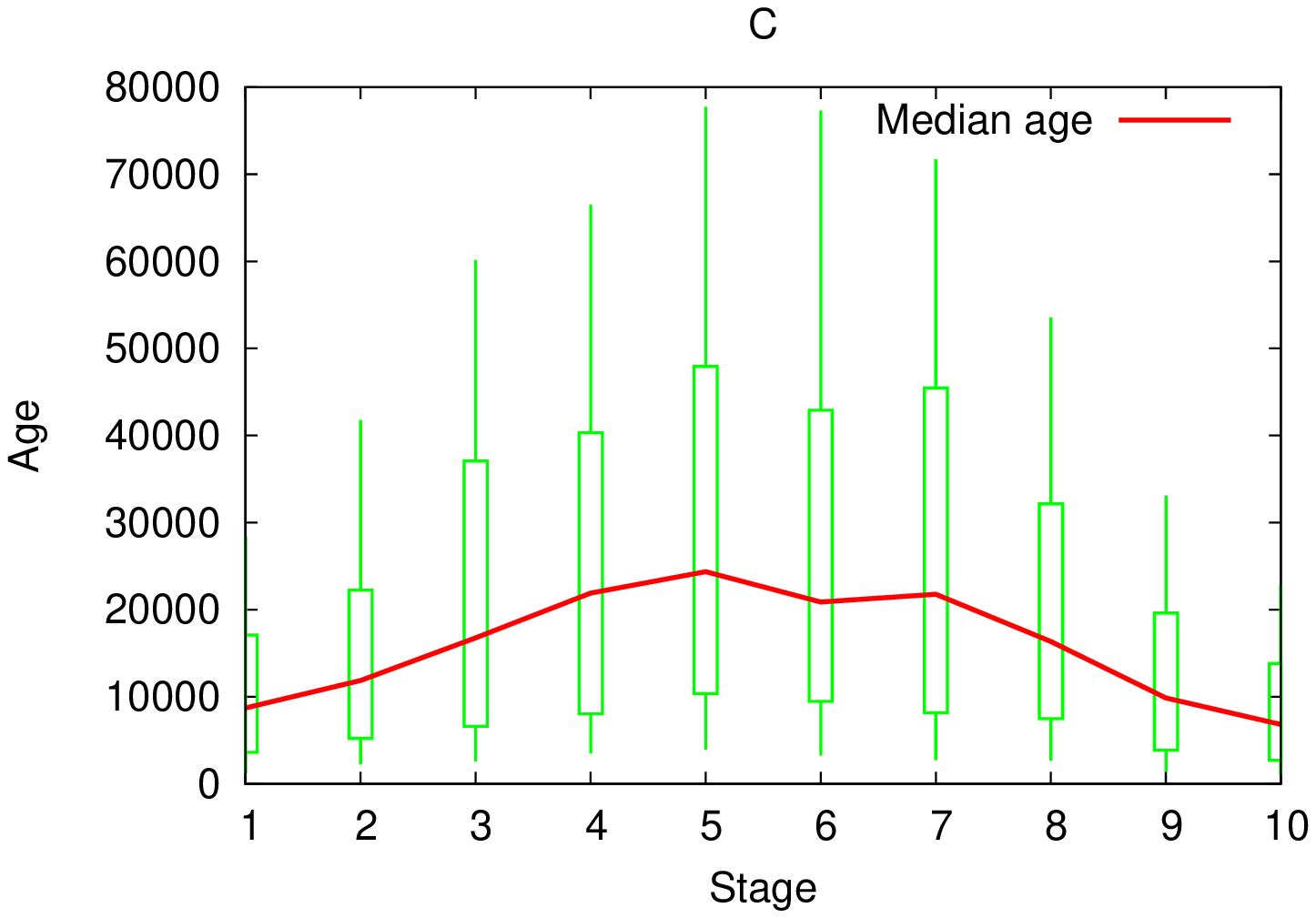}
\caption{Computational results with {\bf Model-4}.
Parameters: 10 runs with different initial populations,
$N$=10 individuals, $L$=10 stages,
specificity function $s(l)$=$l/L$ ($l$=$1\dots L-1$),
$\Gamma$=100 genes at each stage initially,
RF parameter $z$=4, 1,000,000 generations,
probability of RW event $P_{RW}$=$10^{-4}$.
The probability of gene duplication $P_{DP}$ is adjusted dynamically
so that the average DGEN size stays between 700 and 800 genes.
The probability of gene deletion (DL) is $P_{DL}=10^{-6}$.
The red line is the median and the green boxes are the 10th, 25th, 75th,
and 90th percentiles, across all individuals and all runs.
(A) The hourglass score $H$ across evolutionary time.
(B) Lethality probability at each stage.
(C) Age of existing genes at the last generation.}
\label{fig:model-4_results}
\end{figure*}

%\clearpage

\begin{figure*}[!htb]
\centering
\includegraphics[width=70mm,height=80mm,angle=270]{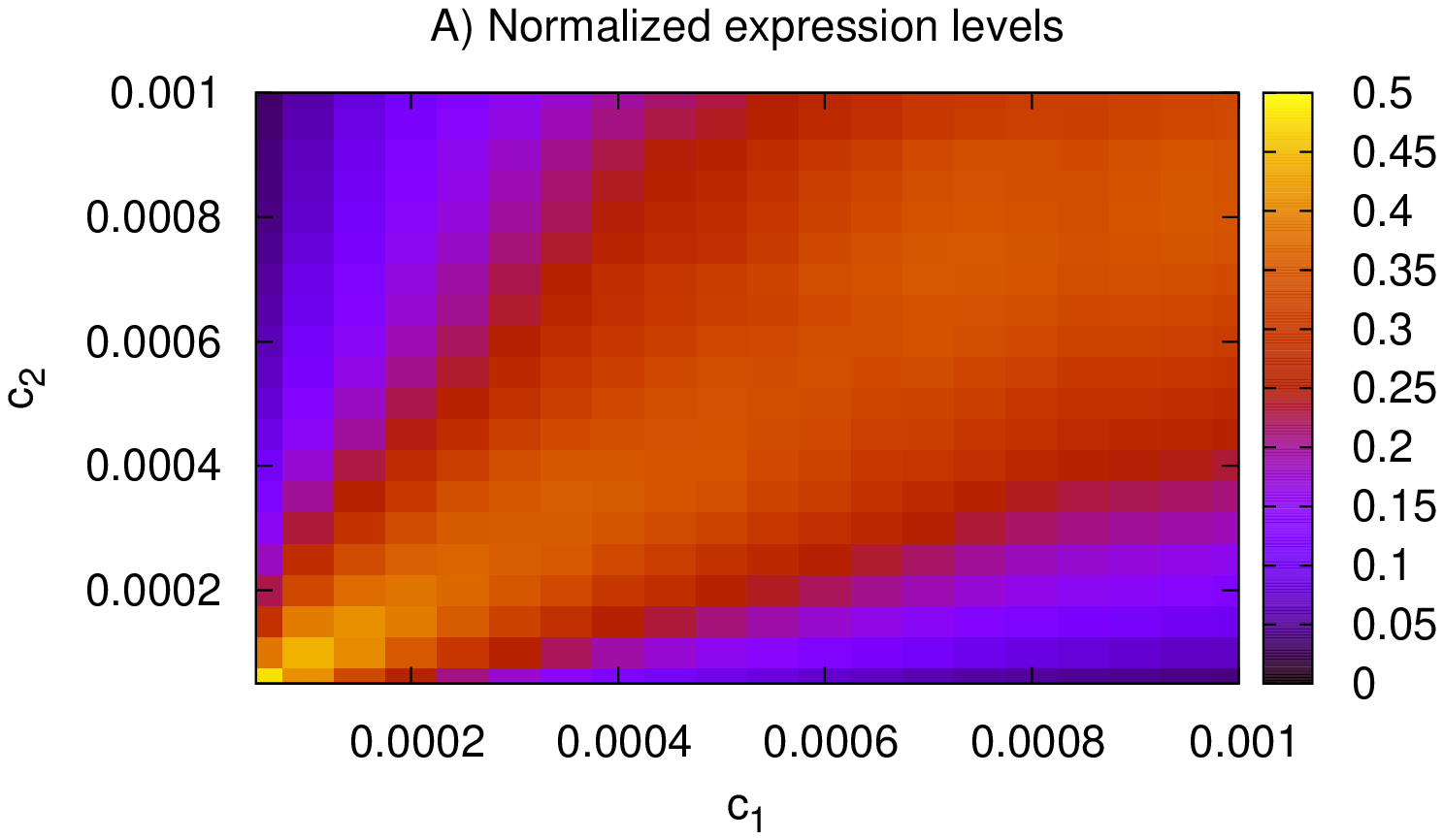}
\includegraphics[width=70mm,height=80mm,angle=270]{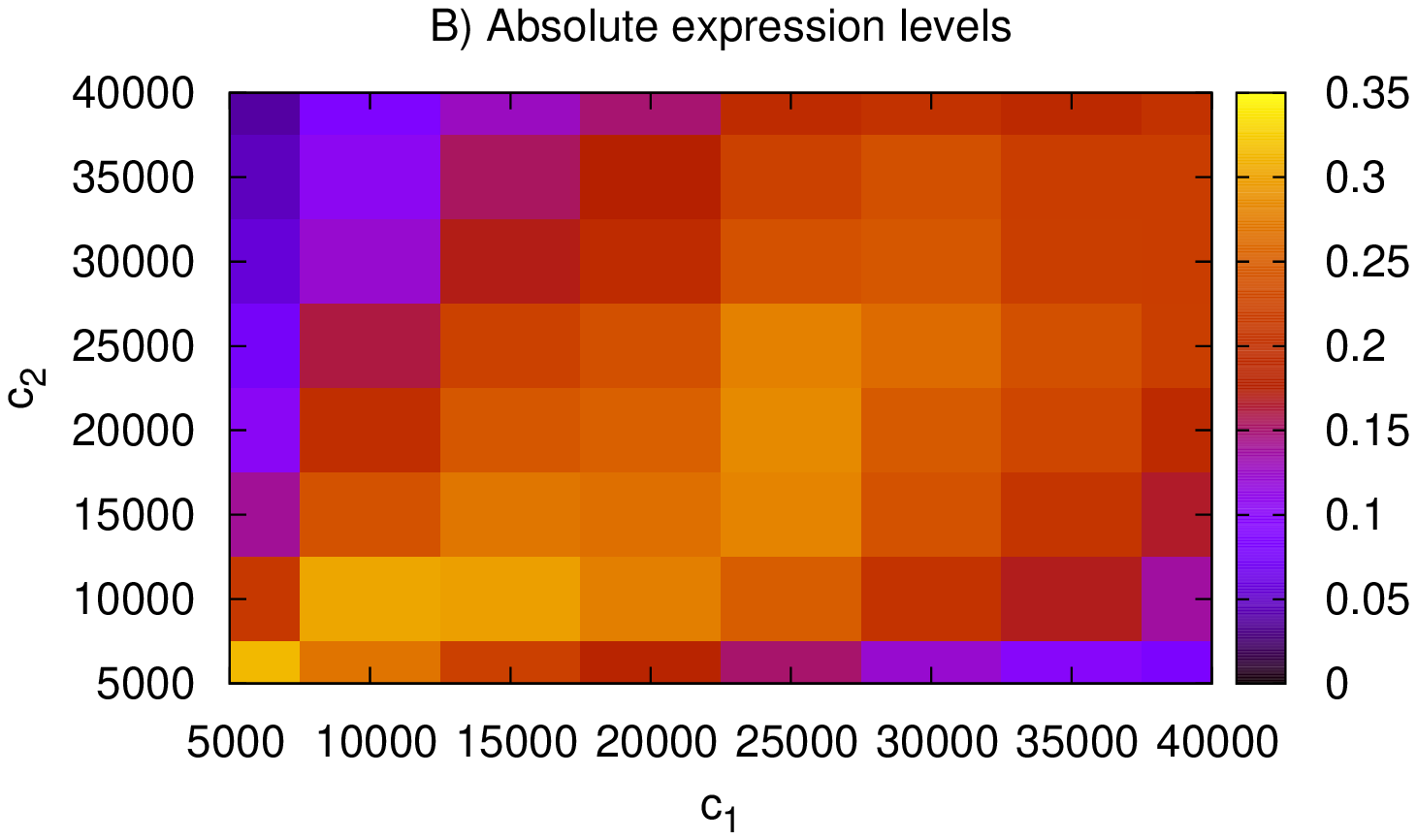}
\caption{
We evaluated the agreement between the two {\em Drosophila} datasets
in terms of the transitioning genes assigned to each stage-pair, 
considering only those genes that appear in both datasets.
Because the appropriate transition threshold may be different at
each dataset, we use a different threshold for each dataset,
say $c_1$ and $c_2$. For each pair $(c_1,c_2)$, we determine
the transitioning genes at each stage-pair with the corresponding dataset 
(i.e., $L-1$ pairs of gene sets), 
and then calculate the average Jaccard similarity across these $L-1$ pairs. 
The Jaccard similarity maps show this average across all stage-pairs for various
threshold pairs $(c_1, c_2)$.
In graph (A) with normalized expression levels, when the two thresholds are roughly equal, the average Jaccard similarity is 
as high as 50\%; this means that about 2/3 of the genes assigned to 
a certain stage-pair using one dataset are also assigned to the same stage-pair
using the other dataset. 
Graph (B) shows a similar Jaccard similarity map for the
case of absolute expression levels.
%The Jaccard similarity map to quantify the consistency of the 
%identified transitioning genes between the two datasets. 
%The map shows the average Jaccard index accross all stage-pairs.
}
\label{fig:jacks}
\end{figure*}

\begin{figure*}[!htb]
\centering
\includegraphics[width=50mm,height=70mm,angle=270]{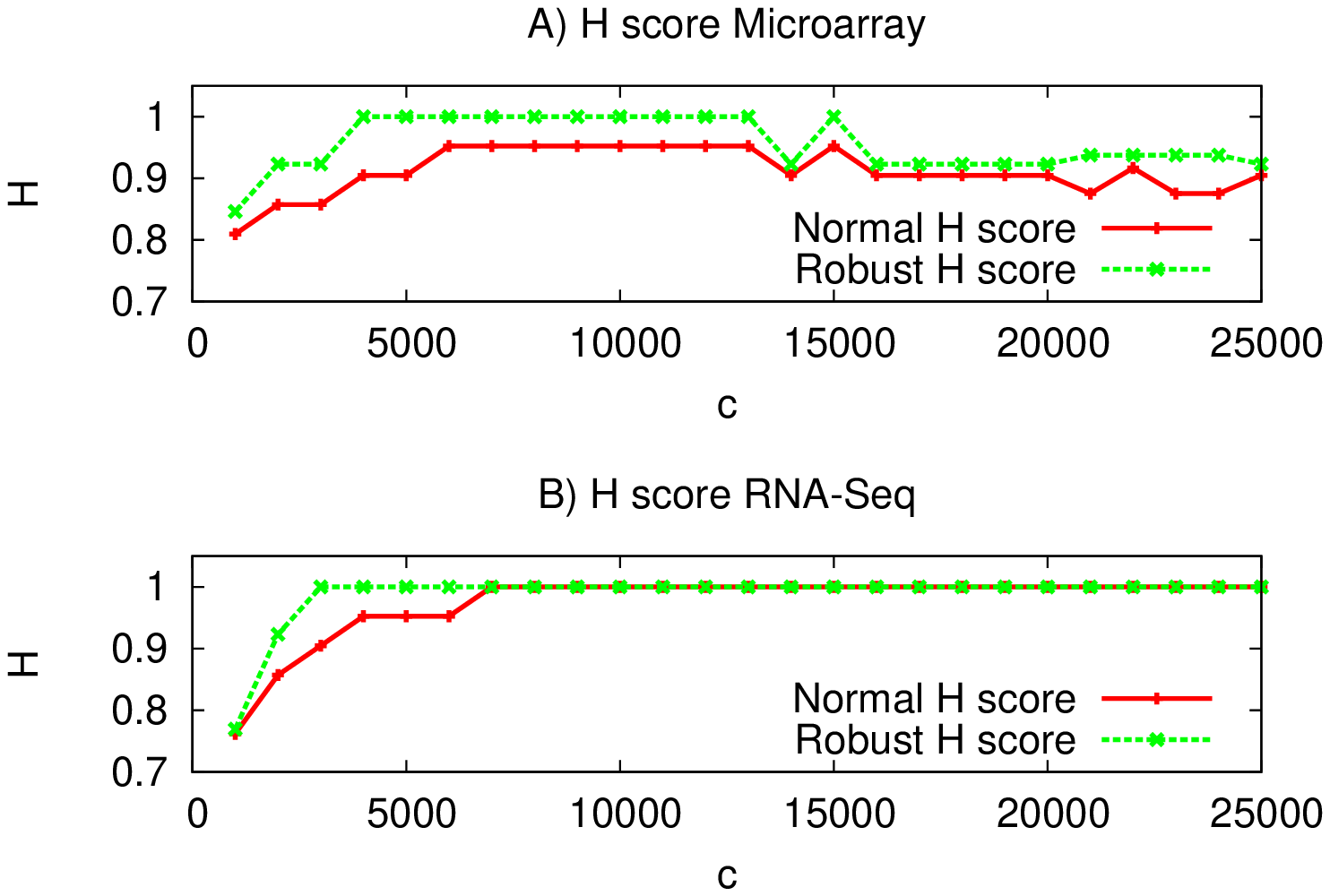}
\includegraphics[width=50mm,height=70mm,angle=270]{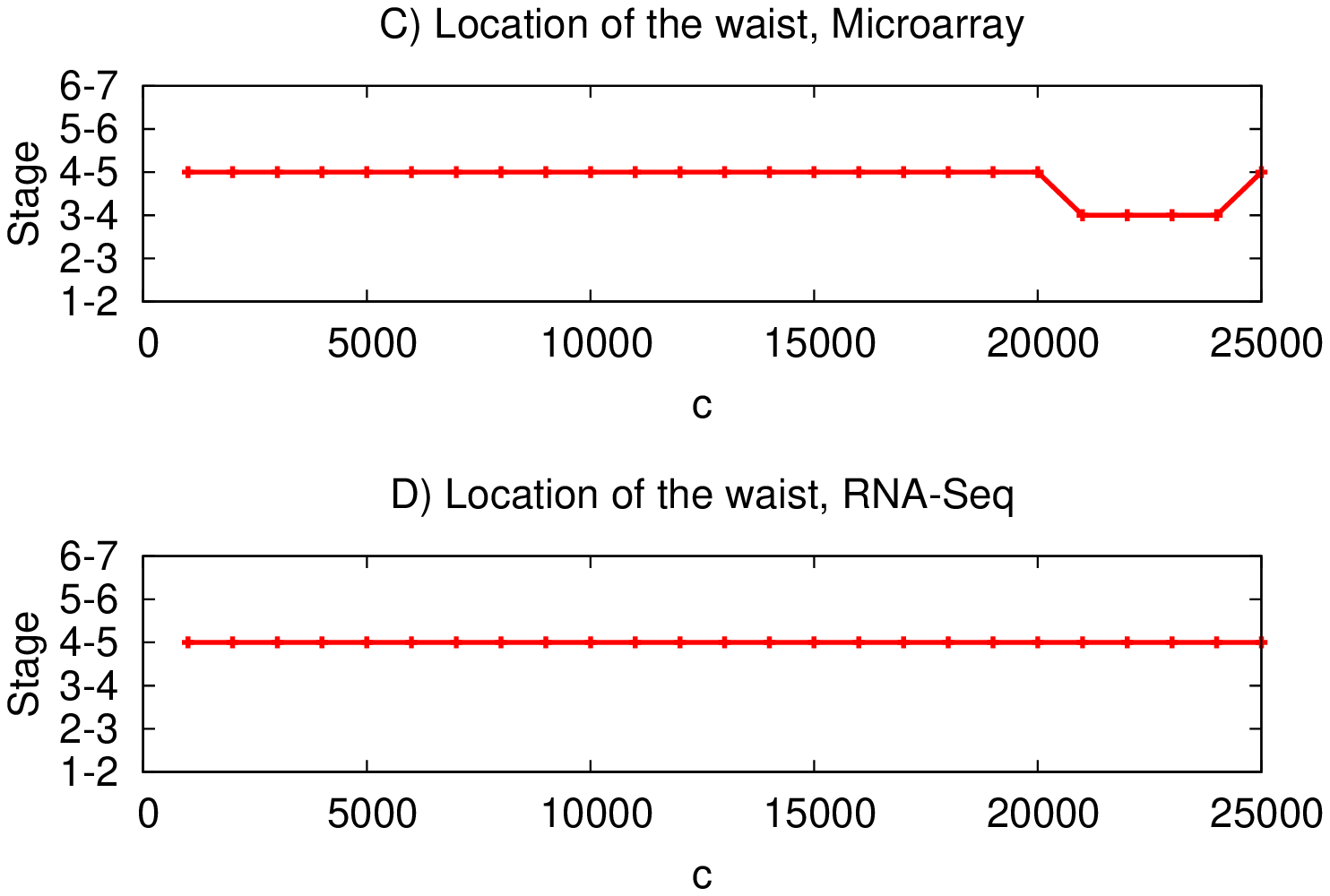}
\includegraphics[width=40mm,height=60mm,angle=270]{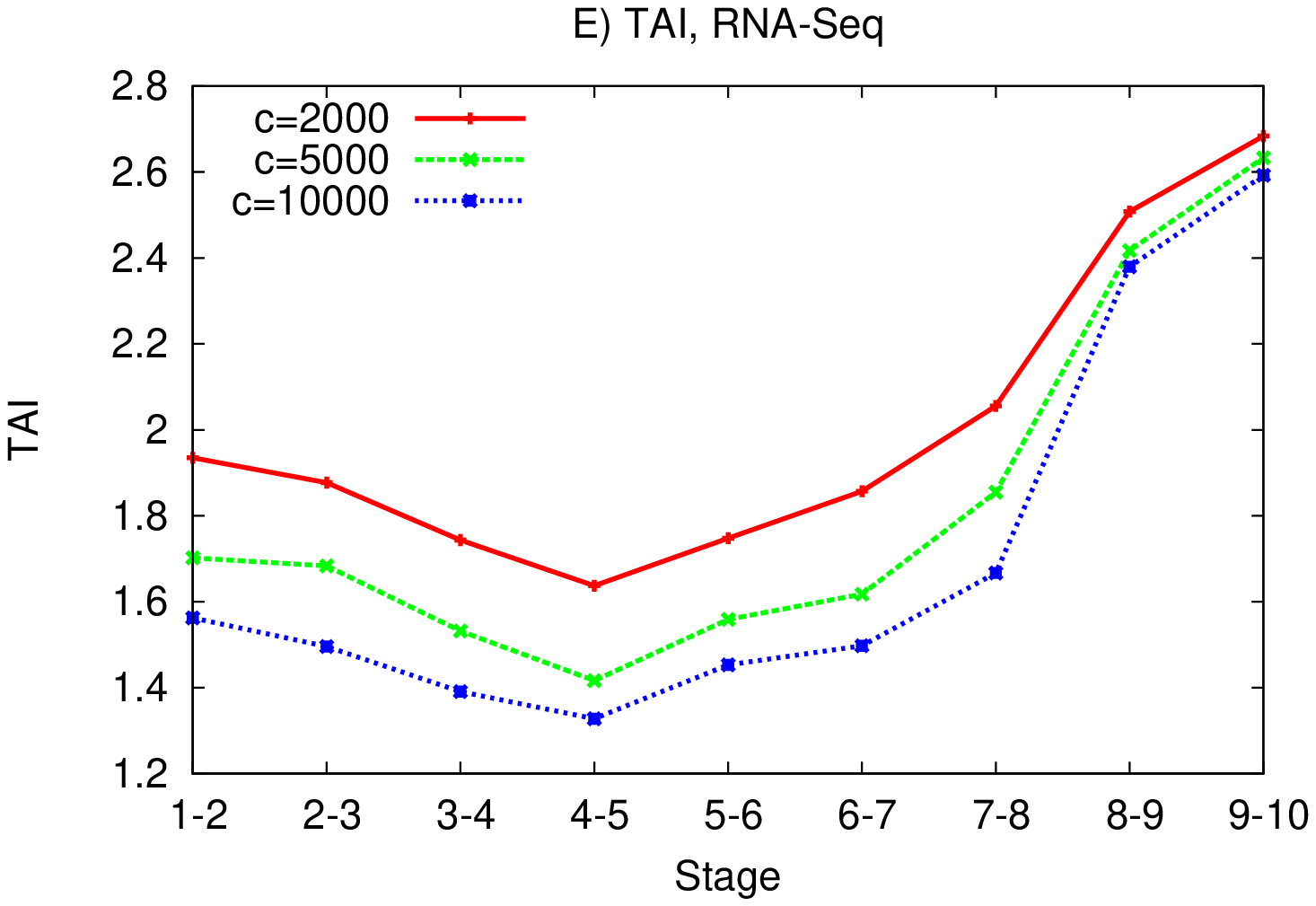}
\includegraphics[width=40mm,height=60mm,angle=270]{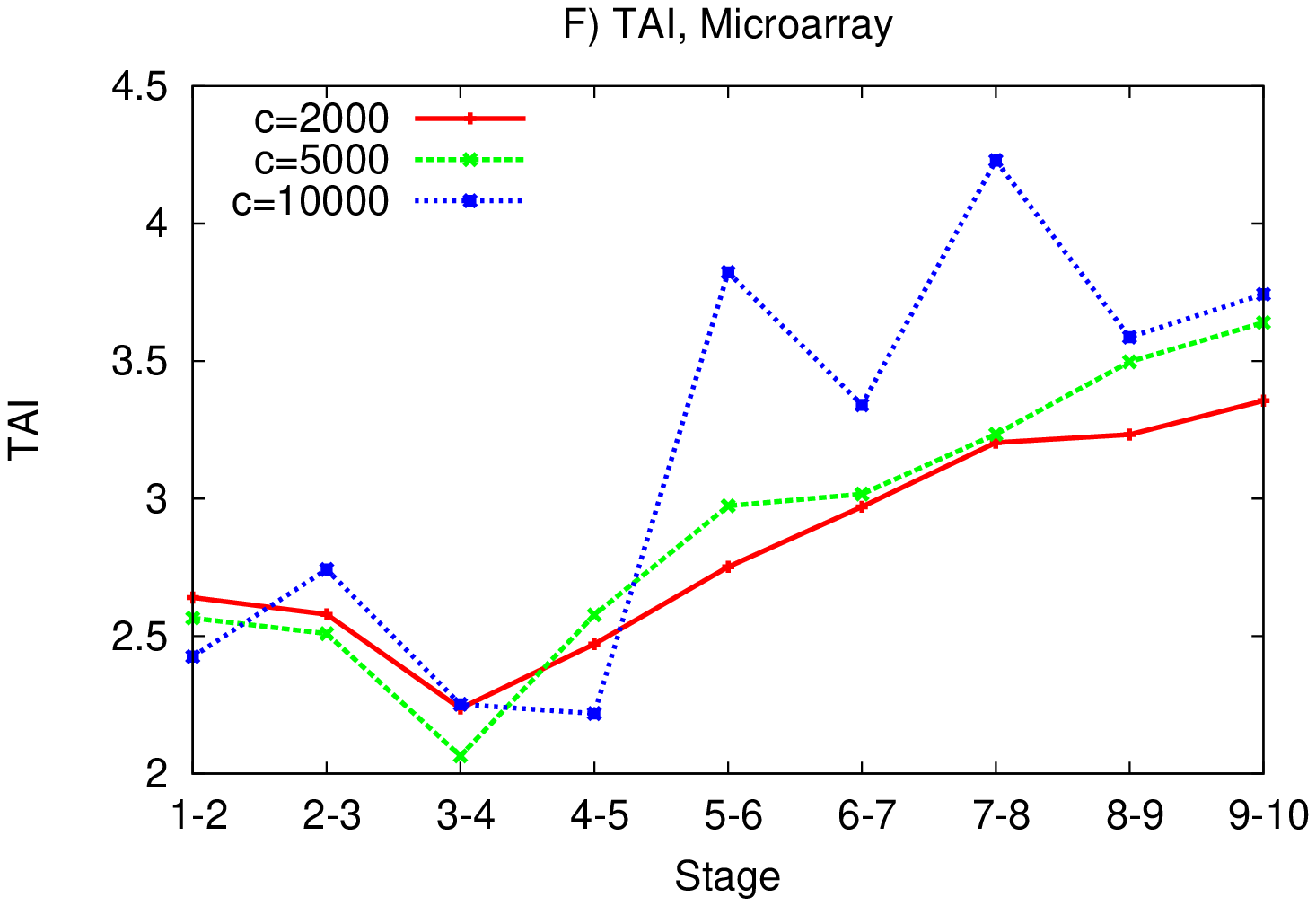}
\includegraphics[width=50mm,height=70mm,angle=270]{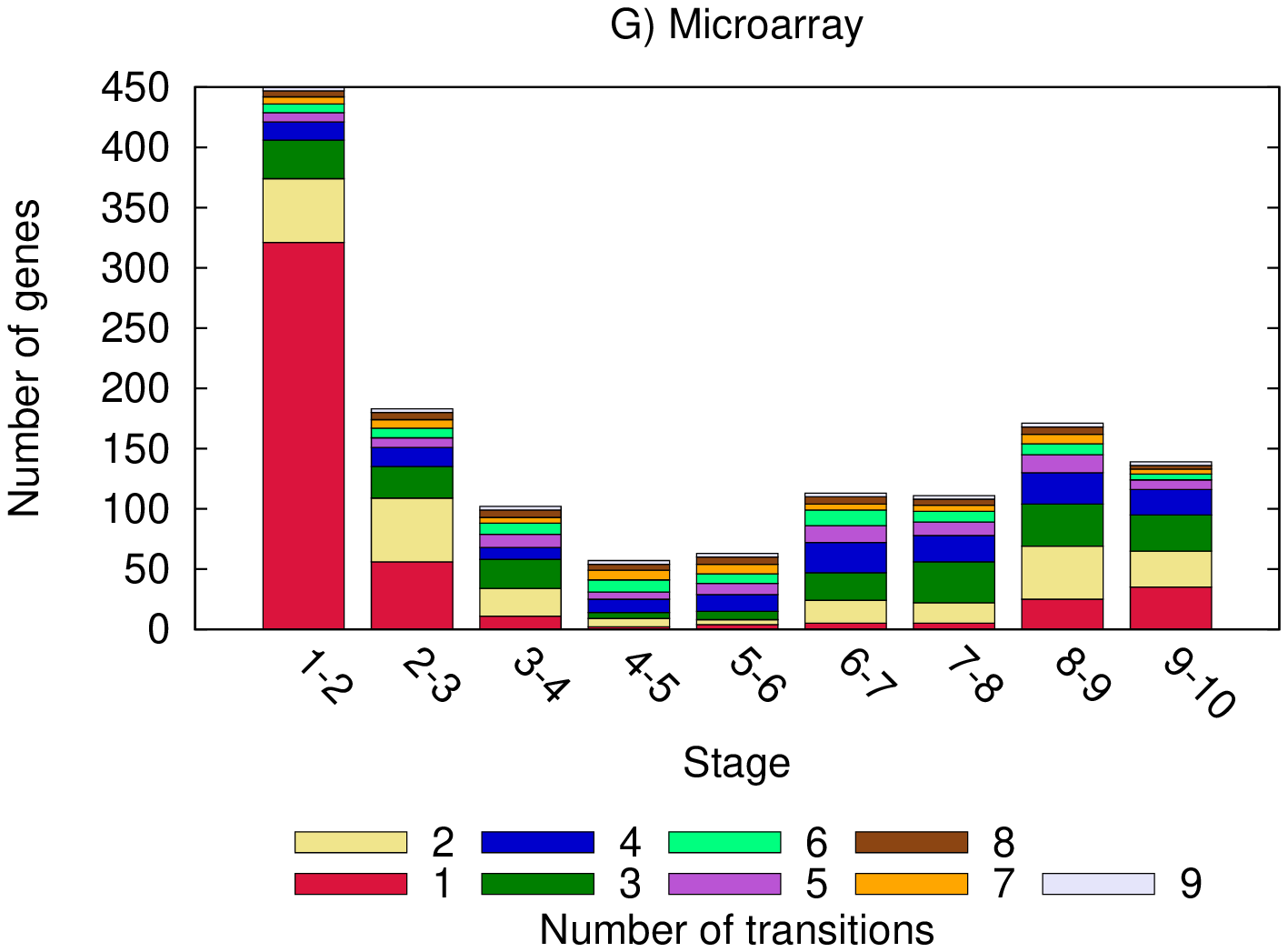}
\includegraphics[width=50mm,height=70mm,angle=270]{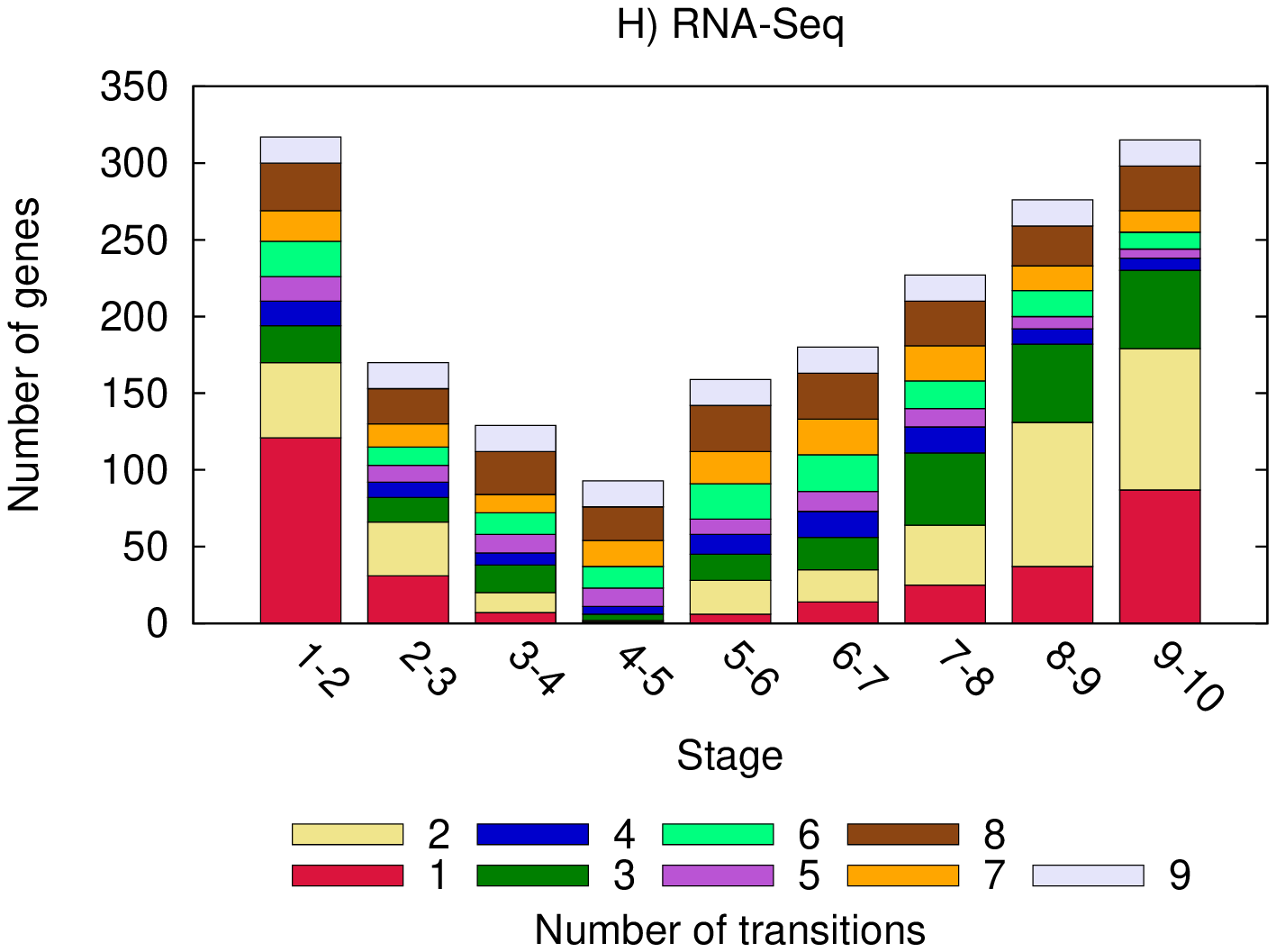}
\vspace{10 pt}
\caption{{\em Drosophila} results using absolute expression levels.
Graphs (A) and (B) show the hourglass score (normal and robust)
as a function of the transition threshold $c$ for the two datasets.
Graphs (C) and (D) show the location of the hourglass waist (stage-pair)
as a function of the transition threshold $c$ for the two datasets.
Graphs (E) and (F) show the Transcriptome Age Index
of transitioning genes for three different values of $c$ (chosen so that the
number of genes with known age index assigned to each stage is at least 10)
for the two datasets.
Graph (G) shows the transitioning genes for the Microarray dataset
with $c$=5000.
The transitioning genes constitute 21\% of all genes in that dataset.
62\% of those genes transition in a single stage-pair. 
Of the remaining, 58\% transition only in consecutive stage-pairs.
Similarly, graph (H) shows the transitioning genes for the RNA-Seq dataset
with $c$=10000. 
The transitioning genes constitute 5\% of all genes in that dataset.
45\% of those genes transition in a single stage-pair. 
Of the remaining, 53\% transition only in consecutive stage-pairs.}
%Graph (G) shows a Jaccard similarity map to quantify the consistency of the
%identified transitioning genes between the two datasets.
%The map shows the average Jaccard index accross all stage-pairs.
\label{fig:data_analysis_results2}
\end{figure*}

\begin{figure*}[!htb]
\centering
\includegraphics[width=50mm,height=70mm,angle=270]{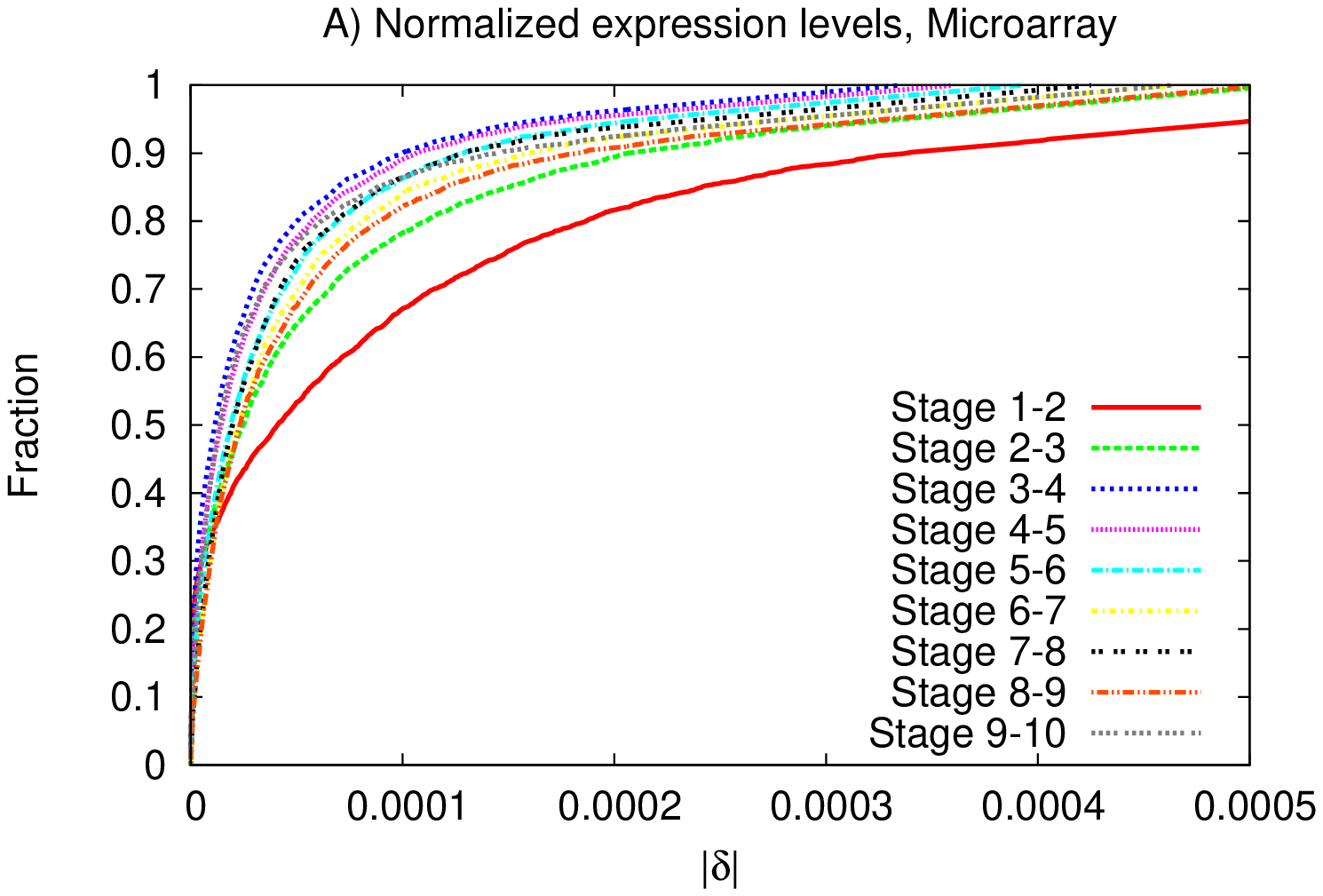}
\includegraphics[width=50mm,height=70mm,angle=270]{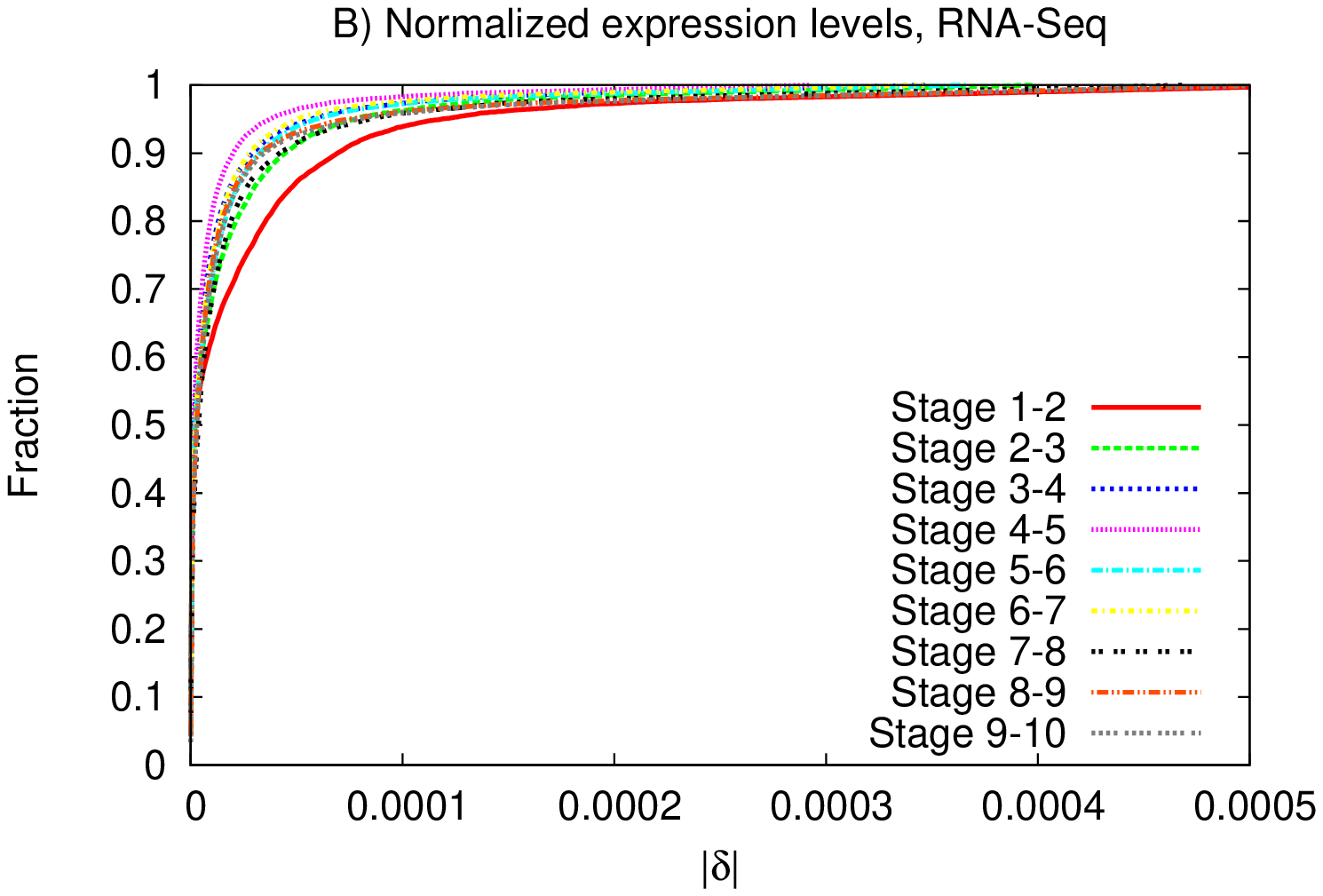}
\includegraphics[width=50mm,height=70mm,angle=270]{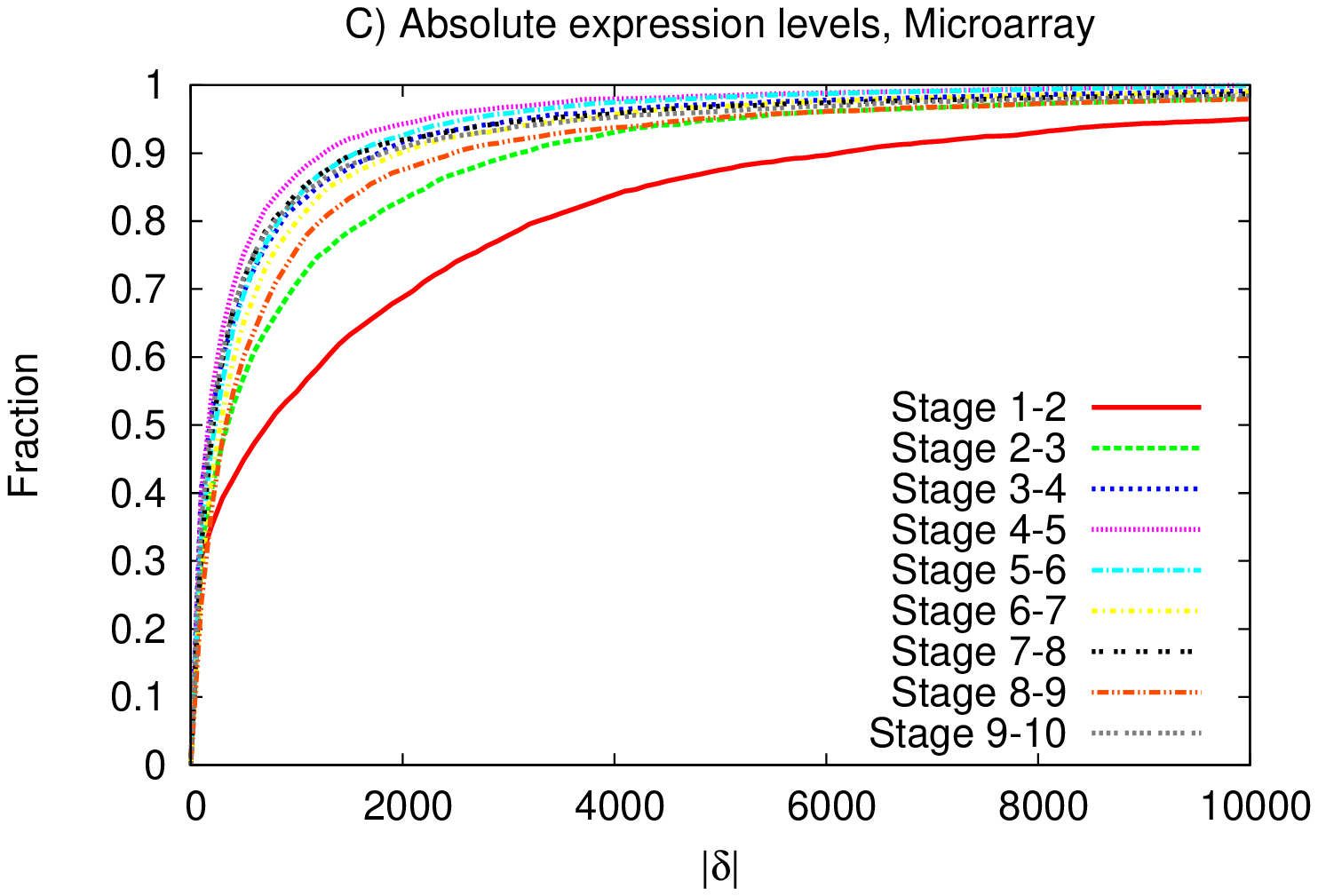}
\includegraphics[width=50mm,height=70mm,angle=270]{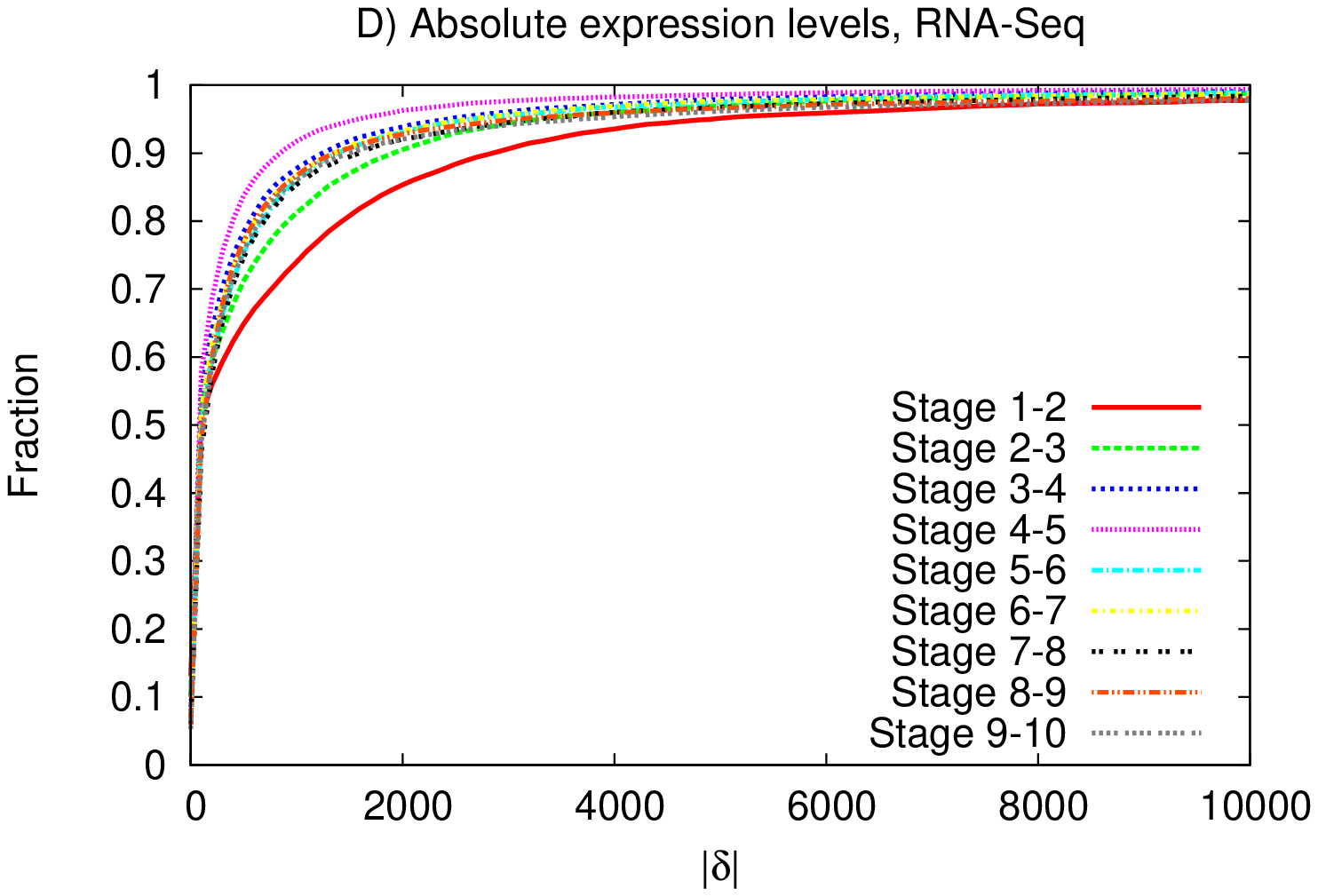}
\caption{{\em Drosophila} data: CDFs of the expression level absolute variations $|\delta|$ across successive stage-pairs. 
(A) normalized expressions, Microarray, (B) normalized expressions, RNA-Seq, 
(C) absolute expressions, Microarray, (D) absolute expressions, RNA-Seq.}
\label{fig:CDFs}
\end{figure*}

\begin{figure*}[!htb]
\centering
\includegraphics[width=50mm,height=70mm,angle=270]{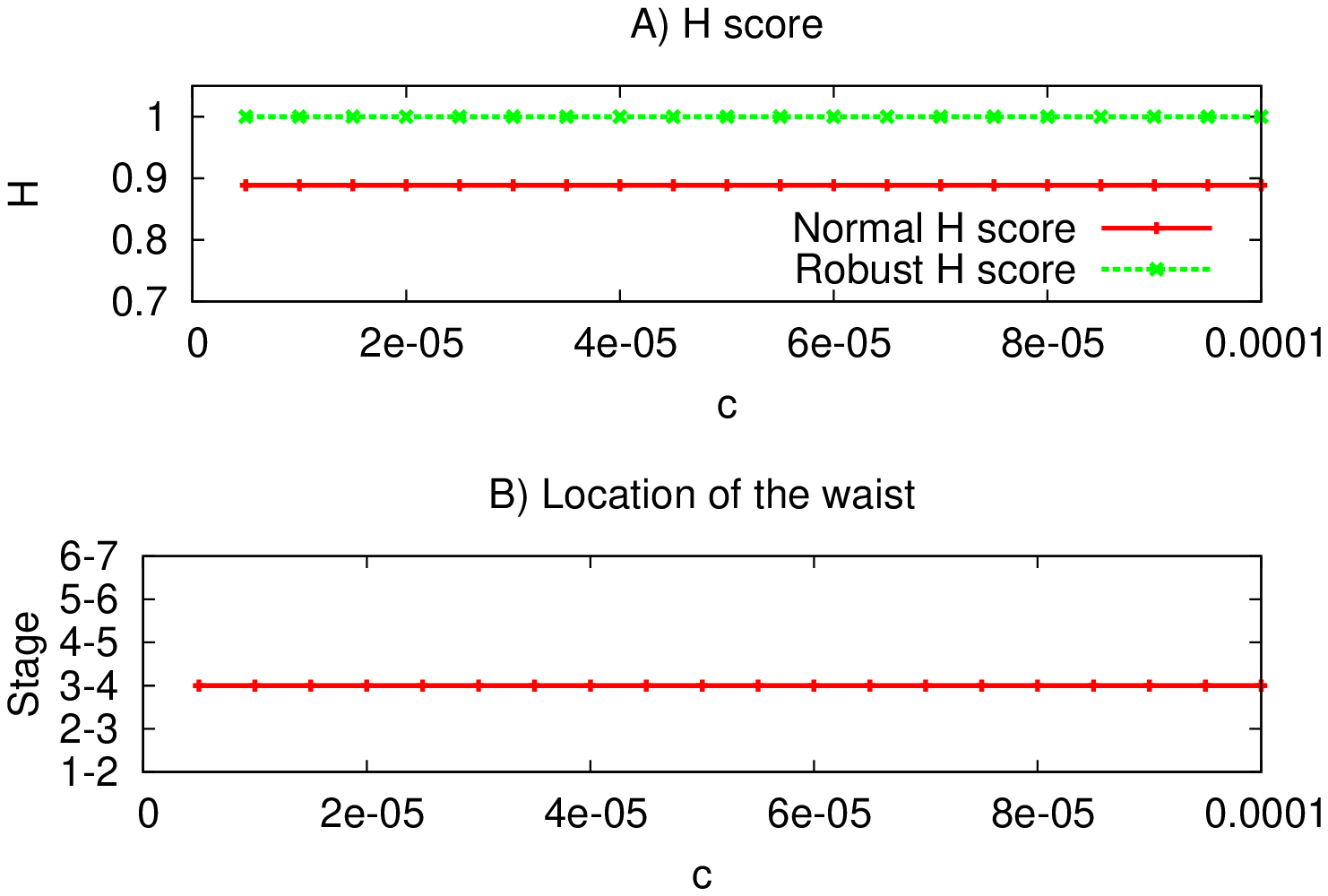}
\includegraphics[width=50mm,height=70mm,angle=270]{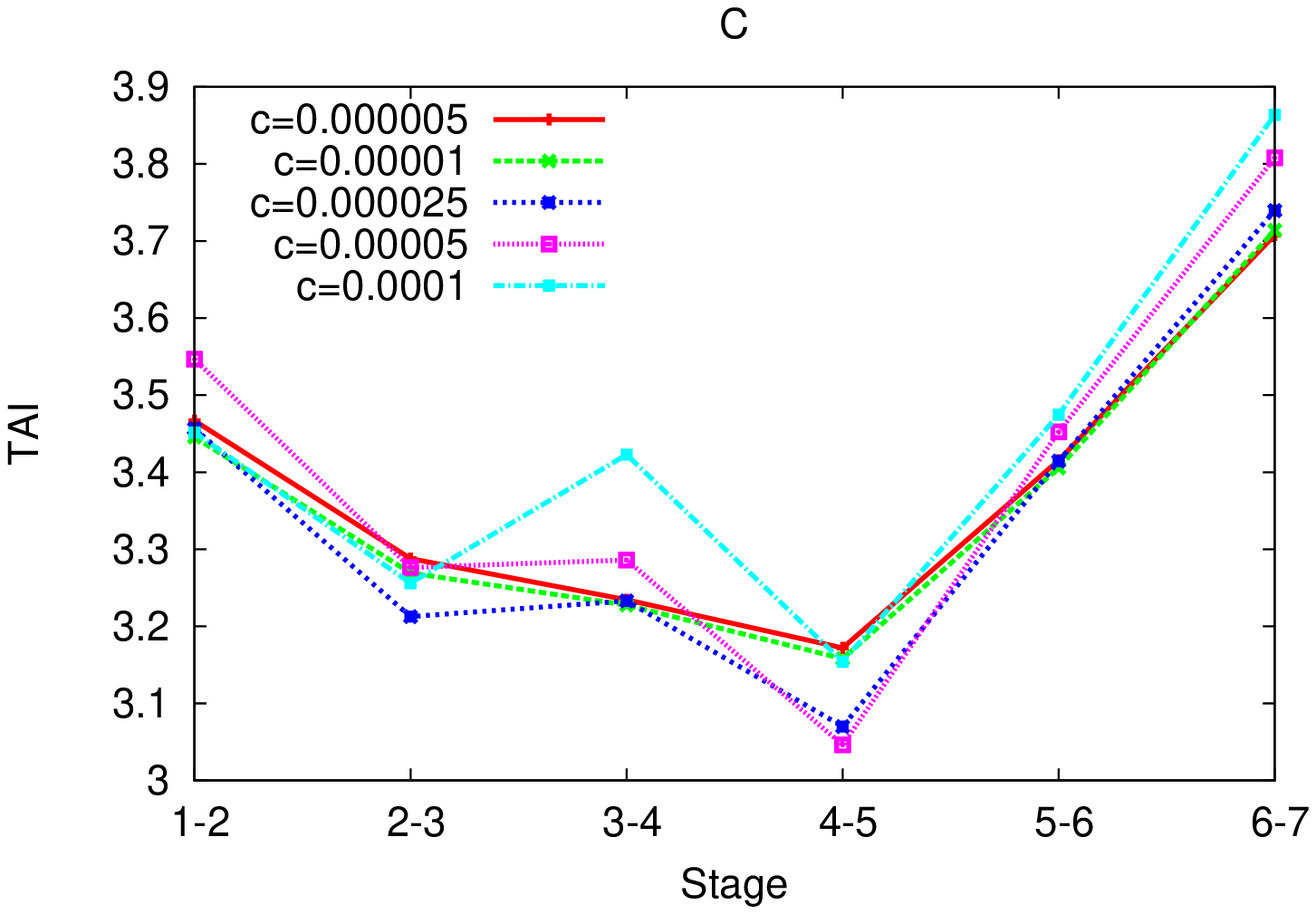}
\includegraphics[width=50mm,height=70mm,angle=270]{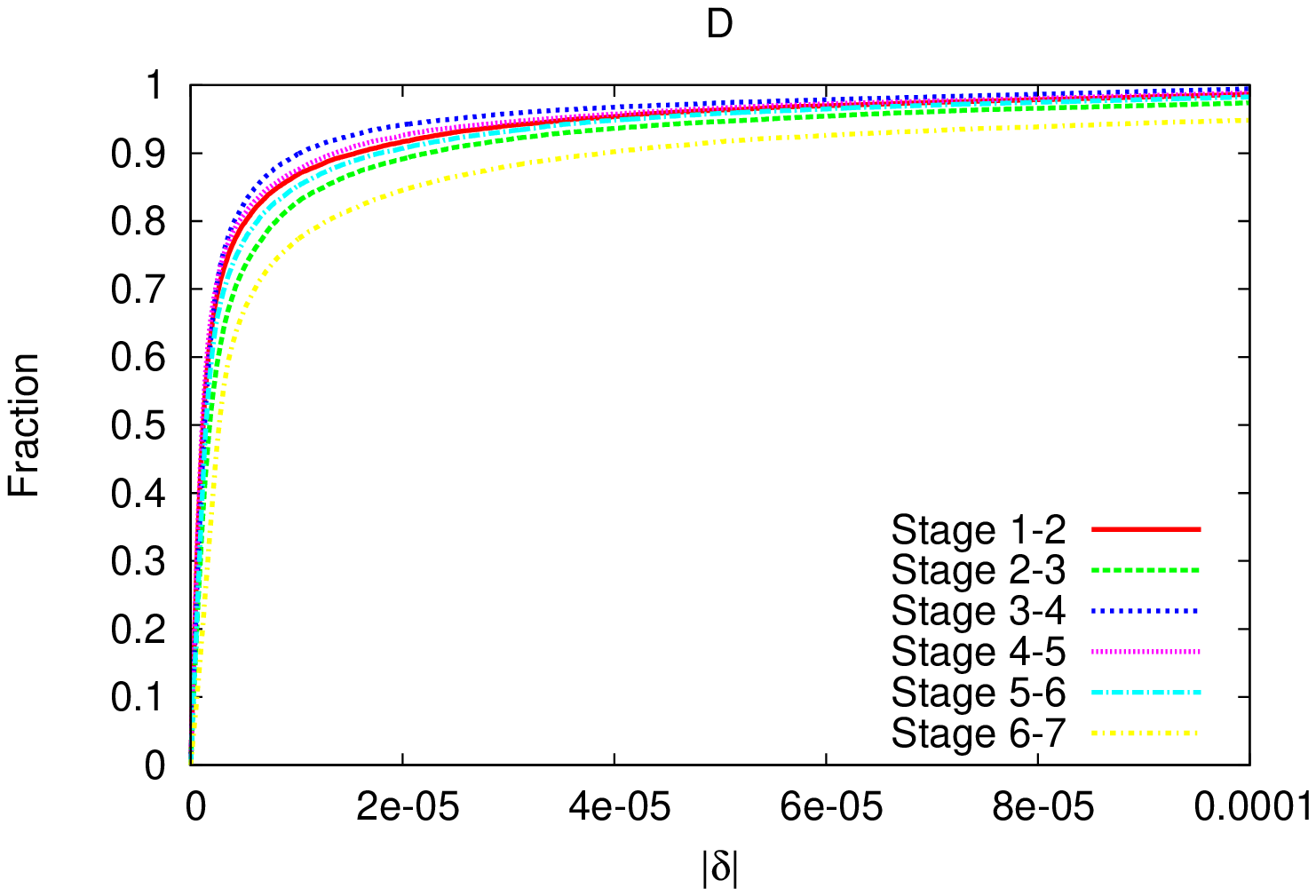}
\includegraphics[width=50mm,height=70mm,angle=270]{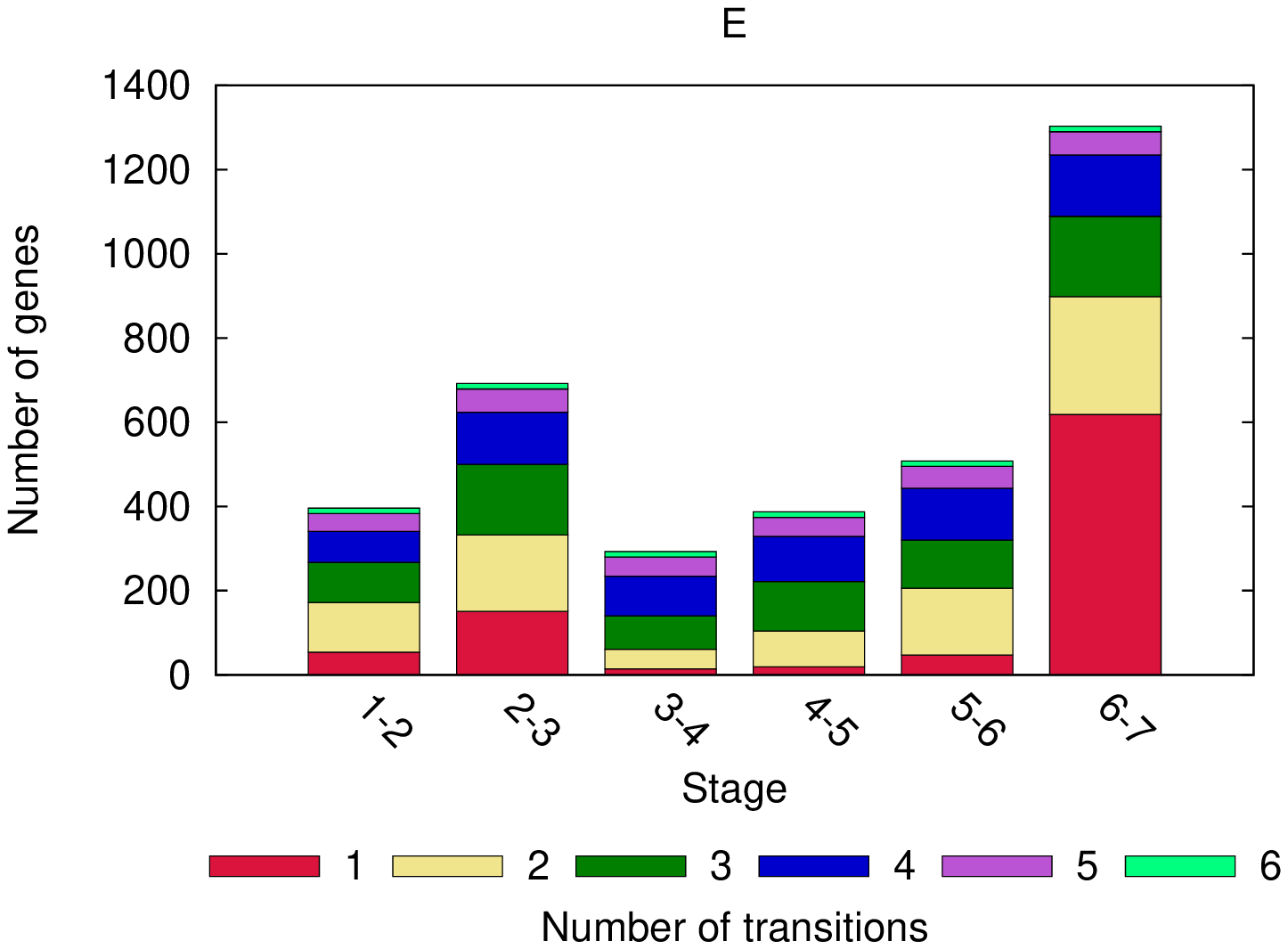}
\vspace{10 pt}
\caption{{\em Arabidopsis thaliana} results  using normalized expression levels.
Data source: Microarray expression levels for 25,207 genes and seven developmental stages \cite{quint2012transcriptomic, xiang2011genome}. 
The transcriptome age values are available from \cite{quint2012transcriptomic}.
Graph (A) shows the hourglass score (normal and robust)
as a function of the transition threshold $c$.
Graph (B) shows the location of the hourglass waist (stage-pair)
as a function of the transition threshold $c$.
Graph (C) shows the Transcriptome Age Index of transitioning genes 
for five different values of $c$ (chosen so that the number of genes with known age index assigned to each stage-pair is at least 290).
Graph (D) shows the CDFs of the expression level absolute variations 
$|\delta|$ across successive stage-pairs. 
Note that in this case the hourglass waist (in terms of number of 
transitioning genes) appears in stage-pair (3,4),
while the oldest genes appear in the next stage-pair. 
Graph (E) shows the transitioning genes with $c$=0.0001.
The transitioning genes constitute 7\% of all genes in that dataset.
49\% of those genes transition in a single stage-pair. 
Of the remaining, 36\% transition only in consecutive stage-pairs.
}
\label{fig:data_analysis_plant1}
\end{figure*}

\begin{figure*}[!htb]
\centering
\includegraphics[width=50mm,height=70mm,angle=270]{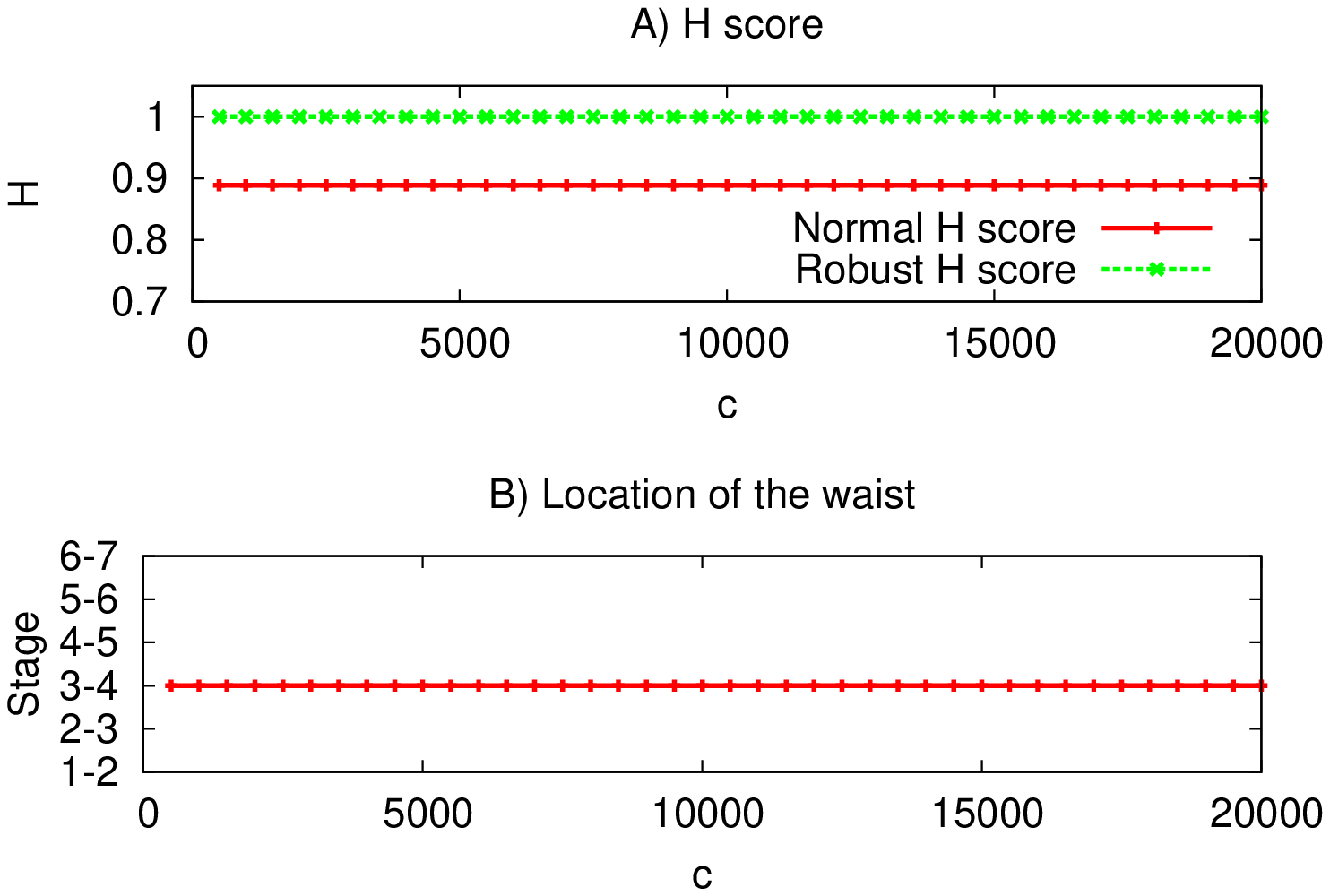}
\includegraphics[width=50mm,height=70mm,angle=270]{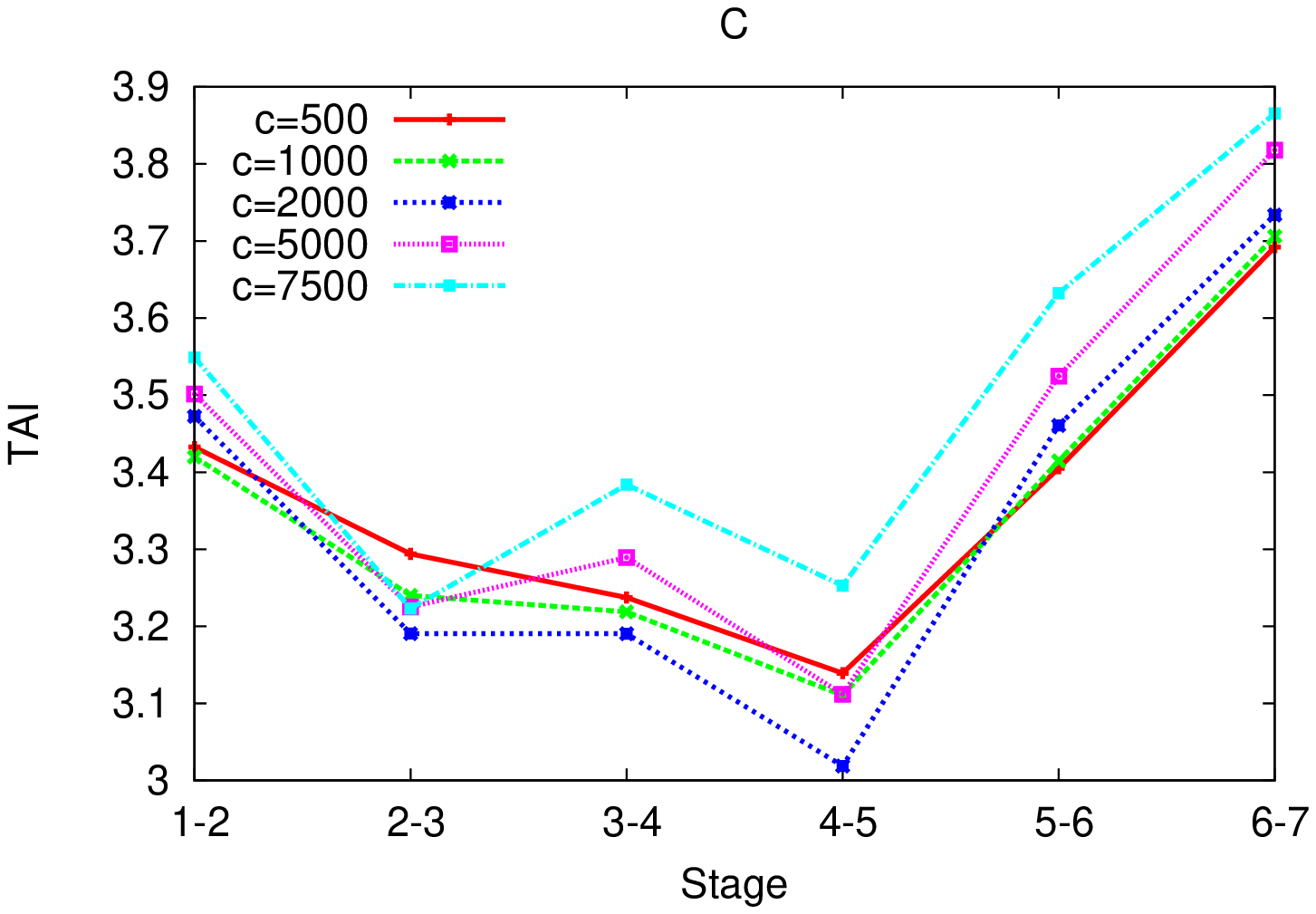}
\includegraphics[width=50mm,height=70mm,angle=270]{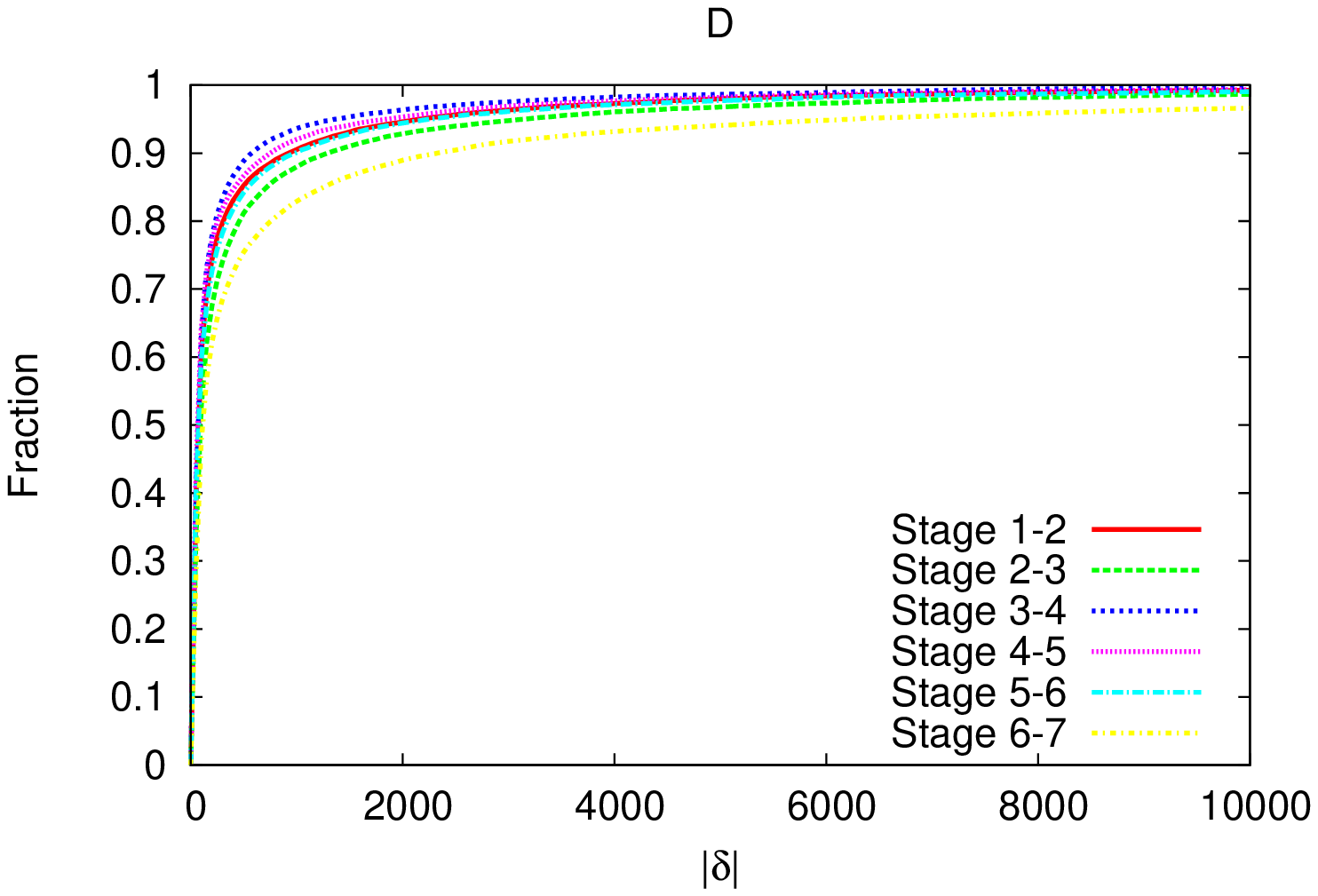}
\includegraphics[width=50mm,height=70mm,angle=270]{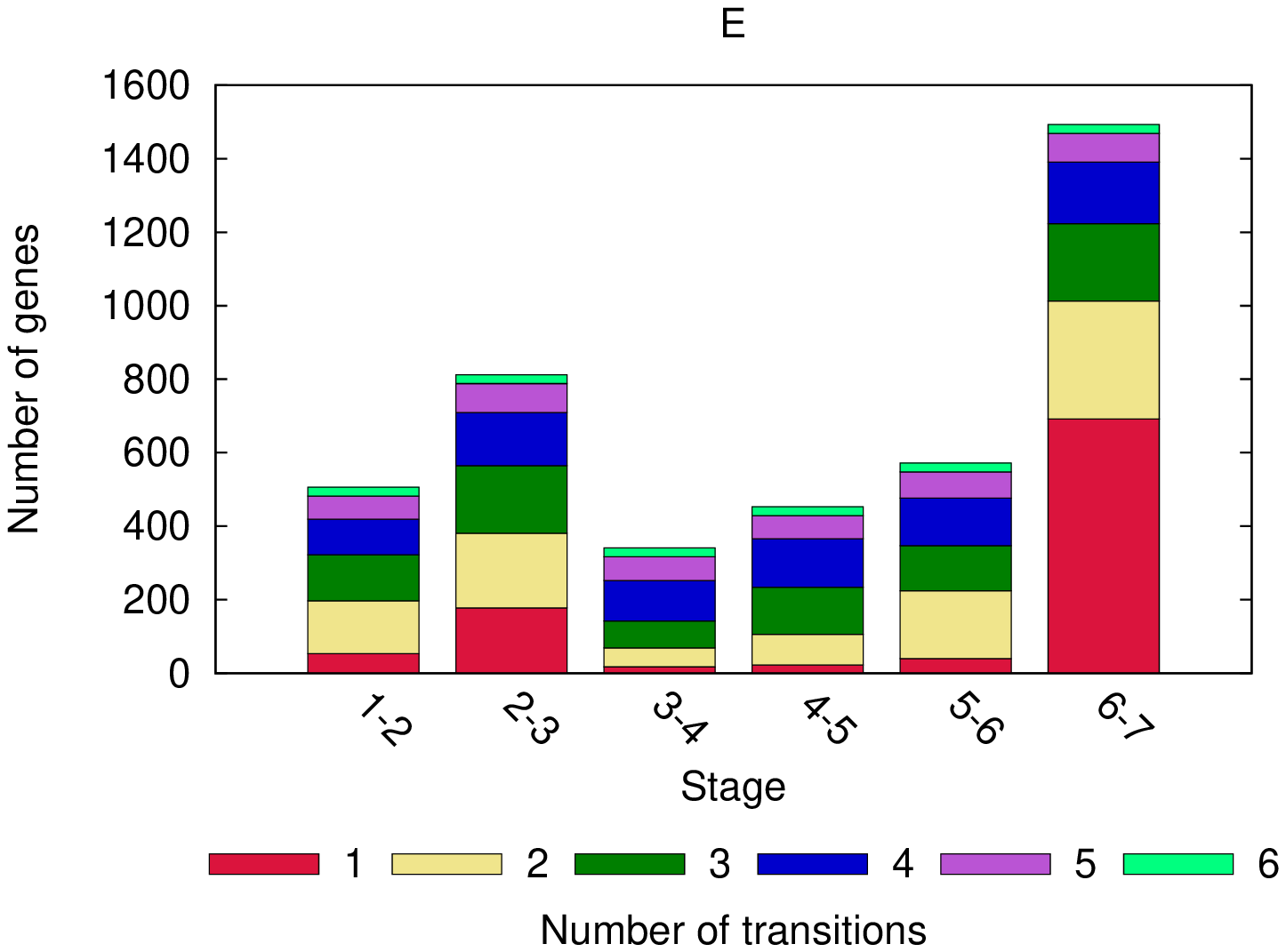}
\vspace{10 pt}
\caption{{\em Arabidopsis thaliana} results  using absolute expression levels.
Graph (A) shows the hourglass score (normal and robust)
as a function of the transition threshold $c$.
Graph (B) shows the location of the hourglass waist (stage-pair)
as a function of the transition threshold $c$.
Graph (C) shows the Transcriptome Age Index of transitioning genes 
for five different values of $c$ (chosen so that the number of genes with known age index assigned to each stage-pair is at least 170).
Graph (D) shows the CDFs of the expression level absolute variations 
$|\delta|$ across successive stage-pairs.
Graph (E) shows the transitioning genes with $c$=5000.
The transitioning genes constitute 8\% of all genes in that dataset.
48\% of those genes transition in a single stage-pair.
Of the remaining, 38\% transition only in consecutive stage-pairs.}
\label{fig:data_analysis_plant2}
\end{figure*}

\begin{figure*}[!htb]
\centering
\includegraphics[width=50mm,height=70mm,angle=270]{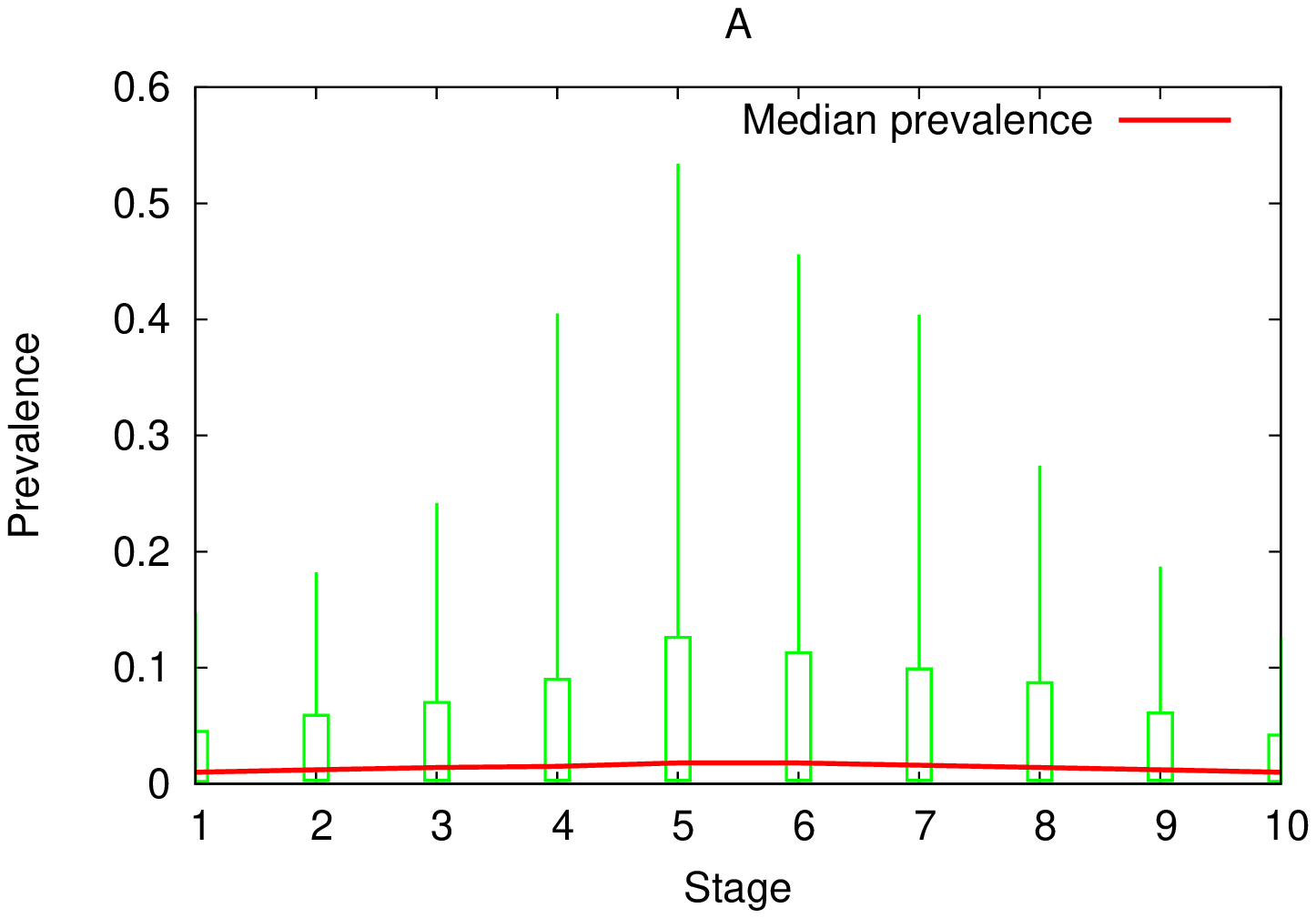}
\includegraphics[width=50mm,height=70mm,angle=270]{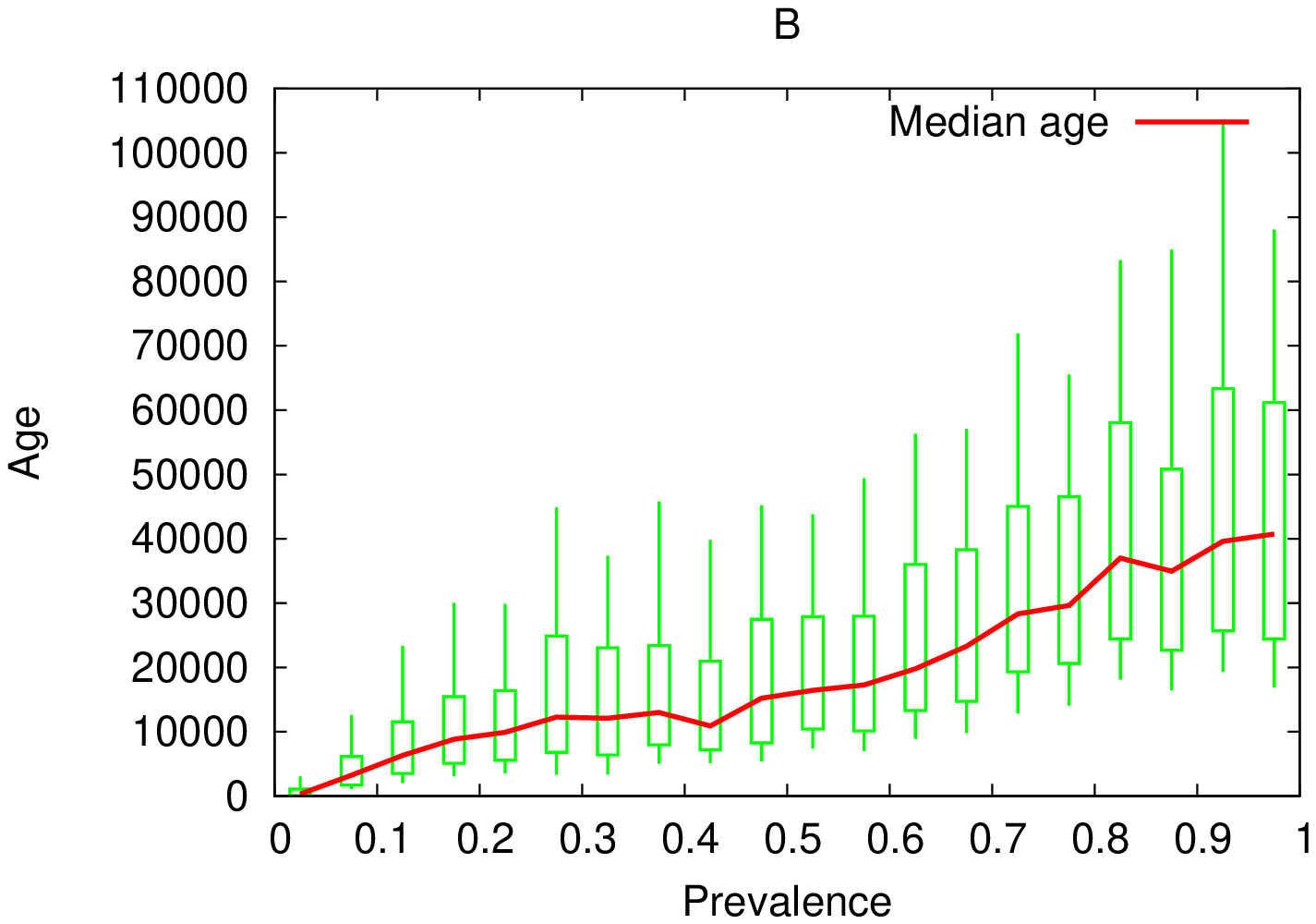}
\caption{The prevalence of a gene $g$ in a population of $N$ individuals is 
the fraction of individuals in which gene $g$ appears. 
These results are obtained using Model-2.
Parameters: 10 runs with different initial populations,
$N$=1000 individuals, $L$=10 stages,
specificity function $s(l)$=$l/L$ ($l$=$1\dots L-1$),
$\Gamma$=100 genes at each stage initially,
RF parameter $z$=4, 500,000 generations,
probability of RW event $P_{RW}$=$10^{-4}$.
The graphs show the median (red lines) and the 10th, 25th, 75th, 
and 90th percentiles (green boxes) for: 
(A) prevalence of genes in each stage after 500,000 generations, 
and (B) gene age as a function of gene prevalence. 
As expected, older genes tend to be more prevalent in the population.}
\label{fig:age_prevalence_results}
\end{figure*}

\begin{figure*}[!htb]
\centering
\includegraphics[width=50mm,height=50mm,angle=270]{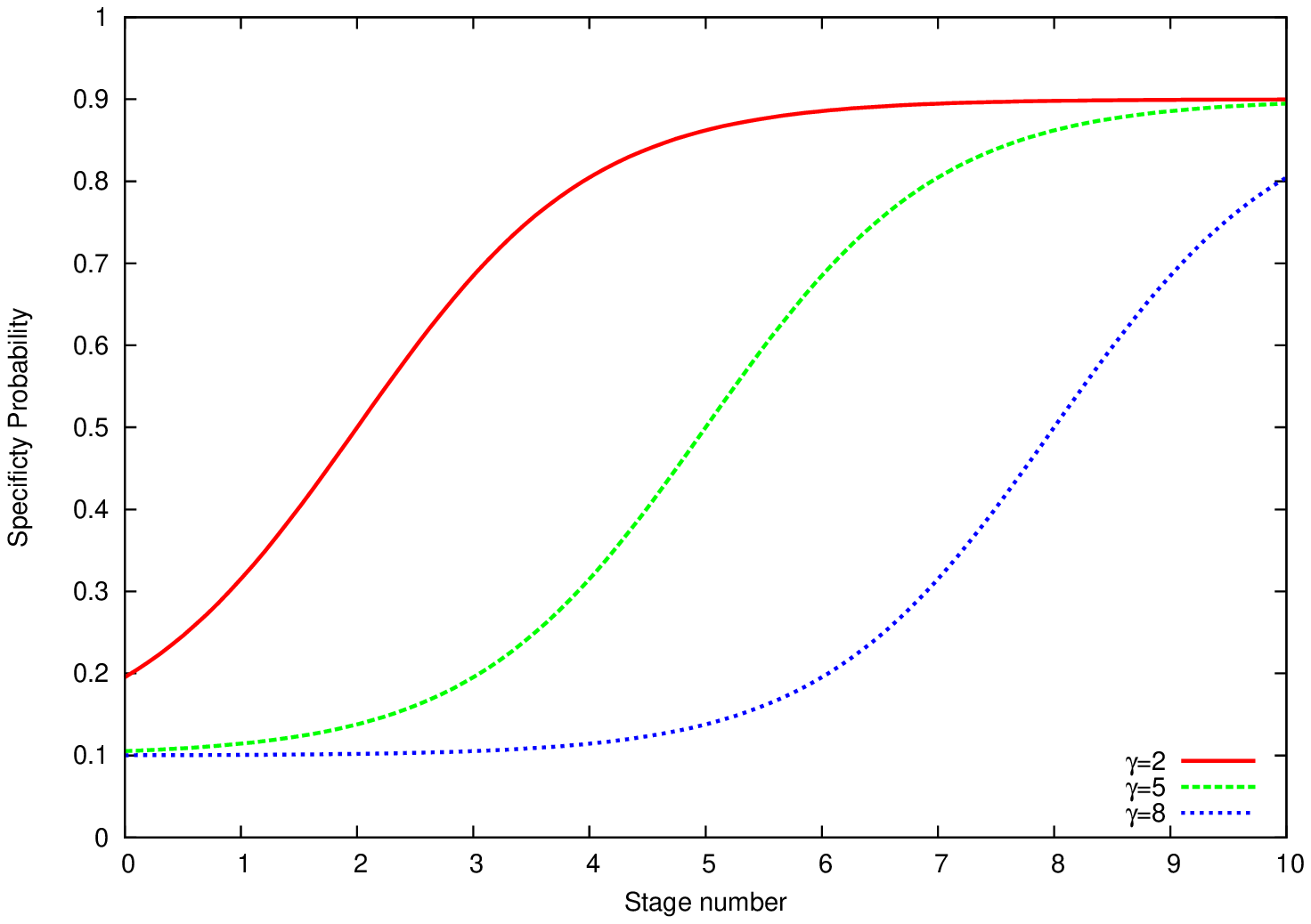}
\caption{A nonlinear specificity function, 
$s(l)=0.9 - \frac{0.8}{1+e^{(\gamma - l)}}$,
for three values of the parameter $\gamma$.
This function allows us to control the stage $\gamma$ at which the
specificity is 50\%.}
\label{fig:sigmoid_specificity}
\end{figure*}

\begin{figure*}[!htb]
\centering
\includegraphics[width=50mm,height=70mm,angle=270]{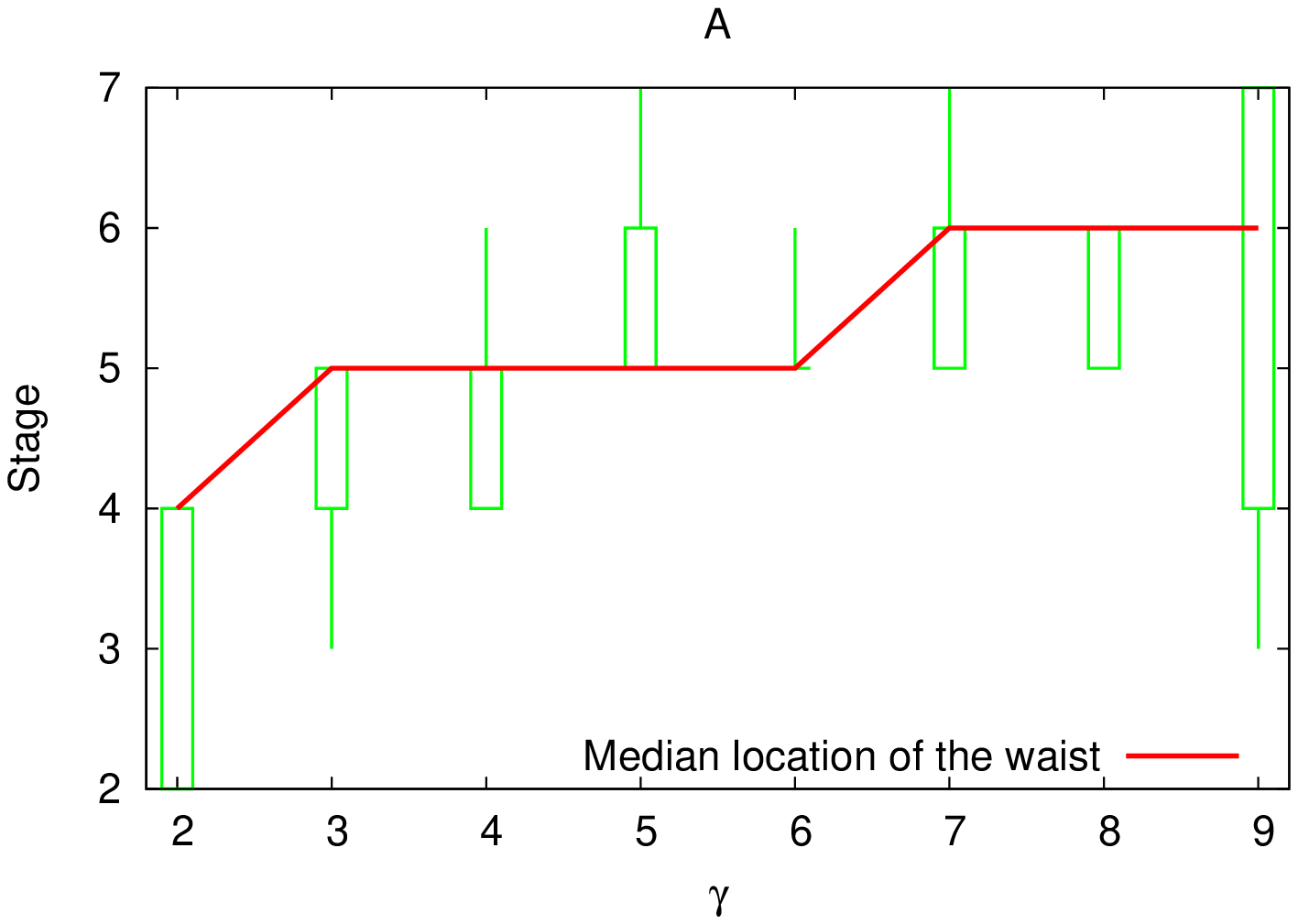}
\includegraphics[width=50mm,height=70mm,angle=270]{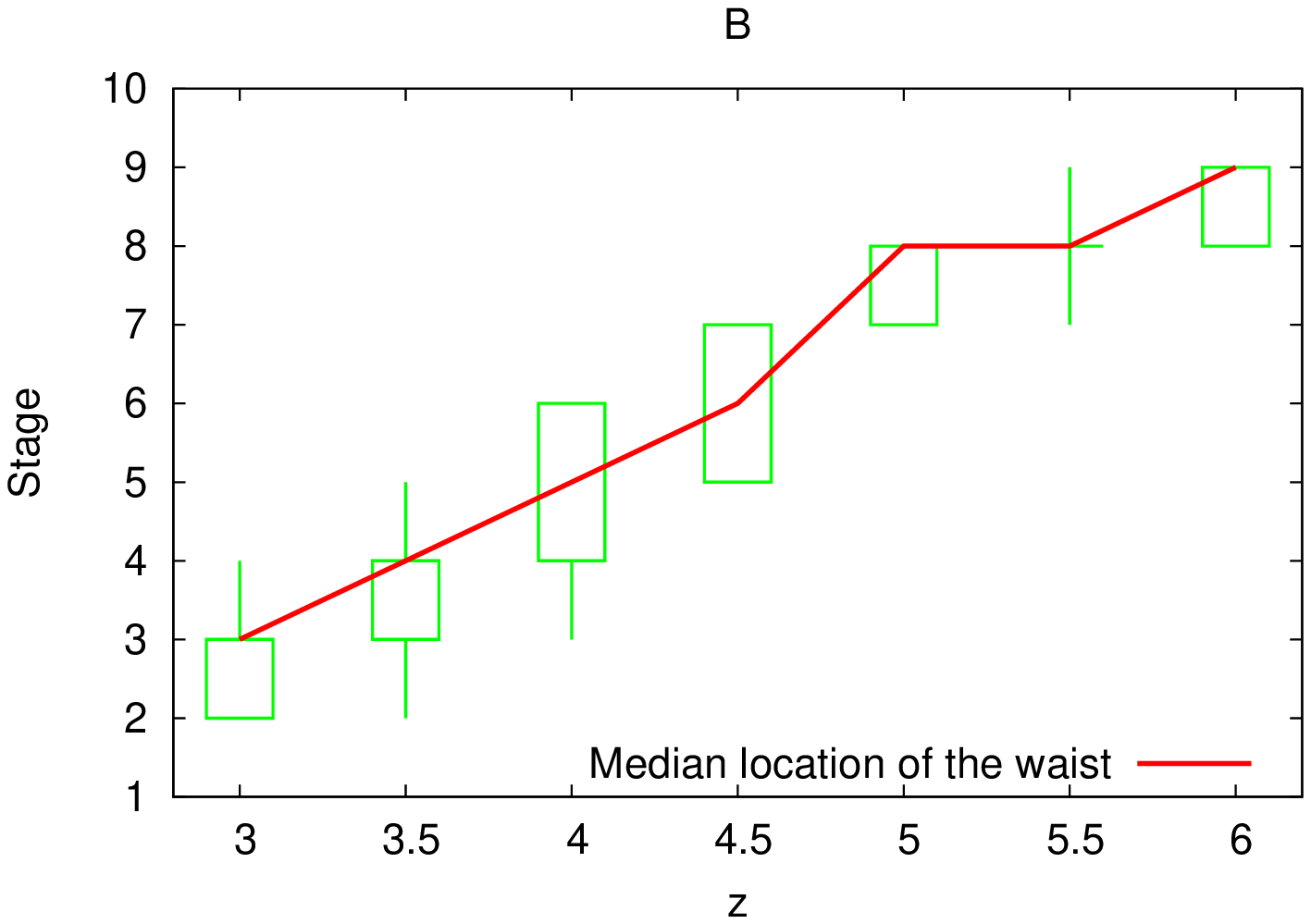}
\caption{We examine the effect of the two model parameters that affect the 
location of the DGEN hourglass waist. 
The first is the specificity function. 
To examine its effect, we use a sigmoid-like mathematical
function that controls the stage $\gamma$ at which the specificity is 50\% 
(see Fig.\ref{fig:sigmoid_specificity}). 
This is the stage with the maximum variance in the number of outgoing 
regulatory edges.
RW events at this stage can cause the largest extent of rewiring 
and so, the highest likelihood of RFs in genes of the next stage.
Graph (A) shows that the location of the hourglass waist is 
``pushed'' towards stage $\gamma$, even though it is not 
always exactly at that stage.
The second way to affect the location of the hourglass waist
is the parameter $z$ that controls the shape of the RF probability. 
Increasing $z$ makes RF events more likely, also increasing the
likelihood of lethal RF cascades. 
Graph (B) shows that as we increase $z$ the hourglass waist 
moves towards later developmental stages. 
These results are obtained using Model-2.
Parameters: 10 runs with different initial populations,
$N$=1000 individuals, $L$=10 stages,
specificity function $s(l)$=$l/L$ ($l$=$1\dots L-1$),
$\Gamma$=100 genes at each stage initially,
RF parameter $z$=4, 500,000 generations,
probability of RW event $P_{RW}$=$10^{-4}$.
The graphs show the median (red lines) and the 10th, 25th, 75th, 
and 90th percentiles (green boxes) of the location of the waist. 
}
\label{fig:waist_location}
\end{figure*}

\begin{figure*}[!htb]
\centering
\includegraphics[width=130mm,height=80mm,angle=0]{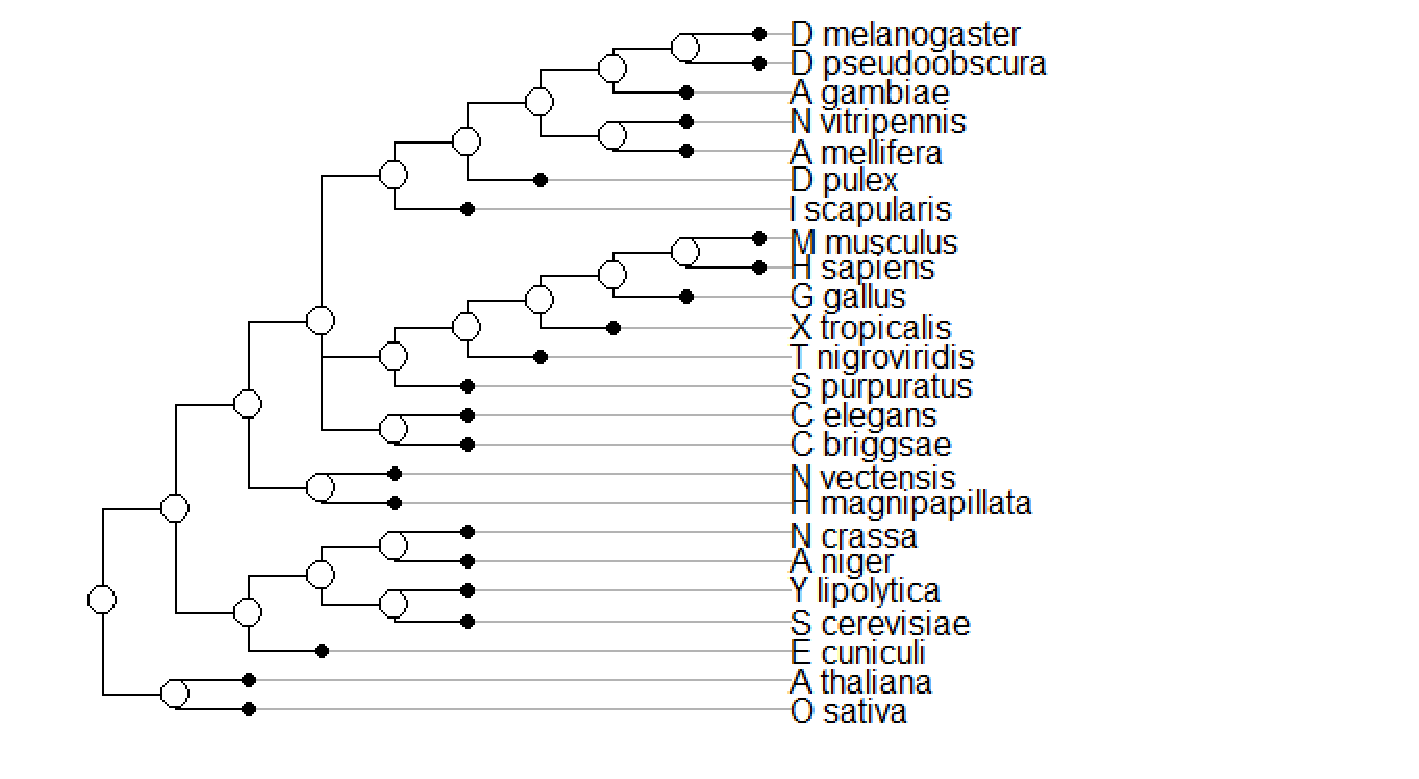}
\caption{The phylotypic tree that we use to calculate the age index of {\em Drosophila}'s genes. Each gene is assigned to one of the following six ages:
Level-1: Common ancestor to Fungi, Plants and Eumetazoa.
Level-2: Common ancestor to Fungi and Eumetazoa.
Level-3: Common ancestor to all Eumetazoa.
Level-4: Common ancestor to all Bilateria.
Level-5: Common ancestor to all Arthropoda.
Level-6: Common ancestor to all Dipteria.}
\label{fig:phylo_tree}
\end{figure*}

\end{document}